\documentclass[a4paper,fleqn,usenatbib]{mnras}

\usepackage{newtxtext,newtxmath}

\usepackage[T1]{fontenc}
\usepackage{ae,aecompl}

\usepackage{graphicx}	
\usepackage[dvipsnames]{xcolor}

\def\clap#1{\hbox to 0pt{\hss#1\hss}}
\def\fdeg{\hbox{$.\!\!^\circ$}}            
\def\farcm{\hbox{$.\mkern-4mu^\prime$}}    
\def\farcs{\hbox{$.\!\!^{\prime\prime}$}}  

\title[NESS Catalogue paper]{The Nearby Evolved Stars Survey (NESS) V: properties of volume-limited samples of Galactic evolved stars}

\author[I. McDonald et al.]{I. McDonald,$^{1}$\thanks{E-mail: iain.mcdonald-2@manchester.ac.uk} 
S. Srinivasan,$^{2}$
P. Scicluna,$^{3}$
O.~C. Jones,$^{4}$
A.~A. Zijlstra,$^{1}$
\newauthor
S.~H.~J. Wallstr\"om,$^{5}$
T.~Danilovich,$^{6,7,5}$
J.~H. He,$^{8,9,10}$
J.~P. Marshall,$^{11}$
\newauthor
J.~Th.~van Loon,$^{12}$
R. Wesson,$^{13}$
F. Kemper,$^{14,15,16}$
A. Trejo-Cruz,$^{2}$
J. Greaves,$^{17}$
\newauthor
T. Dharmawardena,$^{18,19,20}$
J. Cami,$^{21,22,23}$
Hyosun Kim,$^{24}$ 
K.~E.~Kraemer,$^{25}$
\newauthor 
C.~J.~R. Clark,$^{26}$
H. Shinnaga,$^{27,28}$
C. Haswell,$^{29}$
H. Imai,$^{28,30}$
J.~G.~A. Wouterloot,$^{31}$
\newauthor 
A.~J. P\'erez Vidal,$^{32}$
G. Rau, $^{33,34}$
and the NESS collaboration
\\
Author affiliations are listed after the references.
\\
}
\date{Accepted XXX. Received YYY; in original form ZZZ}

\pubyear{2024}

\begin{document}
\label{firstpage}
\pagerange{\pageref{firstpage}--\pageref{lastpage}}
\maketitle
\begin{abstract}
We provide a meta-study of the statistical and individual properties of two volume-complete sets of evolved stars in the Solar Neighbourhood: (1) 852 stars from the Nearby Evolved Stars Survey (NESS), and (2) a partially overlapping set of 507 evolved stars within 300 pc. We also investigate distance determinations to these stars, their luminosity functions and their spatial distribution. \emph{Gaia} {\sc apsis} GSP-Phot {\sc aeneas} temperatures of bright giant stars often appear to be underestimated. Existing literature on AGB stars under-samples both the most and least extreme nearby dust-producing stars. We reproduce the literature star-formation history of the solar neighbourhood, though stellar-evolution models over-predict the number of AGB stars of ages around 500 Myr. The distribution of AGB stars broadly matches the known 300 pc scale height of the Galactic disc and shows concentration in the direction of the Galactic centre. Most dust-producing carbon stars belong to the Galactic thick-disc population. 
\end{abstract}




\section{Introduction}

Asymptotic giant branch (AGB) and red supergiant (RSG) stars are the end states of stars between $\sim$0.8--8 M$_\odot$ and $\sim$8--20 M$_\odot$, respectively. Their mass loss dominates the chemical enrichment of today's Universe as, along with supernovae, they recycle nuclear-processed material back into the interstellar medium (ISM; e.g., \citealt{HofnerOlofsson18}). AGB stars notably enrich He, C, and $s$-process elements \citep[e.g][]{KarakasLattanzio14} and are important sources of interstellar dust (particularly carbon-rich dust); while RSGs especially enrich O, N and $\alpha$-elements below the iron peak. AGB stars also contribute the bulk of a galaxy's infrared light \citep{Maraston+06}. Both AGB and RSG stars are governed by a complex set of interacting physical mechanisms, which makes modelling their evolution and chemical yields challenging. Yet this evolution also determines which stars will undergo supernovae and the set of compact objects that will result. The large physical size of AGB and RSG stars means that binary interactions can become significant, leading to an array of stellar--stellar \citep[e.g.][]{Jorissen2016}, stellar--planetary \citep[e.g.][]{Decin2020} and stellar--compact-object interactions \citep[e.g.][]{Iaconi2017}, and helping determine the mass functions of gravitational-wave sources \citep[e.g.][]{Newton2018}.

Most AGB and RSG stars more luminous than the red-giant-branch (RGB) tip have mass-loss rates of $\dot{M} \approx 10^{-8}$ to $10^{-5}\,{\rm M}_\odot\,{\rm yr}^{-1}$, which exceeds their nuclear-burning rates (hydrogen burning consumes $\approx 10^{-8}$ M$_\odot$ yr$^{-1}$ per 1000\,L$_\odot$ of radiation). Consequently, mass loss controls their evolutionary path \citep{vLGdK+99}. This mass loss arises from a pulsation-enhanced, dust-driven wind \citep[e.g.][]{HofnerOlofsson18}: pulsations levitate material above the photosphere, allowing dust to condense; radiation pressure on this dust drives it from the star. Pulsations appear to dictate whether the star loses mass via such a dusty wind \citep{McDonald18}, but the relationship to overall dust opacity is more complex \citep{MT19}.

At roughly the same time as the dusty wind starts, the third dredge-up phase (3DU) begins: the degenerate helium shell periodically ignites in a thermal pulse, creating convective mixing that brings nuclear-processed material to the surface \citep[e.g.][]{Herwig05}. The thermally pulsating AGB (TP-AGB) begins when stars are slightly more luminous than the RGB tip (brighter still for (super-)\linebreak[2]{}AGB stars up to $\sim$9 M$_\odot$; e.g., \citealt{Bressan12}). Dredged-up carbon can change a star's initially oxygen-dominated chemistry to become carbon-rich, if it achieves C/O\,$>$\,1 by number. This radically changes the dust chemistry and its opacity, changing the properties of mass loss. Thus, the combination of the poorly defined dredge-up efficiency and mass-loss rate are fundamental missing ingredients in our understanding of stellar evolution and the cosmic cycle of matter \citep[e.g.,][]{Iben83,KarakasLattanzio14}.

Progress requires calibration between models \citep[e.g.][]{Bladh19} and observations of stellar winds, but this relies on proxy measures. Mass-loss rates ($\dot{M}$) are best measured at (sub-)\linebreak[2]{}mm wavelengths, using the rotational transitions of the CO molecule \citep{Solomon71}: integrated line intensity is related to $\dot{M}$ \citep[e.g.][]{Loup1993,Knapp1998}; line width is related to the terminal velocity of the stellar wind ($v_\infty$), which can probe momentum-transfer mechanisms \citep[e.g.][]{GVM+16}. The dust-production rate ($\dot{D}$) is determined from mid-infrared spectra and photometry, (assuming a dust mineralogy and grain properties), allowing a gas-to-dust ratio ($\dot{M}/\dot{D}$) to be determined: this probes the dust-condensation efficiency of the wind \citep{GvLZ+17}. Sampling many AGB/RSG stars can trace both the typical evolution of a star of measured properties (e.g., mass), and the range in the wind properties resulting from unobservable properties (magnetic fields, rotation rate, companions, etc.).

CO-line surveys have attempted to probe these relations \citep[e.g.][]{Danilovich15}. However, survey targets have normally been ``cherry-picked'' from a list of well-observed stars, whose properties do not necessarily reflect those of the general population. In contrast, the Nearby Evolved Stars Survey \citep[NESS,][hereafter the NESS Overview]{Scicluna22} is designed to systematically sample nearby stars at different stages of mass loss and evolution.

This paper has three goals that together allow the NESS survey to advance statistical understanding of AGB stars in our Solar Neighbourhood and the wider Galaxy:
\begin{enumerate}
    \item Creating a catalogue of photometry and fundamental parameters for NESS survey stars, and performing a detailed examination of the sample to identify sources that should be rejected or require additional consideration (Section \ref{sec:results}).
    \item Creating a comparison catalogue of evolved stars within 300 pc of the Sun, including a literature search of their properties. \emph{Gaia} Data Release 3 (DR3) \citep{GaiaDR3} has improved distances to many evolved stars, allowing creation of a new, fuller list of evolved stars in the Solar Neighbourhood, which is both larger and more extensive than was available at the inception of the NESS survey.
    \item Combining these catalogues to (a) provide fundamental stellar parameters for nearby AGB and RSG stars (Sections \ref{sec:methods} \& \ref{sec:disc}), (b) re-derive the volume-completeness of the lower mass-loss-rate tiers of the NESS sample and (c) understand both the aggregate and typical properties of evolved stars in the Solar Neighbourhood and wider Galaxy, and their physical distribution (Section \ref{sec:disc2}).
\end{enumerate}


\section{Physical considerations}
\label{sec:general}

\subsection{Defining an evolved star}
\label{apx:more:params}

Stellar samples require careful semantic definition. In this paper, ``evolved star'' refers to any post-main-sequence star with a luminosity of $700 \leq L \leq 200\,000$\,L$_\odot$ and a temperature of $T_{\rm eff} < 5000$\,K. The lower luminosity criterion includes the faintest AGB stars with clear dust production \citep{MZW17}. This includes brighter RGB stars (the RGB tip lies at around $L \sim 2500$\,L$_\odot$ at Galactic metallicity). RGB and AGB stars are normally observationally inseparable, and there may be little physical difference in the characteristics of their stellar surfaces for the same stellar parameters. Despite this, RGB stars appear not to produce dusty winds \citep{BvLM+10}, and it is an unresolved question how mass loss from upper RGB stars differs from AGB stars of similar luminosity.

Similarly, our luminosity range includes not only AGB, but ``super-AGB'' and RSG stars. Accurate separation requires knowing the future of an individual star's stellar evolution. A nominal limit of $M_{\rm bol} = -7.1$\,mag, or $L \approx 55\,000$\,L$_\odot$ is used as the classical AGB limit \citep{Paczynski1971}. The upper luminosity limit of 200\,000\,L$_\odot$ is a soft boundary, beyond which we have considered the luminosity of any object suspicious. \citet{Davies2020} identifies a Galactic limit of $L \approx 158\,000^{+76\,000}_{-35\,000}$\,L$_\odot$, though \citet{deWit2023} finds $L \approx 300\,000$\,L$_\odot$ in the Magellanic Clouds. However, computed luminosities this high are more likely to be erroneous if distances are poorly determined or data poorly fit.

The $T_{\rm eff} < 5000$\,K limit conservatively includes all RGB stars with $L > 700$\,L$_\odot$. Our limits form a box on the Hertzsprung--Russell (H--R) diagram: Padova models \citep{Marigo08,Nguyen2022}\footnote{\url{http://stev.oapd.inaf.it/cgi-bin/cmd_3.8}} indicate solar-metallicity stars with $M \lesssim 4$\,M$_\odot$ enter the box from the bottom (luminosity floor); stars of $4 \lesssim M \lesssim 6$\,M$_\odot$ enter the box from the hotter side, but move out of the box temporarily during their blue loops; while stars of $M \gtrsim 6$\,M$_\odot$ enter the box after crossing the Hertzsprung gap. All RSGs up to $M \sim 20$\,M$_\odot$ should enter this box during their final evolution.

Our limits therefore exclude higher-mass blue and yellow supergiants and Wolf--Rayet stars. We also exclude post-AGB stars (except those in their earliest phases), and central stars of planetary nebulae, which we will refer to separately as ``highly evolved stars''. While the NESS survey includes some such stars, these stars have fundamental differences in mass loss or dust production from the AGB/RSG stars that form the bulk of the NESS survey and stretch the conventional definition of an ``evolved star''. More importantly, the NESS survey does not contain a complete sample of such objects, so we intentionally reject them from this work.

\subsection{Photometric versus spectroscopic temperature}
\label{apx:more:temp}

In normal stellar spectroscopy, the surface temperature of the star is theoretically easy to determine, since the photosphere ($\tau = 1$ layer) is thin, close-to-spherical and largely invariant with wavelength. In this case, photometric colour temperatures will be consistent across the spectrum. Fitting a spectral energy distribution (SED) is effectively simultaneous fitting of many photometric colour temperatures, and gives a single photometric temperature for the star ($T_{\rm phot}$). This photometric temperature should agree with the star's spectroscopic temperature, as derived from the relative depths of atomic and/or molecular lines ($T_{\rm spec}$).

However, as stars evolve and reduce in surface gravity, the atmospheric scale height expands and the $\tau = 1$ layer grows by orders of magnitude, and the surface begins to become less defined \citep[e.g.][]{Hoefner2022}. Large convective cells and surface pulsations lead to temperature gradients on the stellar surface and departure from local thermodynamic equilibrium (LTE). Spectra increasingly depart from stellar atmosphere models, causing progressively larger errors in spectroscopic temperatures, as a static atmospheric model begins to fail to reproduce the stellar spectrum, especially in high-resolution spectroscopy. Subsequent mass loss leads to blanketing of the star by dust, introducing a wavelength-dependent opacity layer that scatters light. The $\tau = 1$ layer therefore becomes wavelength dependent and can expand in parts of the optical and infrared regime by an order of magnitude above the region probed by spectroscopic lines. This leads to departures of the overall SED from the stellar atmosphere model as optical light is reradiated into the infrared \citep[e.g][]{FLH17}. As a consequence, the photometric temperature from the SED and the spectroscopic temperature become progressively further dissociated throughout this evolutionary process, with the photometric temperature of a dust-enshrouded AGB star being up to $\sim\sqrt{10}$ times lower than the spectroscopic temperature.

Once these two temperatures diverge, neither truly reflects the overall star, which lacks a defined surface, and neither can be used to accurately calculate a luminosity. Instead, we must rely upon one of two coarse approximations. The first is to fit a blackbody to the SED, then use this blackbody to calculate a luminosity. This provides some sort of representative temperature when stellar atmosphere models fail to, and can work effectively on SEDs that are poorly sampled and/or with noisy photometry (most dust-enshrouded stars are very-high-amplitude variables). However, a blackbody only provides a good representation to stars that are either lightly obscured by dust, or completely dust-enshrouded. Some AGB stars, particularly those with winds shaped by companions into a disc \citep[cf.][]{Decin2020}, present two-component SEDs, with the AGB star (and/or sometimes its companion) contributing a peak in the optical or near-infrared, and a mid-infrared peak from the circumstellar dust. This is particularly common in post-AGB stars. In these cases, the complex shape of the SED is hard to fit and empirical integration becomes the only option to determine the stellar luminosity. This empirical integration may miss important features of the SED that occur between sampled photometric bands, and requires some assumptions on the underlying spectrum to determine the correct colour-correction and dereddening for each filter.

This paper is, first and foremost, based on information derived from photometric data. Consequently, we report the photometric temperature as default where possible, and report spectroscopic temperatures from the literature as a comparison dataset. The decision of which method is used to determine temperature and luminosity from the photometric data is detailed in Section \ref{sec:disc:merge}.


\section{Methods} 
\label{sec:methods}

This section describes the cross-matching, fitting and parameter extraction for two datasets: the NESS survey and a comparison sample of evolved stars within 300 pc of the Sun. The large number of datasets used means this is a lengthy and technically detailed process, which we devolve to Appendices \ref{apx:more} \& \ref{apx:sources}, retaining here only a summarised version containing factors directly relevant to the scientific results. The reader is specifically directed to Tables \ref{tab:photosources}--\ref{tab:tempsources} for a full list of acronyms and references for the data sources used, and to Appendix \ref{apx:data} for the complete set of results in machine-readable formats.

Some of the NESS sources are not evolved stars but were accidentally included in the survey, and some of the survey sources are too evolved or too massive to meet the evolved-star criteria we invoke here, which are meant to identify AGB/RSG stars. These cases are discussed further in Section \ref{sec:results}. The extended survey was better vetted for sources that were known not to be evolved stars before observation, but we revise the completeness of both the original and extended surveys in Section \ref{sec:disc}.

\subsection{Obtaining \emph{Gaia} DR3 counterparts for NESS sources} 
\label{sec:xmatch}

\begin{table*}
    \centering
    \caption{Tier structures in the NESS survey, with limitations in dust mass-loss rate ($\dot{D}$), distance ($d$) and Galactic latitude ($b$). Star counts include sources that are not evolved stars, detailed in Section \ref{sec:results}.}
    \label{tab:NESS}
    \begin{tabular}{c@{\ }c@{\ }c@{\quad}r@{\ }r@{\ }r@{}}
    \hline
    Tier& \qquad Descriptor & \qquad No. of & Range in $\dot{D}$ & \qquad Distance limit\qquad &  Excluded\\
    \ & \ & sources & (M$_\odot$ yr$^{-1}$) & (pc) & sources\\
    \hline 
    0 & Very low &  19 &  Any  & $<250$ & \qquad $L \leq 1600$\,L$_\odot$, $\delta \geq -30^\circ$ \\
    1 & Low      & 105 &  $<1\times10^{-10}$  & $<300$ & \qquad Stars without 3-$\sigma$ dust excess \\
    2 & Intermediate & 222 & $1\times10^{-10}$ to $3\times10^{-9}$  & $<$600  &  $\left|b\right|<1.5$ for $d>400$\,pc \\
    3 & High     & 324 & $3\times10^{-9}$ to $1\times10^{-7}$  & $<$1200 &  $\left|b\right|<1.5$ for $d>800$\,pc \\
    4 & Extreme  & 182 & $\geq 1\times10^{-7}$  & $<$3000 &   $\left|b\right|<1.5$ for $d>2000$\,pc \\
    \hline
    \end{tabular}
\end{table*}

The complete\footnote{The survey comprises an original survey and an extension, both of which are presented in the NESS Overview paper.} NESS sample consists of five tiers containing 852 sources, summarised in Table \ref{tab:NESS}. The tiers were defined in the NESS Overview paper to be volume-limited samples, complete to a specified dust-mass-loss rate (estimated from {\sc grams} model fits; \citealt{Srinivasan11}) and distance. Sample selection for Tiers 0 and 1 were partly based on the analysis of \citet{MZB12}, which takes distances from the \emph{Hipparcos} catalogue \citep{vanLeeuwen07}; and on \citet{MZW17}, which takes distances from the \emph{Hipparcos--} and \emph{Tycho--Gaia} Astrometric Solutions \citep{MLH15}. Sources in Tiers 2--4 had their distances estimated based on their bolometric luminosity, and excluded regions within 1.5$^\circ$ of the Galactic Plane, due to the potential for source confusion and background contamination.

The NESS survey uses the \emph{IRAS} Point Source Catalogue (PSC; \citealt{IRAS}) as a basis for both its creation and observation. This has poor astrometric resolution, and cross-matching against an optical catalogue with proper-motion data is needed to ensure accurate retrieval of sources in other catalogues. We therefore cross-matched the NESS sample's \emph{IRAS} identifiers to \emph{Gaia} DR3 sources (or alternative optical or near-infrared sources where no \emph{Gaia} DR3 source exists). The differing beam sizes, high proper motion, crowding and obscuration of some sources meant that this was a non-trivial affair that could only be conducted in a semi-automated fashion with significant manual input and checking. Full details are given in Appendix \ref{sec:xmatch:togaia}.

\subsection{A complete sample of nearby evolved stars from \emph{Gaia} DR3}
\label{sec:300pc}

While NESS provides a nominally volume-limited sample of dust-producing evolved stars, a local sample of evolved stars allows us to examine stars that are not producing dust, therefore establishing both the statistical properties of local AGB stars overall, and defining which stars do produce dust. Comparing the two samples allows us to determine how representative local evolved stars are of the wider Galactic population.

We chose to create a list of evolved stars within 300\,pc of the Sun, comprising of:
\begin{itemize}
    \item 1616 \emph{Gaia} DR3 objects with $\varpi \geq 2.5$\,mas, $B_{\rm P} - R_{\rm P} > 1.5$\,mag, $M_{\rm Rp} < -1$\,mag and with either distances in \citet{BJRF+21} of $d < 300$\,pc or (if no distance is listed) $\varpi \geq 3.333$\,mas.
    \item 539 \emph{Hipparcos} stars \citep{vanLeeuwen07} with any one of the following criteria:
    \begin{itemize}
        \item $\varpi \geq 3.333$\,mas, $B_{\rm T} - R_{\rm T} > 1$\,mag and $M_{R_{\rm T}} < -1$\,mag;
        \item $\varpi \geq 3.333$\,mas, and $T_{\rm eff} < 5000$\,K and $L>350$\,L$_\odot$ in \citet{MZW17};
        \item \emph{Gaia} stars with $B_{\rm P} - R_{\rm P} > 1.5$\,mag and no \emph{Gaia} parallax, but with a \emph{Hipparcos} parallax of $\varpi \geq 3.333$\,mas and an inferred $M_{\rm Rp} < -1$\,mag;
        \item the star 5 Psc.
    \end{itemize}
    \item Nine stars not meeting the above criteria but with distances in \citet{ARVDB22} of $<$300 pc.
    \item CW Leo.
    \item IK Tau.
\end{itemize}
Removing duplicates from this dataset left a list of 1880 stars potentially meeting our evolved-star criteria. Details on how this process was performed and the reasoning behind our choice of values in the above list can be found in Appendix \ref{apx:300pc}.

\subsection{A common data reduction framework}
\label{sec:reduce}

To fit the dataset and extract stellar parameters, we employ version 1.1 of the Python Stellar Spectral Energy Distribution (PySSED) routines \citep{McDonald2024}\footnote{Application: \url{https://explore-platform.eu/}}$^,$\footnote{Code and input files used in this work: \url{https://github.com/iain-mcdonald/PySSED}}. In short, {\sc PySSED} will extract and prioritise photometry and ancillary information from pre-selected data sources, automatically reject bad and poorly fitting photometry and, using an appropriate distance and extinction, fit a stellar atmosphere model (in this case, a {\sc bt-settl} model; \citet{AGL+03}) to extract fundamental parameters including temperature and luminosity. {\sc PySSED} takes its filter information from the Spanish Virtual Observatory's Filter Profile Service\footnote{\url{http://svo2.cab.inta-csic.es/theory/fps/}} and, from this information, derives a comparison flux in each filter for each model in the {\sc bt-settl} grid. {\sc PySSED} is run identically for the 300 pc and NESS samples. Full details on the data sources are included in Appendix \ref{apx:sources} for reproducibility.

{\sc PySSED}'s default 3D extinction map \citep{Vergely2022} was used to deredden our collected photometry for interstellar extinction. No attempt is made to account for circumstellar extinction towards the star (see Section \ref{apx:more:temp}).

Attempts to use spectroscopic temperatures as prior constraints in the fit were made, but these were found to poorly represent too many stars, either because the fits differed too much from the SED or because the temperatures themselves were too inaccurate. Consequently, we collect spectroscopic temperatures and report them as ancillary data, but do not use them in our fitting procedure.

High mass-loss rate stars remain poorly fit by stellar atmosphere models: the most obscured stars can sometimes be reasonably well fit with a blackbody to obtain a representative temperature, but most require a different treatment to obtain luminosity. Instead, to ensure the best recovery of stellar properties, we run {\sc PySSED} three times. The first fits a temperature to the interstellar-extinction-corrected SED using the {\sc bt-settl} model atmospheres\footnote{At temperatures below the limits of the {\sc bt-settl} model grid (2000--2300 K, depending on the $\log g$--[Fe/H]--[$\alpha$/Fe] combination), {\sc PySSED} automatically reverts to a blackbody to estimate stellar temperature.}. The second run (performed for the NESS sample only) fits a temperature using a blackbody. The third run (also performed for the NESS sample only) uses trapezoidal integration to produce a luminosity for the star. These will be later combined, based on the relative ability of the model and blackbody to fit the SED (see Section \ref{sec:disc:merge}).

\subsection{Treatment of distances} 
\label{sec:xmatch:distances:paper}

\begin{figure}
\centering
\includegraphics[width=\linewidth]{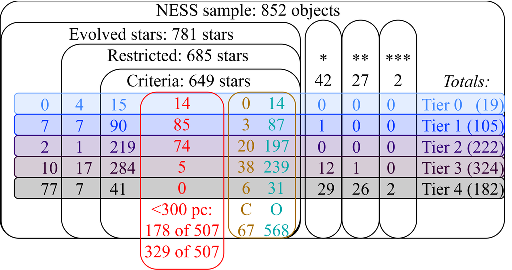}
\caption{Venn diagram of the NESS and 300 pc datasets, separated by tier, identifying stars rejected because they are $^{(\ast)}$not evolved stars, $^{(\ast\ast)}$highly evolved stars, or $^{(\ast\ast\ast)}$sources with unclear classifications, but which are probably not evolved stars. The full NESS dataset of 781 evolved stars is subdivided into 685 stars with valid distances in the ``restricted dataset'' (Section \ref{sec:reduce}), of which 649 stars fit within our evolved-star criteria of $T<5000$\,K and $700 < L < 200\,000$\,L$_\odot$ (Section \ref{sec:disc:merge}). Of these 649, 178 overlap with the 507 stars in the 300 pc comparison sample. The number of carbon- and oxygen-rich stars among these 649 is also given.}
\label{fig:venn}
\end{figure}

Distances remain the primary factor limiting precision measurement of properties of Galactic AGB stars. Parallaxes of evolved stars suffer optical obscuration and variability. Despite their intrinsic brightness, dust obscuration means stars can be very faint in optical surveys, and optical parallaxes can be noisy or missing. For example, optical emission from IRC\,+10216 is dominated not by the star itself, but by light reflected from its dust, hence it is no longer optically a point source \citep[e.g.][]{Kim2021}. \emph{Gaia} DR3 decomposes the star into two different sources (614377930478412032 and 614377930478412544), while the epoch photometry records its variability between 16th and 18th magnitude. Neither source is assigned a parallax.

Surface convection moves stars' photocentric positions, adding an inherent astrometric noise component on timescales of a few months to years, often close to the annual timescale of the parallax signal \citep[e.g.][]{Chiavassa2018}. The parallaxes obtained by \emph{Hipparcos} are therefore too inaccurate for many pulsating AGB stars, while those from \emph{Gaia} DR3 have too short a baseline to reduce these noise components.

\subsubsection{Edge cases}
\label{apx:dist:edge}

Our sample therefore contains edge cases, and some stars may enter or leave our samples as future data changes their fitted distances, luminosities or temperatures. Not all edge cases can reliably be identified.

For the 300 pc sample, we can consider edge cases resulting from the \emph{Gaia} DR3 parallax alone. Of the 1589 stars with $\varpi > 3.333$\,mas, 58 have a parallax with a 1$\sigma$ uncertainty that extends across the 300 pc boundary, as do 54 of the 1941 stars with $2.5 < \varpi < 3.333$\,mas. At 2$\sigma$, 123 and 122 stars cross the 300 pc boundary in each respective direction. Estimating the number of stars at $<n\sigma$ from the boundary to be $\sim 122.5\,n$, $\sim$50 stars may ultimately be placed on the wrong side of the 300 pc boundary, representing roughly a 3 per cent change in the final sample. However, these edge cases will preferentially be fainter stars, close to the luminosity boundary, so the change in the sample that results as stars are moved across this boundary will be larger, also probably by a few per cent. Changes to temperature and luminosity will also occur due to differences in the assumed interstellar reddening: within 300 pc these changes will be negligible, but they become important on scales of the wider NESS sample.

Dealing with these edge cases would add significant complexity to the analysis and, by their nature, it is impossible to judge at present whether they should belong in the sample. Consequently, the stars we list in this sample include only those where the best-estimate distance is within the boundary of the survey and do not take distance uncertainties fully into account.

\subsubsection{Parallax zero-point and statistical bias corrections}
\label{apx:dist:zpt}

The final distance estimates therefore include some potential biases, notably including those of \citet{Malmquist22} and \citet{LK73}. Due to the complex combinations of distances used, we do not attempt to quantify or account for these biases directly, but instead demonstrate their effects in Section \ref{sec:disc}.

The parallax zero-point of \emph{Gaia} DR3 also needs corrections for small distortions, which \citet{LBB+21} describe in terms of stellar colour, magnitude and sky position. This correction works tolerably well for red giants up to $B_P - R_P \approx 3.0$ mag, but is increasingly poorly defined for redder stars. It is also poorly defined for bright stars ($G \lesssim 6$ mag) and reverses direction twice over the range $G = 10.8-13$ mag, rendering correction of variable-star parallaxes impossible without epoch astrometry. Corrections are typically small (tens of $\mu$as; $\sim$1 per cent of the parallax and $\sim 1/6$ of its uncertainty). Most stars we consider are not sufficiently variable to cross these boundaries. We therefore consider it better to apply this inexact correction than not to apply it at all, via the \emph{Gaia} DR3 parallax-to-distance conversions of \citet{BJRF+21}. This also accounts for the Lutz--Kelker bias on statistical samples \citep{LK73}. We only use the geometric distance of \citet{BJRF+21} here, as their photo-geometric distances rely on stellar models that often do not fit our stars well.

\subsubsection{Period--luminosity-relation distances} 
\label{sec:xmatch:distances:pl}

Pulsation period and intrinsic brightness are linked by discrete sequences \citep[e.g.][]{Wood15}. Almost all stars with strong mass loss (NESS Tiers 2, 3 and 4) pulsate most strongly in the fundamental pulsation mode \citep{McDonald16,MT19}, and are the most likely to lack accurate distances by other methods. Assuming that these stars are on the fundamental pulsation mode, we can use a $P-L$ relation (in this case, the period--$K_{\rm s}$-band relationship of \citet{RMF+10}) to estimate a distance to the star.

The $P-L$ relation has finite width: \citet{RMF+10} report a width of 0.293 mag at $K_{\rm s}$-band, creating a distance uncertainty of 14 per cent (plus any uncertainty in mean $K_{\rm s}$-band magnitude). The method still breaks down for the most optically obscured stars, which suffer from significant dust absorption, even at $K_{\rm s}$-band.

Table \ref{tab:persources} shows our sources of pulsation period. Higher priority is generally reserved for surveys with longer observations, which are likely to more accurately recover these long-period pulsations. Larger surveys were also more frequently given higher priority, given their potential for more homogeneous data.

\subsubsection{Bolometric-luminosity distances} 
\label{sec:xmatch:distances:mbol}

The NESS sample was created by assuming that the luminosity function of cool evolved stars in the solar neighbourhood closely approximates the LMC sample of \citet{Riebel12}. Their luminosities were corrected for the geometry of the LMC as published in \citet{Haschke2012}, with over 10\,000 random samples taken from the uncertainty distribution for the luminosity of each star, with kernel density smoothing used to convert these into a luminosity function of LMC evolved stars.

Riebel et al.'s LMC sample is mostly defined from the global LMC population by cuts in the $J-K$ colour--magnitude diagram that select both RSGs and AGB stars more luminous than the RGB tip. This leads to a rapid tapering of the luminosity function below $L \sim 3000$\,L$_\odot$, and a hard cutoff at $L = 1000$\,L$_\odot$. However, some fainter stars on the fundamental and first-overtone pulsations are included, as these are largely populated by AGB but not RGB stars, leading to a slightly smoother cutoff. The resulting luminosity function has a median of 6200\,L$_\odot$, and a 16th to 84th percentile range of 2300--9000\,L$_\odot$. We use this to provide a luminosity-based distance to all the stars in the sample.

The $^{+45}_{-63}$ per cent uncertainty in luminosity translates into a distance uncertainty of $^{+20}_{-39}$ per cent. This makes it the most-accurate method of determining distances to optically obscured stars for which no period is known. However, it does assume stars are as luminous as the LMC median, therefore performs badly on stars both less-luminous AGB and luminous RSG stars. It may also create a global distance bias if the LMC median luminosity differs markedly from our Galaxy's. However, the NESS Overview paper \citep{Scicluna22} showed good agreement with parallax distances from \emph{Gaia} DR3, and that the 16th--84th centile range of scatter about the 1:1 correlation ($\sim$25 per cent) for individual stars was very close to the expected $^{+20}_{-39}$ per cent. The new SEDs in this work allow us to both improve the SED quality and perform an interstellar reddening correction.

\subsubsection{``Restricted'' and ``unrestricted'' datasets} 
\label{sec:xmatch:distances:restriction}

Our full list of distance inputs are given in Table \ref{tab:distsources}. This prioritises \citet{ARVDB22}: a bespoke catalogue of AGB-star distances considered the most reliable as it includes data from both maser and optical parallax measures. If this does not exist, then the \emph{Gaia}-based distances of \citet{BJRF+21} are used instead, merged with the \emph{Hipparcos} parallax distances if available. If these do not exist, then the raw \emph{Gaia} DR3 parallax is used to provide a distance. This set of parallax-based distances is used for all stars in the 300 pc sample and most stars in the NESS sample. We refer to it as the ``\emph{restricted dataset}'' and use this when comparing the NESS and 300 pc samples.

Some NESS stars lack accurate parallax distances (statistics in Figure \ref{fig:venn}). Consequently, we add the other literature sources of parallaxes, and kinematic and other distance measures (full list in Table \ref{tab:distsources}). Finally, we compute two new distance estimates based on period--luminosity distances and bolometric-luminosity distances (see below). We refer to this as the ``\emph{unrestricted dataset}'', which we use in the revision of the NESS tiers and discussion, unless otherwise stated.

These two additional methods can refine distances to NESS targets, but have the potential to bias or invalidate our analysis: see Appendix \ref{apx:dist} for a comparative discussion.

\subsection{Treatment of errors}
\label{apx:more:errors}

Any statistical sample of stars defined by distance or luminosity criteria is liable to be incomplete, and new data will cause the set of sampled stars to change. This, along with the complexities of evolved-star distance estimation, means we opt for a maximum-likelihood estimator of distance, rather than a probabilistic analysis. 

Conventionally, errors in SED-fitted temperatures and luminosities would be derived from errors in the underlying photometry. However, these formal errors ignore the ``unknown unknowns'', which dominate the uncertainties. In the well-fit sources of the 300 pc sample, the reduced $\chi^2$ reaches a minimum of $\chi^2_{\rm r} \approx 2-5$ for a few sources, but is typically around $\chi^2_{\rm r} \sim 30$.

\citet[][their section 3.9]{McDonald2024} provide the primary unquantified sources of errors. Broadly speaking, they are:
\begin{itemize}
    \item Lack of accounting for stellar blending, photometric zero-point errors, filter profile errors, artefacts and unflagged problems.
    \item Uncertainties in reddening correction, e.g., errors in the 3D extinction maps, distance errors, and errors in the slope of the reddening law and associated colour corrections.
    \item Unquantified errors in the underlying stellar models.
\end{itemize}

However, there are particular aspects that are important for our samples, and the NESS sample especially, namely:
\begin{itemize}
    \item Poorly quantified distances, and the high reduced unit-weight error (RUWE) of the \emph{Gaia} parallax measurements. These affect star's luminosity and (to a much smaller effect) the surface-gravity estimates needed to choose comparison stellar atmosphere models.
    \item Source variability in single-epoch photometry. This is mitigated by averaging over a large number of filters/catalogues and by prioritising catalogues with photometric averages.
    \item Variability-induced Malmquist bias. A survey may only detect variable stars during the bright part of their pulsation cycle, leading to a reporting bias. The median reported flux from a set of stars in a survey may be brighter than the median actual flux.
    \item Departures from local thermodynamic equilibrium (LTE).
    \item Poor representation of the SED by a stellar atmosphere model, e.g., due to circumstellar dust and binary companions, which are not included in the fitting procedure. This has a dominant effect towards the extreme tiers of the NESS sample.
\end{itemize}
Rather than guess at errors that may be wildly inappropriate, we determine it best not to assess errors on parameters at this stage.

\citet{McDonald2024} gives indicative errors of a few per cent in temperature for a random sample of Galactic giant stars, and correspondingly a few per cent in luminosity when distance is well known and extinction low (see their figures 8, 11 and 12). This may be representative for the less-variable, less-dusty stars in the lower NESS tiers but, for the reasons cited above, the uncertainty in the higher tiers of NESS will be substantially more than this.


\section{Results} 
\label{sec:results}

Appendix \ref{apx:data} contains descriptors for the final tables of computed and collated stellar parameters for both the 300 pc and NESS samples.

\subsection{300 pc sample}
\label{sec:xmatch:fromgaia}

Stars were rejected from the final dataset if they did not meet the distance, temperature and luminosity criteria. 
These rejections leave a final, complete sample of 507 luminous evolved stars within 300 pc of the Sun. The large reduction in the number of sources comes primarily from the application of the $L > 700$ L$_\odot$ limit, applied to the conservative magnitude limits we used to select stars based on their \emph{Gaia} and \emph{Hipparcos} photometry.
We return to this dataset and compare it to the NESS sample in the discussion.

\subsection{NESS: rejected sources}
\label{sec:rejects}

To identify NESS sources that were not evolved stars, we progressed through the following screening processes:
\begin{itemize}
        \item During the manual inspection of cross-matches (Appendix \ref{sec:xmatch:togaia}), visual inspection of images in {\sc aladin} identified extended sources, such as resolved PNe and proto-PNe (PPNe), parts of other nebulae, knots of interstellar medium and young stars in clusters.
        \item NESS targets were passed through {\sc simbad} to identify other names and collect basic information. A variety of objects that were clearly not evolved stars were removed (e.g., $\alpha$ Cen).
        \item Literature on objects with {\sc simbad} spectral types of earlier than K0 was retained only if there was an AGB-like component, i.e., if the spectral classification was in error, the system was a binary, or if the spectral class was strongly variable.
        \item Objects with primary {\sc simbad} classifications of post-AGB, PN or PPN had their images, SEDs and literature data scrutinised. Objects were assigned to be highly evolved stars if we agreed with these classifications. We relied heavily on the use of the 5000\,K criterion to separate AGB from post-AGB objects, which retains objects that may be classified as PPN or very young post-AGB stars, but which have yet to properly leave the AGB.
        \item Any object with a fitted temperature over 5000\,K, or a fitted luminosity of $<$700\,L$_\odot$ or $>$200\,000\,L$_\odot$ was also selected for detailed investigation. Sources were removed if they were not consistent with evolved stars. In effect, this imparts criteria of 5000 K and K0 as a division between RSGs and YSGs.
        \item Any object with a double-peaked SED or very badly fitting SED was also checked (see Section \ref{sec:disc:merge}).
\end{itemize}
As well as these processes, an extensive manual investigation of sources was performed during the fine-tuning of the processes described above, and sources which returned unexpected parameters or had unusual or badly fitting SEDs were investigated. A list of objects with special requirements (e.g., due to nearby confusion) is provided in Appendix \ref{apx:reject:complex}.

We uncovered 71 NESS sources that are unsuitable for inclusion in this analysis (see Figure \ref{fig:venn}, a list in Table \ref{tab:rejects}, and details in Appendix \ref{apx:reject}). In summary, these rejects include 42 sources that do not appear to be evolved stars, and 27 sources highly evolved objects (resolved planetary nebulae and known post-AGB stars that have passed the evolutionary stage of being mass-losing AGB or RSG stars). Three sources were identified as having unclear classifications: \emph{IRAS}\,13428--6232 (PM\,2-14) and \emph{IRAS}\,16437--4510 were rejected, but \emph{IRAS}\,17205--3418 retained. Additionally, one duplicate source was identified in the original NESS source list (\emph{IRAS}\,19597+3327, 19597+3327A). However, the NESS survey therefore still consists of 852 unique pointings, because the co-ordinates for both identifiers are identical, and the source is listed among the 71 rejected sources.

This reduces the number of NESS sources considered here from the original 852 to 781 in the unrestricted dataset. Excluding the stars with unknown distance (therefore unknown luminosity and only a very crude $\dot{D}$), the restricted dataset is reduced to 685 stars. Further restrictions to remove sources not meeting our temperature and luminosity criteria for an evolved star are made in Section \ref{sec:disc:merge}.

\begin{table}
    \caption{\emph{IRAS} numbers of the NESS sources not meeting the evolved-star criteria of this work.}
    \label{tab:rejects}
    \centering
    \begin{tabular}{@{}c@{\quad}c@{\quad}c@{\quad}c@{\quad}c@{}}
        \hline\hline
        \multicolumn{5}{c}{\emph{Not evolved stars}}\\
05377+3548 &
05388--0147 &
05389--6908 &
05401--6940 &
06050--0623 \\
06491--0654 &
07422+2808 &
09572--5636 &
09576--5644 &
09578--5649 \\
10431--5925 &
11202--5305 &
11254--6244 &
11260--6241 &
11266--6249 \\
13416--6243 &
14050--6056 &
14359--6037 &
15141--5625 &
16124--5110 \\
16434--4545 &
16545--4012 &
16555--4237 &
16557--4002 &
17326--3324 \\
17423--2855 &
17441--2822 &
17590--2337 &
18008--2425 &
18072--1954 \\
18155--1206 &
18288--0207 &
18585--3701 &
18595+0107 &
19117+1107 \\
19597+3327A &
20002+3322 &
20081+2720 &
22133+5837 &
22540+6146 \\
22544+6141 &
22548+6147 \\
        \multicolumn{5}{c}{\emph{Highly evolved objects (post-AGB stars, planetary nebulae, etc.)}}\\
04395+3601 &
05251--1244 &
08011--3627 &
09256--6324 &
10197--5750 \\
14562--5406 &
15445--5449 &
06176--1036 &
15452--5459 &
16133--5151 \\
16594--4656 &
17103--3702 &
17150--3224 &
17163--3907 &
17251--3505 \\
17347--3139 &
17427--3010 &
17441--2411 &
18450--0148 &
18458--0213 \\
19244+1115 &
19327+3024 &
19374+2359 &
20028+3910 &
20547+0247 \\
21282+5050 &
23541+7031 \\
        \multicolumn{5}{c}{\emph{Uncertain classifications, probably not evolved stars}}\\
13428--6232 &
16437--4510 \\
    \hline
    \multicolumn{5}{p{0.95\columnwidth}}{Note the duplicate source \emph{IRAS}\,19597+3327 was also rejected, and \emph{IRAS}\,19597+3327\,A retained.}\\
    \hline
    \end{tabular}
\end{table}


\section{NESS survey completeness and biases} 
\label{sec:disc}

A comparison of different distance-estimation and fitting methods is given in Appendix \ref{apx:dist}.

\subsection{Revised distances}
\label{sec:disc:distcomp}

\begin{figure}
\centering
\includegraphics[height=\linewidth,angle=-90]{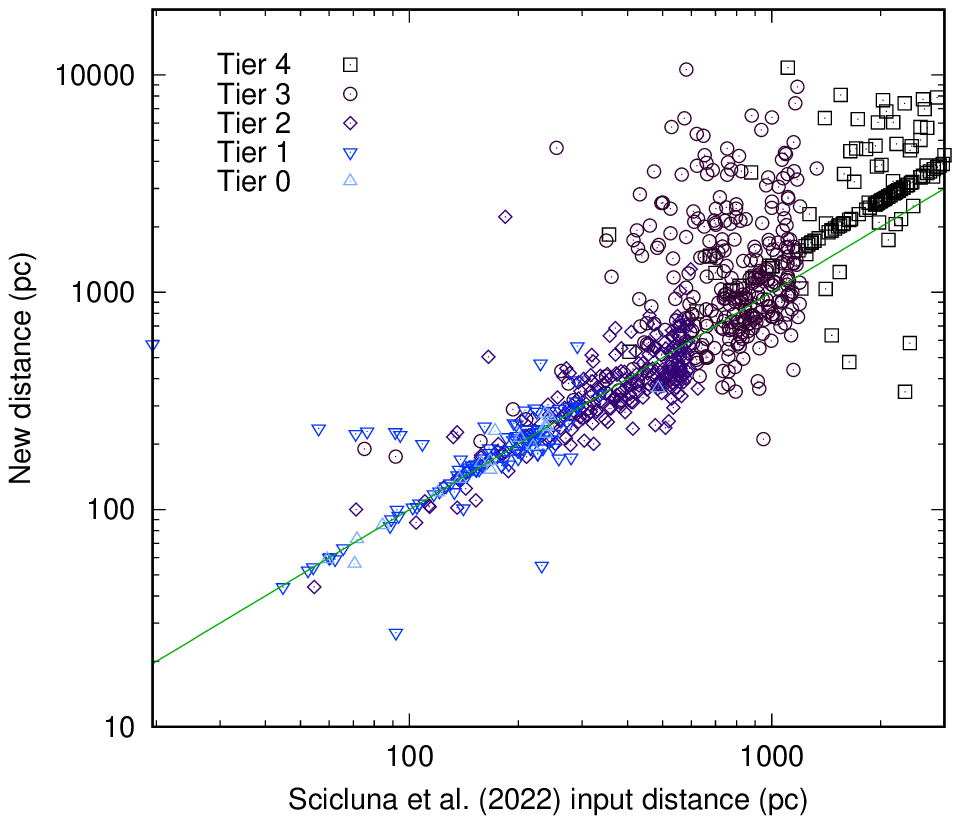}
\caption{Comparison of distances used in the original NESS survey definitions (see the NESS Overview paper) and revised distances used in this paper. A line of equality is shown in green.}
\label{fig:newdist}
\end{figure}

Figure \ref{fig:newdist} shows the revised distances resulting from a combination of luminosity-, period--luminosity- and parallax-based distances.

The vast majority of the new distances come from \emph{Gaia} parallaxes, and the majority of those without \emph{Gaia}-based distances are distant, extreme sources that also lack well-defined pulsation periods, so remain on the (albeit now slightly offset) diagonal line at large distances. Consequently, the majority of the points that scatter a long way from the line of equality do so because \emph{Gaia} DR3 distances are now being adopted instead of either the \emph{Hipparcos}/\emph{Tycho--Gaia} parallax distances (used to define Tiers 0 and 1) or the luminosity distances (used to define Tiers 2, 3 and 4).

\subsection{Merging datasets and sources with out-of-bounds parameters}
\label{sec:disc:merge}

\begin{figure*}
\centering
\includegraphics[height=0.9\linewidth,angle=-90]{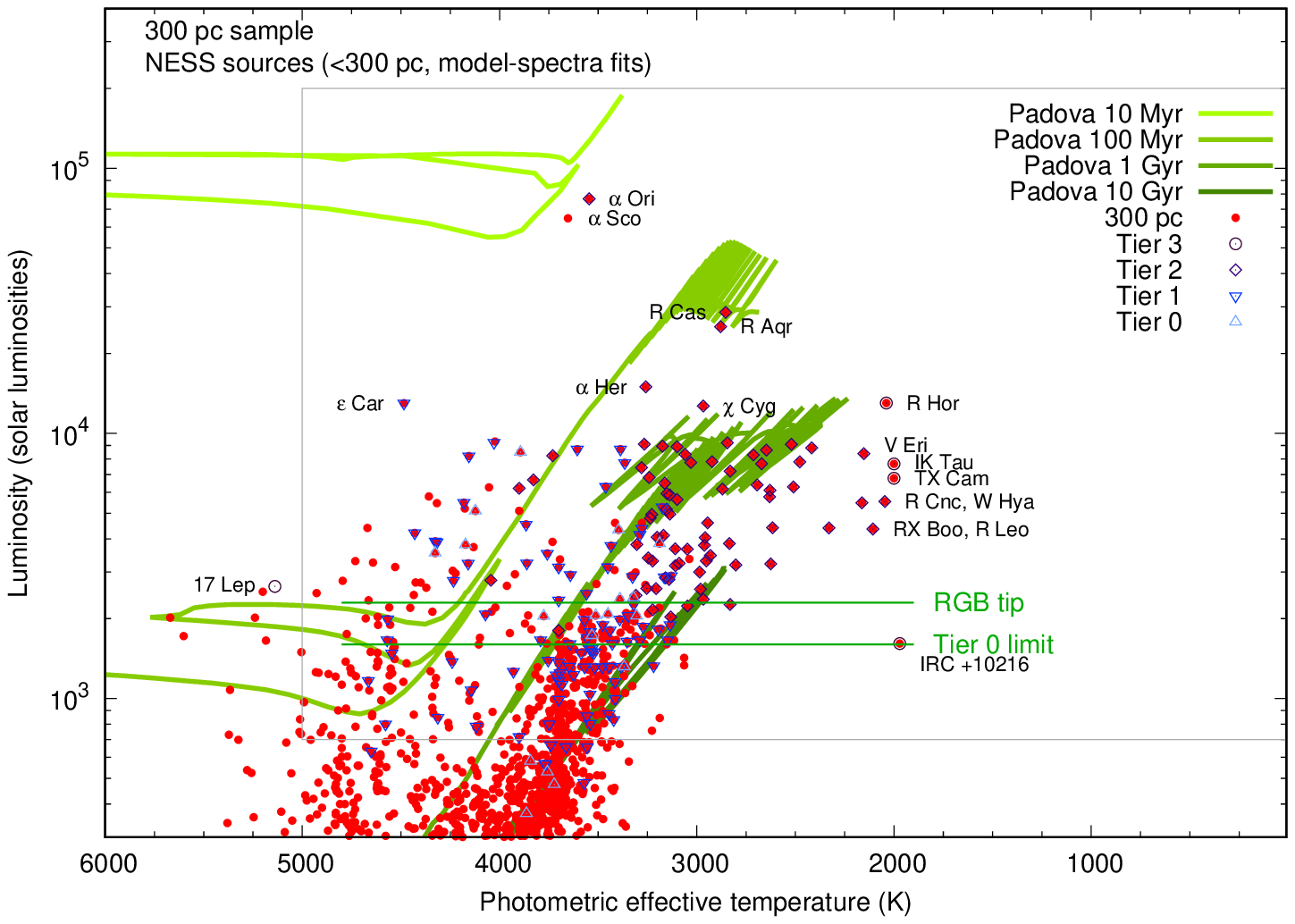}
\includegraphics[height=0.9\linewidth,angle=-90]{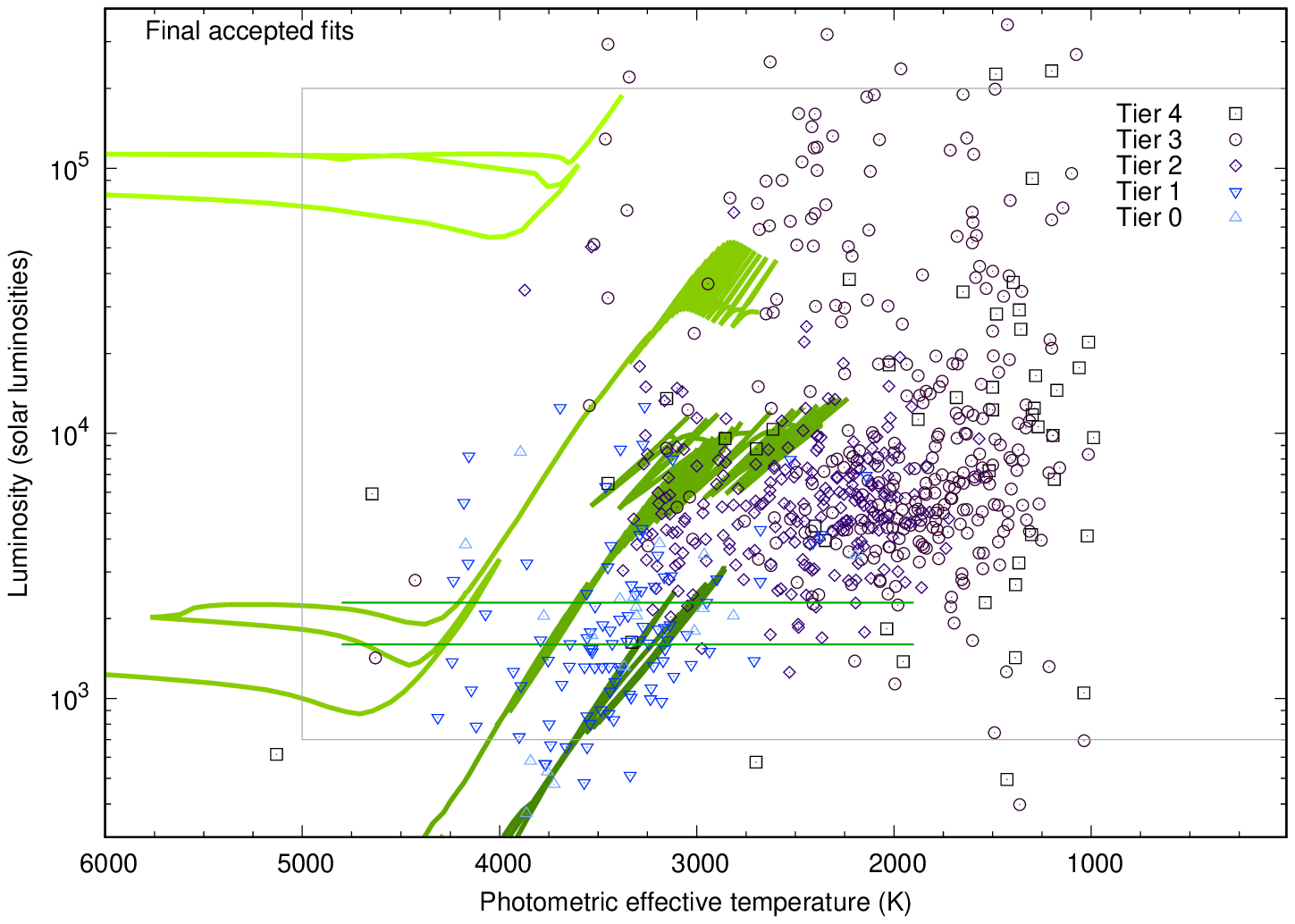}
\caption{H--R diagrams of our samples. \emph{Upper panel:} sources within 300 pc; \emph{lower panel:} the NESS sample combined from stellar-model and blackbody fits. \emph{Gaia} DR3 sources within 300 pc of the Sun (see Section \ref{sec:300pc}) are shown as solid red points; NESS sources within this sample are shown in blue/purple open symbols as indicated on the plots. Solar-metallicity Padova isochrones at different ages are shown: sources to the right of these may have strong dust production. The grey boundary shows the temperature (5000 K) and luminosity (700 / 200\,000 L$_\odot$) bounds denoting evolved stars for the purposes of this work. The short green lines at 1600 and 2300 L$_\odot$ respectively mark the nominal luminosity cutoff for Tier 0 and the approximate location of the RGB tip. Outlying stars are named.}
\label{fig:300pc_hrd}
\end{figure*}

Different methods can result in very different luminosities, particularly for more-extreme stars whose SEDs depart strongly from a typical stellar SED (see Appendix \ref{apx:dist:lumtypes}). However, regardless of the method, stars can be found at luminosities that are either too small or too large to be physically reasonable for evolved stars. Model atmospheres generally reproduce the expected properties better than blackbody fits, with the exception of some stars in Tier 3 and most stars in Tier 4.

To determine which reduction method we should use to most-reliably determine final parameters for different stars, we introduce a new goodness-of-fit parameter, $GOF$, which is defined based on the ratio of observed to modelled flux ($F_{\rm o} / F_{\rm m}$), and the fractional error in the observed flux, $\Delta F_{\rm o} / F_{\rm o}$, such that:
\begin{equation}
    GOF = \mathrm{Median}\left|\frac{F_{\rm o} / F_{\rm m} - 1}{\Delta F_{\rm o} / F_{\rm o}}\right| .
\end{equation}

Using this $GOF$ statistic for our model-derived and blackbody fits, we adopt the following criteria:
\begin{enumerate}
    \item The model-derived parameters are used by default.
    \item The blackbody-derived temperature and trapezoid-integrated luminosity are used if one of the following criteria are met:
    \begin{enumerate}
        \item $GOF_{\rm bb} / GOF_{\rm model} < 0.5$, or
        \item $GOF_{\rm bb} / GOF_{\rm model} < 1$ and the model-fitted and blackbody-fitted temperature and luminosity fulfil any one of the following criteria:
        \begin{enumerate}
            \item $T_{\rm model} < 2500$\,K,
            \item $T_{\rm bb} < 2000$\,K,
            \item $T_{\rm model} > 5000$\,K,
            \item $L_{\rm model} < 700$\,L$_\odot$ but $L_{\rm bb} > 700$\,L$_\odot$, or
            \item $L_{\rm model} > 200\,000$\,L$_\odot$ but $L_{\rm bb} < 200\,000$\,L$_\odot$.
        \end{enumerate}
    \end{enumerate}
\end{enumerate}
The combined fits for this \emph{``merged dataset''} are shown in the H--R diagrams in Figure \ref{fig:300pc_hrd} (and separately in Figure \ref{fig:bb-model-hrd}).

A median luminosity of sources with parallax-based distances of 5396 L$_\odot$ is found. An indicative error range for individual sources, based on the central 68 per cent of this data, is 2453--14\,976\,L$_\odot$. This revised median luminosity above allows us to also revise our luminosity-based distances: these distances are reported in the digitised tables (Appendix \ref{apx:data}) but are not used in this analysis to avoid distance estimation becoming an iterative problem. 

Of the 685 non-rejected sources with distances, 19 sources have luminosities below 700\,L$_\odot$ (of which one, \emph{IRAS}\,16383--4626 also has a temperature of $>$5000\,K), while 17 have luminosities above 200\,000\,L$_\odot$, the observed upper luminosity limit for RSG stars (discussed individually in Appendix \ref{apx:reject:criteria}). Stars in these categories tend to include less-evolved stars with shorter updated distances, extremely luminous supergiants like $\mu$ Cep, and stars whose distances (therefore luminosities and interstellar extinctions) have suspected errors or underestimated uncertainties. Some stars fit more than one of these criteria.

Removing these sources leaves 649 sources that meet physically plausible criteria for evolved stars and have a distance estimate that does not rely on a luminosity-based or period-based distance. Since we consider one or more measured properties of the removed stars to be in error, we base our summary statistics on the remaining, ``criteria-matching'' stars, which represent the dataset used throughout the remainder of this paper unless otherwise specified.

\subsection{Completeness of NESS Tiers 0 and 1}
\label{sec:disc:complete}

\begin{table}
    \centering
    \caption{Summary completeness statistics of the NESS sample's unrestricted dataset, showing a revised tier set using updated stellar properties.}
    \label{tab:completeness}
    \begin{tabular}{@{}c@{\ }l@{\quad}r@{\quad}r@{\ \ }r@{\ \ }r@{\ \ }r@{\ \ }r@{\ \ }r@{\quad}r@{\quad}r@{}}
    \hline\hline
    Tier& Descriptor & Source & ($\times$) & ($-$) & ($\uparrow$) & ($\downarrow$) & ($\uparrow\uparrow$) & ($\downarrow\downarrow$) & Revised & ($+$)\\
    \ & \ & count & \ & \ & \ & \ & \ & \ & count & \ \\
    \hline 
      0 & Very low            &  19 &  0 &   7 &  1 & -- & -- &  2 &  13 & 5\\
      1 & Low                 & 105 &  1 &   3 &  9 & -- &  0 &  0 &  92 & 5\\
      2 & \rlap{Intermediate} & 222 &  0 &  23 &  1 &  2 &  9 & 25 & 230 & --\\
      3 & High                & 324 & 13 & 145 &  4 & 25 &  2 &  5 & 144 & --\\
      4 & Extreme             & 182 & 57 &  50 & -- &  5 &  4 & -- &  74 & --\\
    {\it Sum} & -- & {\it 852} & {\it 71} & {\it 228} & {\it 15} & {\it 32} & {\it 34} & {\it 13} & {\it 553} & {\it 10}\\
    \hline
    \multicolumn{11}{p{0.43\textwidth}}{The source count reflects the NESS sample, modified for rejected sources ($\times$), removals due to revised distances ($-$), objects moving out of the stated tier (either by moving up to higher tiers ($\uparrow$) or moving down to lower tiers ($\downarrow$)), objects moving into the stated tier (by moving up ($\uparrow\uparrow$) or down ($\downarrow\downarrow$) tiers), with the source count in the revised source list as the penultimate column. The final column ($+$) gives possible additions from the 300 pc sample. Blank items (--) are either not assessed or are impossible movements.}\\
    \hline
    \end{tabular}
\end{table}

\begin{figure}
\centering
\includegraphics[height=\linewidth,angle=-90]{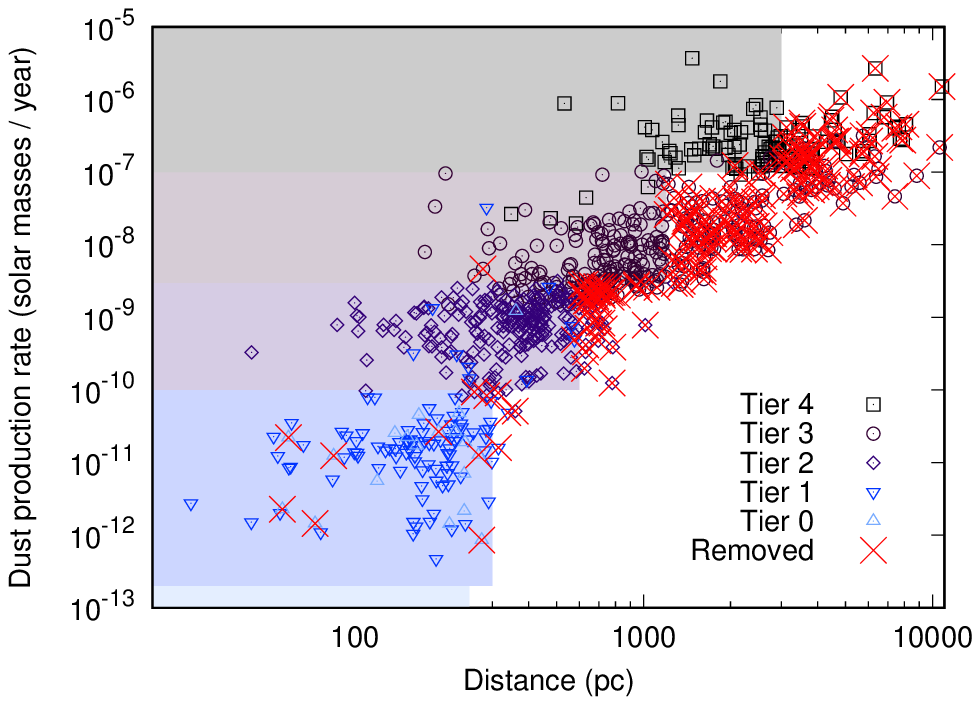}
\caption{Revised tiering from distance and DPR changes. Filled boxes show tier boundaries. Note that the boundary for tier 0 is a luminosity-based boundary, rather than a $\dot{D}$ boundary, resulting in some objects with negligible mass loss being removed that are otherwise within the tier boundaries.}
\label{fig:dDdot}
\end{figure}

The completeness of the NESS tiers affects the survey's ability to make fully accurate estimates of the volume-limited return by AGB stars to the Solar Neighbourhood. Many distances to AGB stars remain uncertain by more than a factor of two, particularly in the upper tiers 2, 3 and 4, where selection biases are also important (see above and Section \ref{apx:bias}).

For the lower tiers, 0 and 1, our 300 pc sample of evolved stars provides a unique dataset to examine the completeness of NESS Tiers 0 and 1. Figure \ref{fig:300pc_hrd} shows the H--R diagram of sources in the sample and NESS sources within 300 pc for comparison.

A summary of the possible inclusions and exclusions found in the remainder of this section are listed in Table \ref{tab:completeness} and shown in Figure \ref{fig:dDdot}. A full list of NESS sources and their revised tiers, and a by-tier version of Figure \ref{fig:dDdot} are given in Figure \ref{fig:dDdottiers}.

\subsubsection{Checking completeness of the 300 pc sample}

All but one of the sources in the restricted NESS dataset with fitted distances of $<$300 pc has a counterpart in the 300 pc sample (although, due to the way the different datasets were put together, sometimes with a different primary identifier, e.g., a \emph{Hipparcos} identifier instead of the \emph{Gaia} DR3 identifier).

The single exception is the symbiotic binary star 17 Lep (\emph{IRAS}\,06027--1628). This is an A-type main-sequence star with a probable early-M-type giant companion. The colour of this system was too blue to be selected for the 300 pc sample, and the model fit in any case produces a fit that exceeds our 5000\,K temperature limit. Consequently, this is one of the systems that falls into our restricted dataset, but is not included in the criteria-matching dataset. Since {\sc PySSED} is not set up to deal with equal-luminosity but unequal-temperature binaries well, we cannot trust the properties of this system as recorded in the above analysis either (as is true for the other stars outwith the criteria-matching dataset).

\subsubsection{Summary}

Table \ref{tab:completeness} also suggests some stark changes to the NESS catalogue: 228 of the 781 non-rejected sources are removed in the revised list, and 47 end up in a new tier (see Appendix \ref{apx:newtier} for details).

However, it must also be stressed that these updated criteria are also estimates: they still contain an (albeit reduced) level of bias, and they may incorrectly remove individual sources. Overall, we expect these revised criteria to give a picture closer to the truth. In the following discussion, we will refer to this list of 553 remaining, re-tiered sources as the ``\emph{revised source list}'' to accompany the restricted and unrestricted datasets defined previously.

\subsection{Selection biases in the NESS sample}
\label{apx:bias}

NESS Tiers 0 and 1 suffer from the usual Lutz--Kelker bias (Section \ref{apx:dist:zpt}) but, because these are nearby stars, we expect such biases to be comparatively small.

In contrast, NESS Tiers 2, 3 and 4 were selected purely on luminosity-based distances, and dust mass-loss rates based on those distances, median luminosity and infrared-based $\dot{D}$. This introduces more severe and complex biases into these higher tiers.

Most notably, luminosity-based distances strongly select intrinsically luminous stars. In theory, any source with $n$ times the assumed median luminosity should be included if it is within $\sqrt{n}$ times the tier's distance boundary. Given the NESS Tier 2--4 distance limits are substantially greater than the scale height of the Galactic disc, a radius increased by a factor of $\sqrt{n}$ should sample $n$ times more stars. If the brightest RSGs reach $L \sim 200\,000$\,L$_\odot$ \citep{Davies2020}, thus $n \sim 32$, this could theoretically lead to an overabundance of the brightest supergiants in Tier 4 by up to a factor of $\sim$32, out to distances of $\sim$17 kpc.

In practice, interstellar extinction and source crowding limits our view of RSGs on the far side of the Galaxy. Furthermore, the assumed dust-production rate, $\dot{D}$, scales with assumed distance as $\dot{D} \propto d$. If a luminous, distant RSG with an extreme mass-loss rate (here $\dot{D} \geq 10^{-7}$\,M$_\odot$\,yr$^{-1}$) is brought into the NESS sample by artificially reducing its distance, then its $\dot{D}$ will also be reduced, and a portion of these stars will fall out of the extreme tiers because they do not meet the tier's minimum $\dot{D}$ criterion.

The systematic inclusion of these intrinsically luminous stars means they are substantially over-represented in the NESS survey. Conversely, low-luminosity stars will be preferentially absent because their distances and $\dot{D}$ will be under-estimated. Given $\dot{D}$ in general increases with luminosity and the number of stars per annulus increases with distance (due to the larger volume contained therein), and given the NESS $\dot{D}$ tier limits rise more than linearly with distance, we find many more stars drop out of NESS tiers (or are demoted to less-extreme tiers) than move into them (or move up; see Appendix \ref{apx:newtier} for detail). This is fortunate, as we can reduce biases in our nominally volume-limited survey mostly by removing stars found to be problematic, rather than identifying and observing many new stars that we have missed.

Extreme tiers also (intentionally) select stars with high apparent $\dot{D}$, as generated by the {\sc grams} models \citep{Srinivasan2011}. In reality, $\dot{D}$ is a measurement of infrared excess and assumes a spherical geometry. Aspherical mass loss is relatively common, but most strongly manifests itself as an equatorial density enhancement or disc, thought to be most frequently generated by a binary companion \citep[e.g][]{Decin2020}. Face-on discs will have a similar $\dot{D}$, as the star will not appear significantly dust-enshrouded, but will still show infrared excess. Edge-on discs, however, will be modelled with a significantly larger $\dot{D}$, as the star will be heavily dust-enshrouded. The invocation of a spherical geometry will therefore over-estimate $\dot{D}$. In this way, the NESS survey is unfortunately also biased to edge-on discs and (more broadly) equatorial density enhancements of dust around stars.


\section{Statistical properties of the samples} 
\label{sec:disc2}

\subsection{Comparison of the NESS temperatures and spectroscopic surveys}

\begin{figure}
\centering
\includegraphics[height=\linewidth,angle=-90]{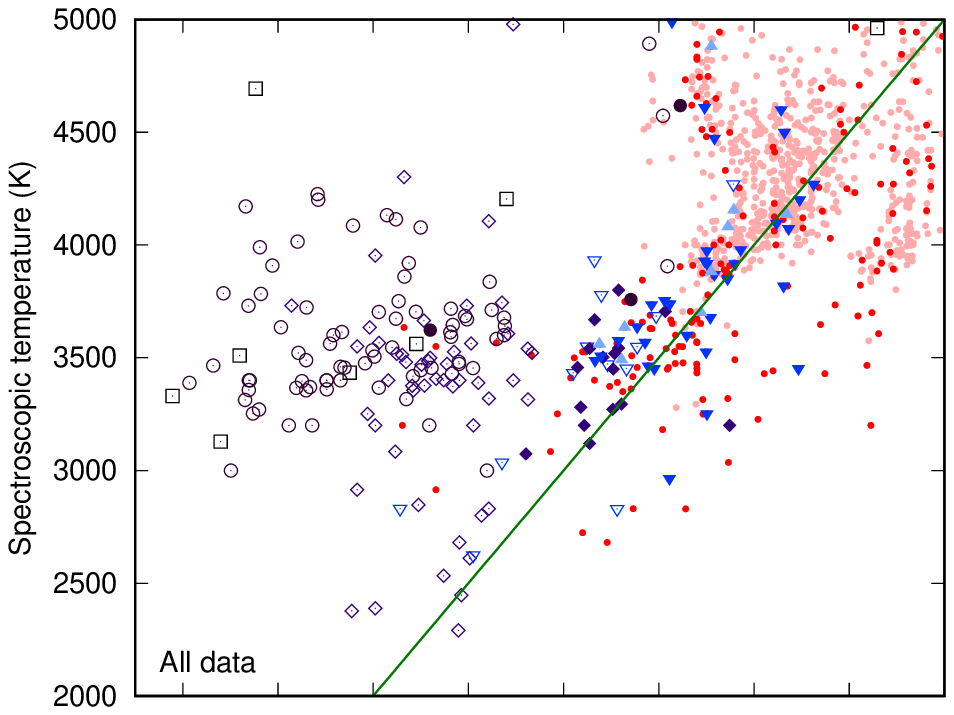}
\includegraphics[height=\linewidth,angle=-90]{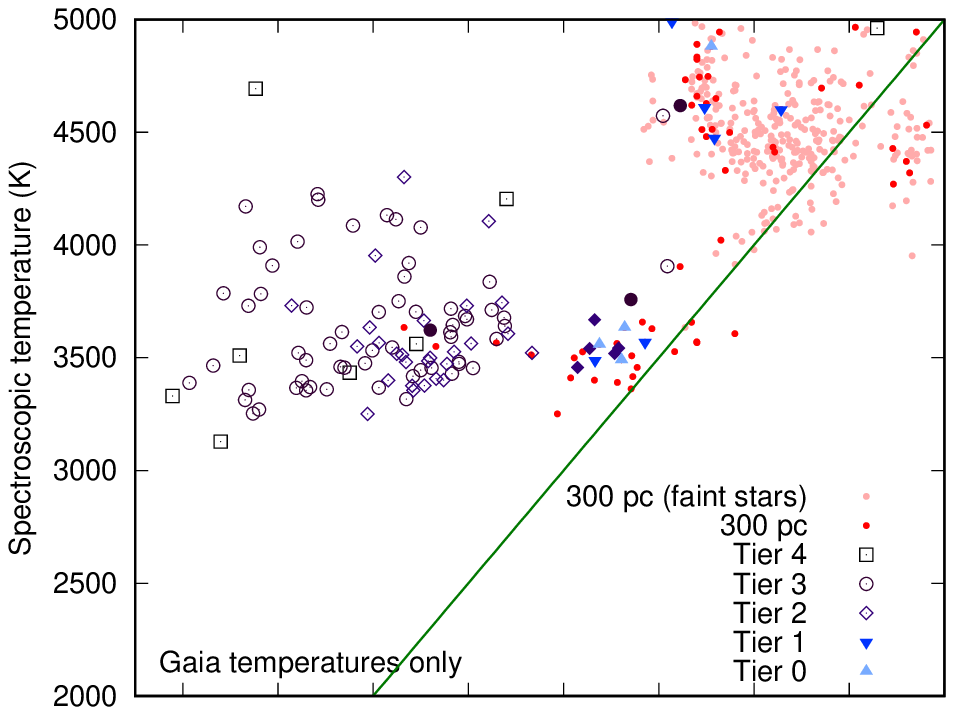}
\includegraphics[height=\linewidth,angle=-90]{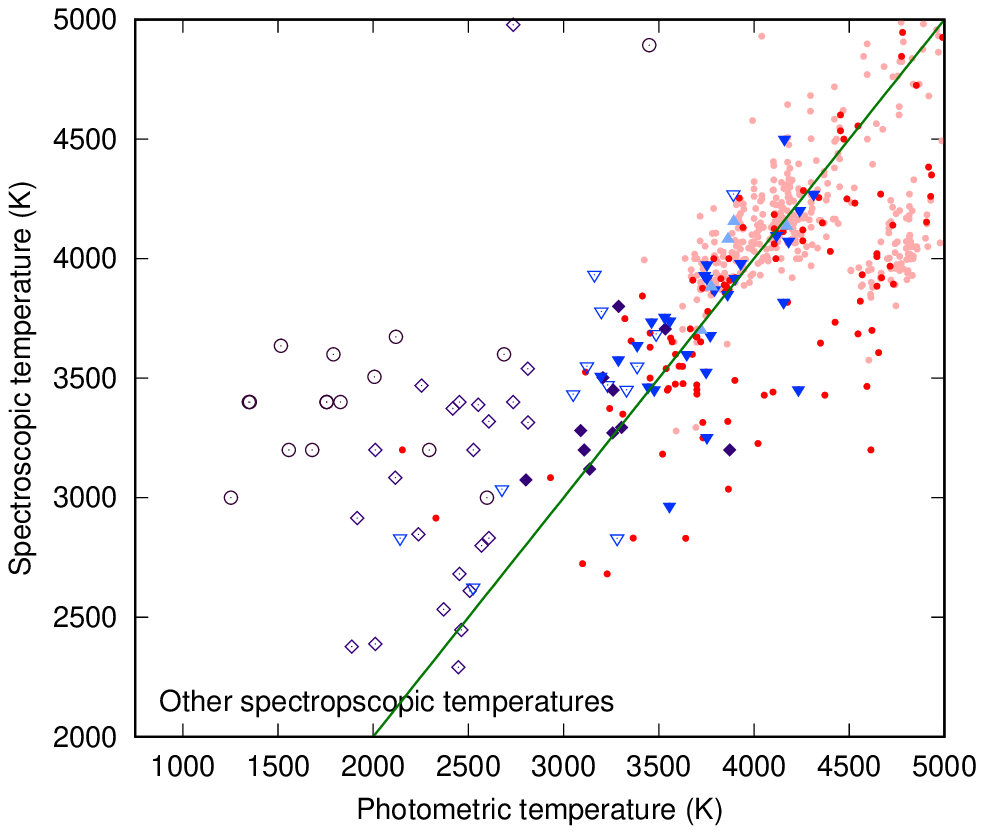}
\caption{Comparison between photometric temperatures from this work's SED fitting and spectroscopic temperatures from the literature. As in previous plots, points show the 300 pc sample and NESS tiers. However, in this plot, filled points show stellar-atmosphere model fits have been used to fit the SEDs, while hollow points show stars where blackbody fits were used. Pink objects show stars in the 300 pc analysis that are not in the 300 pc sample, with either $L<700$\,L$_\odot$ or $d>300$\,pc. The diagonal green line shows parity agreement between the two temperature measures. \emph{Top panel:} all data; \emph{middle panel:} only \emph{Gaia} {\sc apsis} GSP-Phot {\sc aeneas} temperatures; \emph{bottom panel:} all other spectroscopic temperatures.}
\label{fig:Tspec}
\end{figure}

Comparison of the photometric (SED-fitted) temperatures produced by {\sc PySSED} to literature temperatures derived from spectra gives us an opportunity to both: (1) test the accuracy of both methods and (2) test the strength of optical obscuration around dust-producing stars. Figure 11 of \citet{McDonald2024} shows that {\sc PySSED} can typically reproduce the spectroscopic temperature of stars derived from XShooter spectra \citep{Arentsen2019} to within a few percent: taking only stars below the 5000\,K bound of our paper's remit, $T_{\rm spec} - T_{\rm phot} = 29$\,K (--146 to 302)\,K\footnote{Statistical distributions in this section are often poorly approximated by normal distributions. Unless otherwise stated, we quote median values where possible. For error/uncertainty estimates, we cite the range encompassing the central 68.3 per cent of data points in brackets.}, which sets an approximate expectation for an accurate recovery.

\subsubsection{300 pc sample}
\label{sec:tspec300}

\begin{table}
    \caption{Comparison of literature spectral types to measured photometric temperatures.}
    \label{tab:sptypetemp}
    \centering
    \begin{tabular}{lc@{\ \ }cc@{\ \ }c}
        \hline\hline
        \multicolumn{1}{c}{Spectral}    & \multicolumn{4}{c}{Median and 68 per cent intervals (K)} \\
        \multicolumn{1}{c}{type}    & \multicolumn{2}{c}{300 pc}    & \multicolumn{2}{c}{NESS} \\
        \hline
K0 & 4394 & ($4005-4844$) & --- & (---)\\
K1 & 4341 & ($4177-4578$) & --- & (---)\\
K2 & 4219 & ($4046-4450$) & 4159 & ($3871-4171$)\\
K3 & 4175 & ($3950-4740$) & 3929 & ($3891-4118$)\\
K4 & 4104 & ($3893-4764$) & 3740 & ($3740-3893$)\\
K5 & 4014 & ($3747-4731$) & --- & (---)\\
M0 & 3753 & ($3660-3911$) & 3753 & ($3612-3770$)\\
M1 & 3700 & ($3669-4531$) & --- & (---)\\
M2 & 3692 & ($3605-4546$) & 3170 & ($2492-3534$)\\
M3 & 3570 & ($3414-3828$) & 3162 & ($1957-3428$)\\
M4 & 3454 & ($3363-3641$) & 3231 & ($2133-3476$)\\
M5 & 3288 & ($3162-3700$) & 2736 & ($2369-3258$)\\
M6 & 3164 & ($3037-3731$) & 2514 & ($1828-3137$)\\
M7 & 2831 & ($2153-3115$) & 2100 & ($1755-2451$)\\
M8 & --- & (---) & 1849 & ($1352-2171$)\\
M9 & --- & (---) & 1401 & ($1253-2123$)\\
 \hline
    \end{tabular}
\end{table}

\begin{table}
    \caption{Comparison of non-\emph{Gaia} literature spectral types to measured spectroscopic temperatures.}
    \label{tab:sptypetemp2}
    \centering
    \begin{tabular}{lc@{\ \ }cc@{\ \ }c}
        \hline\hline
        \multicolumn{1}{c}{Spectral}    & \multicolumn{4}{c}{Median and 68 per cent intervals (K)} \\
        \multicolumn{1}{c}{type}    & \multicolumn{2}{c}{300 pc}    & \multicolumn{2}{c}{NESS} \\
        \hline
K0 & 4410 & ($4233-4730$) & --- & (---)\\
K1 & 4296 & ($4160-4500$) & --- & (---)\\
K2 & 4210 & ($4032-4616$) & 4135 & ($3200-4202$)\\
K3 & 4126 & ($4000-4270$) & --- & (---)\\
K4 & 4066 & ($3930-4233$) & --- & (---)\\
K5 & 4000 & ($3886-4134$) & --- & (---)\\
M0 & 3870 & ($3679-4000$) & 3700 & ($3252-3918$)\\
M1 & 3762 & ($3200-3999$) & --- & (---)\\
M2 & 3672 & ($3477-3994$) & 3600 & ($2965-3706$)\\
M3 & 3652 & ($3452-3736$) & 3673 & ($3000-3800$)\\
M4 & 3472 & ($3182-3688$) & 3452 & ($2830-3637$)\\
M5 & 3350 & ($3271-3433$) & 3400 & ($3271-3577$)\\
M6 & 3281 & ($3120-3442$) & 3294 & ($3120-3469$)\\
M7 & 3084 & ($2915-3200$) & 3200 & ($2915-3635$)\\
M8 & --- & (---) & --- & (---)\\
M9 & --- & (---) & 3400 & ($3000-5076$)\\
 \hline
    \end{tabular}
\end{table}

\begin{table}
    \caption{Comparison of \emph{Gaia} literature spectral types to measured spectroscopic temperatures.}
    \label{tab:sptypetemp3}
    \centering
    \begin{tabular}{lc@{\ \ }cc@{\ \ }c}
        \hline\hline
        \multicolumn{1}{c}{Spectral}    & \multicolumn{4}{c}{Median and 68 per cent intervals (K)} \\
        \multicolumn{1}{c}{type}    & \multicolumn{2}{c}{300 pc}    & \multicolumn{2}{c}{NESS} \\
        \hline
K0 & 4596 & ($4511-4773$) & --- & (---)\\
K1 & 4616 & ($4335-4672$) & --- & (---)\\
K2 & 4485 & ($4326-4743$) & --- & (---)\\
K3 & 4455 & ($4333-4596$) & --- & (---)\\
K4 & 4405 & ($4158-4587$) & --- & (---)\\
K5 & 4525 & ($4359-4863$) & --- & (---)\\
M0 & 4596 & ($3657-4915$) & --- & (---)\\
M1 & 4832 & ($4709-5171$) & --- & (---)\\
M2 & 5130 & ($4649-5396$) & 3678 & ($3640-3717$)\\
M3 & 3904 & ($3569-5257$) & 3569 & ($3453-3757$)\\
M4 & --- & (---) & 3561 & ($3518-3665$)\\
M5 & 3499 & ($3390-3606$) & 3559 & ($3429-3684$)\\
M6 & 3541 & ($3361-3634$) & 3521 & ($3459-3634$)\\
M7 & 3526 & ($3512-3550$) & 3526 & ($3366-3613$)\\
M8 & --- & (---) & 3377 & ($3355-3456$)\\
M9 & --- & (---) & 3388 & ($3252-3730$)\\ \hline
    \end{tabular}
\end{table}

Figure \ref{fig:Tspec} compares the photometric temperatures derived in this work with spectroscopic temperatures from literature data (see Table \ref{tab:tempsources}). The 300 pc sample, which (having few dusty stars) should not be significantly affected by either circumstellar dust or errors in interstellar reddening corrections, still has an enormous scatter of $39\ (-534 $ to $ 434)$\,K, giving a Pearson correlation co-efficient between the two temperatures of only $R = 0.54$.

Errors may arise from the {\sc PySSED} SED fitting, spectroscopic temperatures, and intrinsic variability in the ``surface'' temperatures of the stars as they pulsate. 

Spectral temperatures derived from single-epoch observations are affected by stellar variability. Conversely, our SEDs are comprised of multi-epoch observations, so we expect our temperature estimates to be the more accurate. Supporting evidence for this can be seen in Figure \ref{fig:300pc_hrd}, top panel, which shows a scatter in the width of the giant branch of $\sim\pm 200$\,K, some of which will be intrinsic.

We can identify problems in spectroscopic temperatures by considering \emph{Gaia} and other literature separately (see bottom two panels of Figure \ref{fig:Tspec}). Non-\emph{Gaia} temperatures show a difference from {\sc PySSED} of --24\ (--658 to 224)\,K and $R = 0.64$. This scatter is still far in excess of the expected (--146 to 302)\,K error indicated above. Figure \ref{fig:Tspec}, bottom panel, shows relatively good recovery of temperatures for most objects, but a long tail of objects exists where spectroscopic temperatures are considerably warmer than the fitted photometric temperatures. We also display stars analysed as part of the 300 pc sample's construction, but either too faint or marginally too distant to qualify. These resolve this long tail into a sequence of stars offset below the parity line by $\sim$700\,K.

This offset sequence comes mostly from the PASTEL meta-catalogue, with original sources deriving from older (1980s/1990s) publications. These predate important advances in modelling M-star spectra, such as accurate TiO line lists. We, therefore, consider these earlier literature temperatures outdated.

As a further comparison, Table \ref{tab:sptypetemp} compares the literature spectral types of the 300 pc sample stars to the effective temperatures found from SED fitting against stellar atmosphere models. These stars are mostly dustless, except for the latest spectral types, so are not strongly affected by the above comparisons. The literature spectral types are taken from {\sc simbad}, with the spectral type stripped down to the first recognisable letter--digit pair (e.g., detailed values like K5.5 become truncated to K5, ranges like M2--M6e become M2, other designators like ``O-rich'' or ``C-rich' are ignored). Spectral types with $\leq$2 entries are left blank. These photometric temperatures can be compared to non-\emph{Gaia} and \emph{Gaia} spectroscopic temperatures (Tables \ref{tab:sptypetemp2} and \ref{tab:sptypetemp3}): the difference between the 300 pc and NESS samples within each table, and the growing difference in temperature between photometric and non-\emph{Gaia} spectroscopic temperatures between tables, both demonstrating the cooling effect of circumstellar dust on temperatures derived from SED fitting.

These three tables can be compared to literature conversion tables \citep[e.g.][]{FPT+94} to demonstrate that the \emph{Gaia} temperatures for late-type stars are systematically high (cf., Section \ref{sec:tspec300}). The offset between \emph{Gaia} {\sc apsis} GSP-Phot {\sc aeneas} spectral temperatures and {\sc PySSED} photometric temperatures is 334\ (7 to 1031)\,K and $R = 0.68$. The scatter in temperatures (root-mean-square, rms = 500\,K) is much larger the expected 200--300\,K \citep[cf.][their table 5]{Andrae2018}, and the offset is considerably larger than the typical tens of Kelvin.

Figure \ref{fig:Tspec}, middle panel, identifies a group of stars in the 300 pc sample (19 of 186) where {\sc PySSED}'s photometric temperature is 3500--4000\,K, wheras \emph{Gaia}'s spectroscopic temperature is 4250--5000\,K. These stars have {\sc simbad} spectral classifications of K3--M8, with most being K5--M2: classes much more consistent with the {\sc PySSED} temperatures than \emph{Gaia}, strongly suggesting poor recovery of surface temperatures by \emph{Gaia} in giant stars around this temperature. If we expand the sample to include fainter stars, as before, we find that \emph{Gaia} analysis effectively avoids assigning temperatures much lower than $\sim$4200\,K to giant stars in the solar neighbourhood, though no clear reason for this was resolved. \citet{GaiaApsis} makes no direct indication of how \emph{Gaia} {\sc apsis} models AGB stars, though the stellar evolution models used by GSP-Phot only extend as far as the RGB tip. Alternatively, the difference may be related to saturation limits within \emph{Gaia}. In either case, these effects are concerning for the use of \emph{Gaia} spectroscopic temperatures for giant-branch stars. 

As a check, we can also difference the \emph{Gaia} and non-\emph{Gaia} temperatures in a similar way, which provides an offset of 304 (38 to 589)\,K, with the \emph{Gaia} temperatures being higher, and $R = 0.85$, again indicating that \emph{Gaia} is the source of the disagreement.

The substantial differences between temperature estimates in the 300 pc sample mean that it is not realistic to use them to measure interstellar or circumstellar reddening, or $\dot{D}$.

\subsubsection{NESS and the effects of dust-enshrouding}

Figure \ref{fig:Tspec} shows that NESS sources fitted with model atmospheres (generally those in Tiers 0, 1 and 2) occupy similar regions of the diagram to the 300 pc sample, though they naturally tend to the cooler temperatures of the upper RGB and especially AGB, due to their sampling. A few cooler stars, mostly from Tier 2, have well-agreeing temperatures of 2000$\sim$3000\,K: these come from the {\sc pastel} catalogue, with more than half from the carbon-star data of \citet{Lambert86}.

The more extreme stars from Tiers 3 and 4 tend to lie to the left-hand side of the diagram (see Section \ref{apx:more:temp} for explanation), with spectral temperatures of $\sim$3500\,K, but where {\sc PySSED} fits much cooler blackbodies of 1000$\sim$2500\,K due to the dust that enshrouds them. As expected, the more extreme the tier, the further from the parity line the stars fall. Using $R^2 \propto T^4$, we can estimate that the SED-averaged $\tau=1$ opacity layer lies at approximately 2--12 times the spectroscopic stellar radius. This roughly corresponds to the range of distances from the star that different dust species begin to condense, as predicted to begin by models \citep[e.g.][]{BH12}. We note that scatter to the left or right of the line can occur if interstellar extinction is under- or over-corrected, respectively.

\subsection{Galactic AGB-star luminosity functions and dust production}
\label{sec:disc:300pc}

\begin{figure}
\centering
\includegraphics[height=\linewidth,angle=-90]{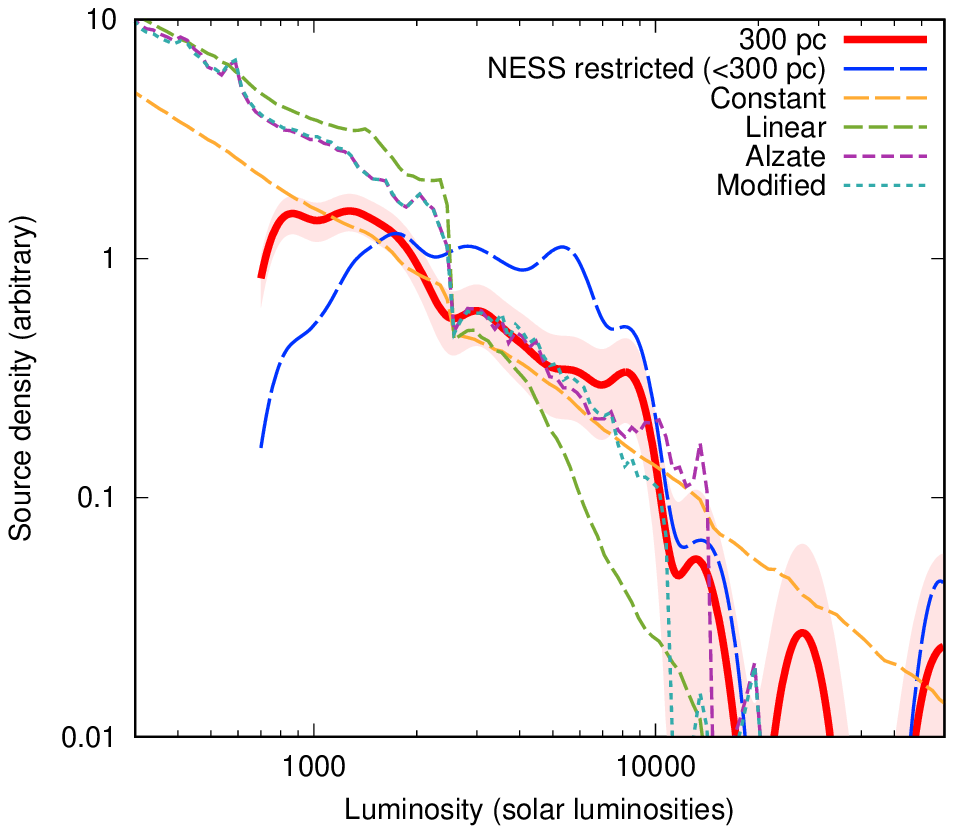}
\caption{Luminosity function of the 300 pc sample. A Gaussian kernel of width 0.05 dex was used to smooth the data. Also shown are luminosity functions from the {\sc parsec} stellar evolution models, assuming different SFHs (see text).}
\label{fig:lumfn}
\end{figure}

\subsubsection{The 300 pc sample: tracing star-formation history effects}
\label{sec:lumfnsfh}

Figure \ref{fig:lumfn} shows the luminosity function of the 300 pc sample. The shape of this luminosity function is dictated primarily by the initial-mass function (IMF) and stellar evolution, but has important second-order effects from the Galaxy's star-formation history (SFH). If we convolve the luminosity function from a stellar evolution model with the SFH of the solar neighbourhood, we should obtain the 300 pc sample. The AGB/RSG luminosity function can therefore, in theory, be used to measure the SFH of the solar neighbourhood \citep[cf.][and follow-on works]{Saremi2021}. In practice, uncertainties in the data limits us to comparison against pre-estimated SFHs.

Local stars formed in a wider volume. Peculiar velocities of stars in the solar neighbourhood are typically 8--15\,km\,s$^{-1}$ for stars 10$^8$--10$^9$ years in age and 25--50\,km\,s$^{-1}$ for stars 10$^9$--$5 \times 10^9$ years in age \citep{Griv2009}. A young star can cross the entire 300 pc sphere in 39--73 Myr, while older (more numerous) stars cross it in 12--23 Myr. All but the youngest, most massive stars in the 300 pc sample therefore diffused here from other parts of the Galaxy (though mostly those close to the solar circle), and thus represent the star-formation history of a wider swathe of the Galaxy.

We choose the {\sc parsec} stellar evolution tracks as a comparative stellar evolution model \citep{Bressan12,Pastorelli2019}, with solar metallicity and default settings. For our test SFH, we use \citet{Alzate2021}, using their $S^{15}_{100}$ sample with their Grid C isochrones, a \citet{Kroupa01} initial-mass function (IMF) and their $\sigma_i = 0.075$, as presented in their Figure 13(d). This local SFH peaks towards older ($\sim$10 Gyr) populations, but has smaller peaks at intermediate ($\sim$5 Gyr, $\sim$2 Gyr) timescales.

Figure \ref{fig:lumfn} first shows examples of luminosity functions generated for a constant star-formation rate (SFR) and a SFR that decreases linearly with time. A declining SFR decreases the relative number of bright AGB stars as the average time spent above the RGB tip decreases. The RGB tip also becomes more pronounced, as the smearing of the RGB tip in luminosity among a heterogeneous population transitions to a single, almost fixed luminosity in a population of a narrowly bracketed age. Compared to these two functions, the 300 pc sample more closely follows the constant SFH, with the RGB tip being less pronounced and the AGB above it being less steep. However, above $\sim$10\,000 L$_\odot$, there is a much more rapid fall-off of stars that better approximates the linear model.

The \citet{Alzate2021} SFH provides a reasonable overall prediction of the AGB luminosity function. However, it over-predicts the step at the RGB tip and the observed $\sim$10\,000 L$_\odot$ step occurs at a higher luminosity ($\sim$14\,000 L$_\odot$). The RGB-tip step is mostly controlled by older populations ($\gtrsim$10 Gyr). However, since the luminosity functions from the {\sc parsec} luminosity functions do not include our $T < 5000$\,K criterion, it is difficult to accurately model. The higher-luminosity step is controlled by the youngest burst of star-formation, which dictates the maximum AGB mass and the highest luminosity that stars will reach. This can be better replicated if populations of $\sim$1.0 Gyr are reduced. The final function in Figure \ref{fig:lumfn} shows a modification to the \citet{Alzate2021} SFH, removing the contribution from the 500 Myr bin, which better reproduces the observed function.

Alternatively, AGB stars of $\sim$500 Myr in age could be incorrectly modelled in {\sc parsec}: 10\,000\,L$_\odot$ is roughly the upper luminosity limit for carbon stars (Section \ref{sec:lumfnc}). An imprecise treatment of mass loss around this boundary (which also depends on details of atmospheric chemistry) could incorrectly predict of the luminosity function in this regime. The slight excess of AGB stars at $\sim$8000 L$_\odot$ compared to the {\sc parsec} model could mean that the highest-mass carbon stars (or the lowest-mass stars undergoing hot bottom burning) are not quite attaining the luminosities expected.

\subsubsection{Dust production and the NESS sample}
\label{sec:lumfndust}

\begin{table}
    \centering
    \caption{Fraction of dusty stars in different luminosity ranges}
    \label{tab:dusty}
    \begin{tabular}{ccc}
    \hline\hline
    Luminosity & Counts & Percentage\\
    (L$_\odot$) & \ & \ \\
    \hline 
    700--2000  & 67 of 353 & 19 \\
    2000--3000 & 34 of 59  & 58 \\
    3000--5000 & 36 of 59  & 61 \\
    5000--10\,000 & 37 of 50 & 74 \\
    $>$10\,000    & 3 of 9  & 33 \\
    \hline
    \end{tabular}
\end{table}

\begin{figure}
\centering
\includegraphics[height=\linewidth,angle=-90]{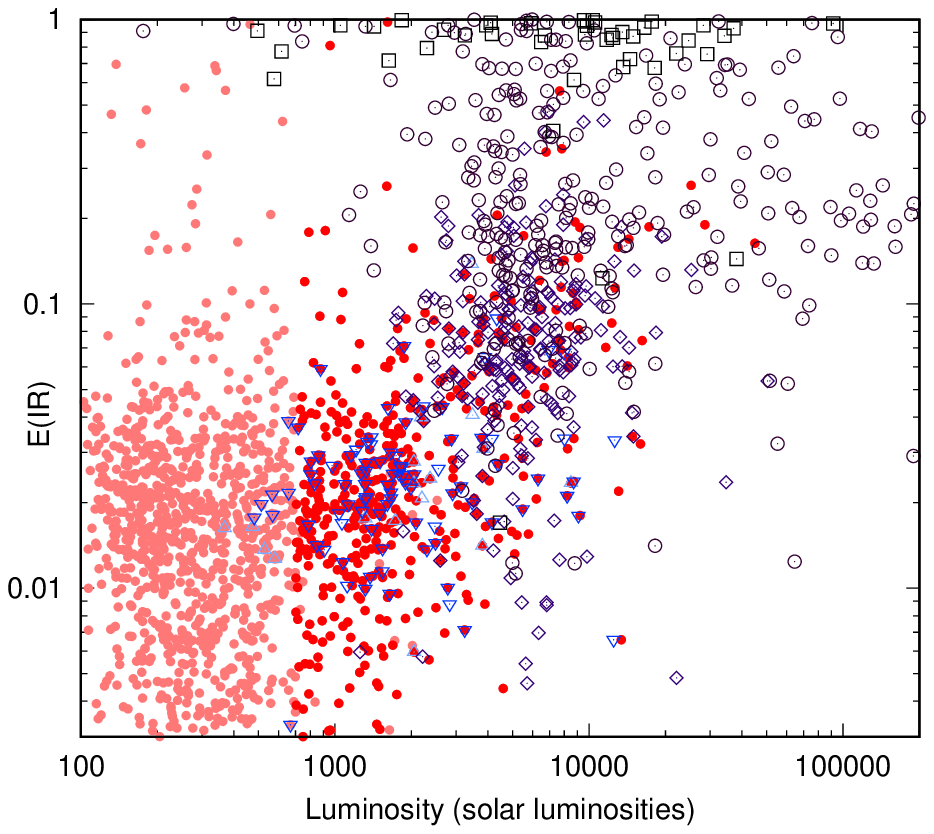}
\caption{An approximation of the fraction of stellar luminosity reprocessed by circumstellar dust, $E_{\rm IR}$. Colours and point shapes are as in previous plots. The pink symbols show sampled giant stars within 300 pc that did not meet the temperature and luminosity criteria for inclusion in the 300 pc sample.}
\label{fig:eir}
\end{figure}

\begin{figure}
\centering
\includegraphics[height=\linewidth,angle=-90]{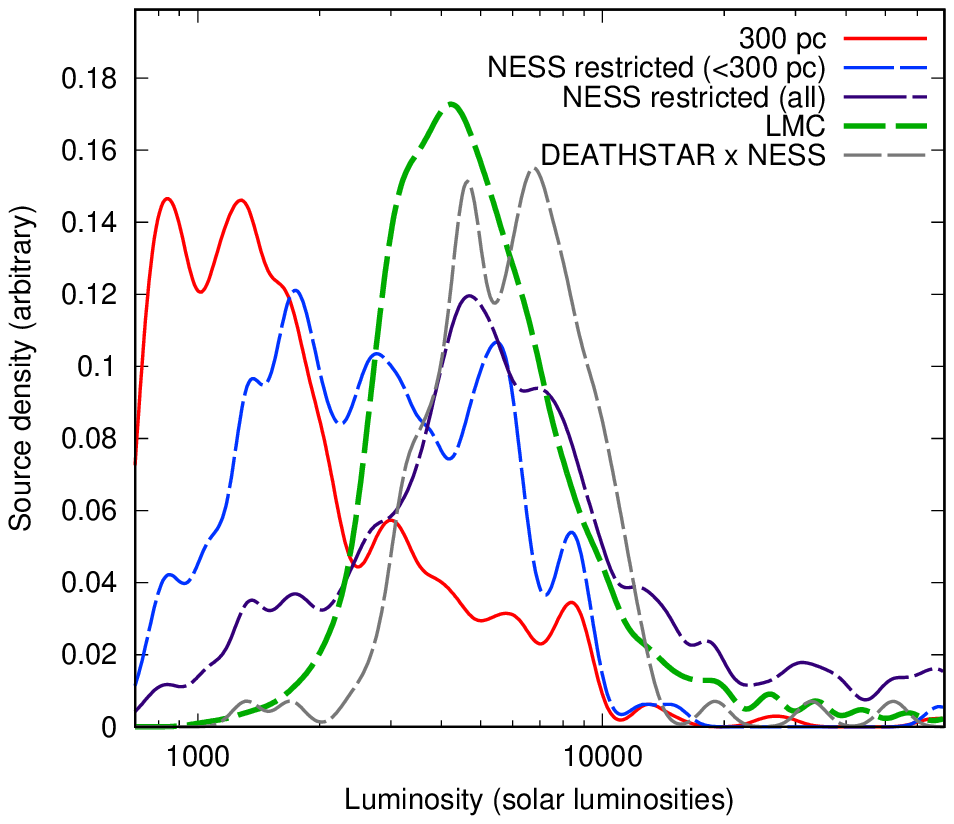}
\caption{Luminosity functions of the 300 pc sample, NESS stars within 300 pc, the entire NESS sample and the overlap between the DEATHSTAR and NESS surveys.}
\label{fig:deathstar}
\end{figure}

\begin{table}
    \centering
    \caption{Average values and central 68 per cent and 95 per cent intervals (68\% CI / 95\% CI) for $E_{\rm IR}$ in different tiers.}
    \label{tab:eir}
    \begin{tabular}{cccc}
    \hline\hline
    Tier & Mean & 68\% CI & 95\% CI\\
    \hline 
    300\,pc ($<$700\,L$_\odot$) & 0.010 & \llap{$-$}0.001--0.029 & $-$0.027--0.081 \\
    300\,pc ($>$700\,L$_\odot$) & 0.019 & 0.003--0.053 & $-$0.020--0.174 \\
    Tier 0 & 0.018 & 0.006--0.031 & --- \\
    Tier 1 & 0.023 & 0.010--0.033 & $-$0.010--0.069 \\
    Tier 2 & 0.065 & 0.026--0.120 & $-$0.003--0.261 \\
    Tier 3 & 0.208 & 0.089--0.584 & \ \,0.022--0.964 \\
    Tier 4 & 0.880 & 0.618--0.969 & 0.017$^1$--0.995 \\
    \hline
    \multicolumn{4}{p{0.9\columnwidth}}{$^1$Two unphysical values have been removed and treated as errors.}\\
    \hline
    \end{tabular}
\end{table}

NESS effectively sub-samples the local distribution of \emph{dusty} AGB stars. Comparing the NESS and 300 pc samples therefore probes dust production at different luminosities. However, obtaining a true and complete luminosity function for the NESS survey is currently impossible. A lack of distance estimates affects the extreme tiers of sources, though has only minor ($\sim$10 per cent) effects in lower tiers (cf., Table \ref{tab:completeness}). When coupled with the enhanced difficulty in obtaining stellar parameters for extreme dust-producing stars (Sections \ref{apx:more:temp} and \ref{apx:more:errors}), this means adding the extreme tier to a luminosity function of NESS sources is only an approximate process. By using the NESS restricted sample here, we avoid stars with highly uncertain luminosities at the expense of incompleteness.

The small numbers in NESS tiers allow us only to define the fraction of dusty stars in some key luminosity ranges (Table \ref{tab:dusty}).
Except for the very brightest stars (whose small numbers make them unreliable), the fraction of AGB stars showing noticeable dust production at a given luminosity remains relatively constant above the RGB tip, rising only slowly from $\sim$60 to $\sim$75 per cent.

We can also introduce $E_{\rm IR}$, the fraction of stellar light reprocessed into the infrared by dust around the star, which is related to the SED-averaged optical depth of the circumstellar dust, $\tau$, as $\tau = -\ln(1-E_{\rm IR})$. This can be approximated by taking the modelled SED longwards of 2\,$\mu$m, and separating it into two components. From the first point in the remaining SED ($F_2$ at $\lambda_2$), we can estimate the central star's contribution to the infrared SED by assuming a Rayleigh--Jeans law of $F \propto \lambda^{-2}$. We can then determine the luminosity of dust, $L_{\rm dust}$ as the integral of the remaining SED:
\begin{equation}
L_{\rm dust} = \int_{\lambda = 2\,\mu {\rm m}}^\infty F_\lambda - F_2 \left(\frac{\lambda}{\lambda_2}\right)^{-2} d\lambda .
\end{equation}
This allows us to compute
\begin{equation}
E_{\rm IR} = L_{\rm dust} / L ,
\end{equation}
where $L$ is the total luminosity of the star integrated across all wavelengths. We show the computed values for $E_{\rm IR}$ in Figure \ref{fig:eir}.

Dustless stars typically have an $E_{\rm IR}$ of a few per cent, driven by the 2.2\,$\mu$m CO band, which places the $K_{\rm s}$ band below much of the IR SED and therefore offsets $E_{\rm IR}$ slightly from zero. Median values and intervals for each tier, plus stars within 300 pc below $L = 700$\,L$_\odot$, can be seen in Table \ref{tab:eir}. The few stars in Tier 0 are broadly consistent with the dustless stars in the 300 pc sample. Tier 1 stars have a slightly higher mean and 68 per cent interval, but cannot be isolated from dustless stars. Tiers 2, 3 and 4 stars have progressively higher $E_{\rm IR}$, commensurate with higher dust production, with values of unity (complete dust obscuration) being increasingly more common.

Considering all stars in Figure \ref{fig:eir}, $E_{\rm IR}$ discontinuously jumps from a ``dustless'' few per cent to $E_{\rm IR} \approx 0.1$ just above the RGB tip ($L \approx 2500$\,L$_\odot$). Investigation of individual stars suggests that this sudden jump corresponds to the major increase in dust production \citep{MZ16b} caused when stars begin long-secondary-period (sequence D) oscillations \citep{MT19}. While we retain discussions of pulsation for future work, most Galactic stars appear to go through this transition between about 2000 and 5000 L$_\odot$. Pulsation-sequence transitions occur earlier at lower masses \citep{Trabucchi2021} and the AGB lifetime, IMF and SFH dictate\footnote{Integration of the Padova isochrones convolved with the modified \citet{Alzate2021} SFH indicate $\sim$54 per cent of AGB stars brighter than the RGB tip should have $M_{\rm init} < 1.6$\,M$_\odot$.} that the lowest mass AGB stars ($M_{\rm init} \approx 0.8-1.6$\,M$_\odot$, $M_{\rm RGB\ tip} \approx 0.65-1.5$\,M$_\odot$) should be most numerous. It is therefore surprising that the onset of dust production in Galactic stars typically occurs at luminosities much higher than in globular clusters \citep[700--2300\,L$_\odot$][]{BvLM+10,MBvLZ11}, where RGB-tip masses are only slightly lower ($M_{\rm init} \approx 0.8-0.9$\,M$_\odot$, $M_{\rm RGB\ tip} \approx 0.60-0.70$\,M$_\odot$; \citealt{TML+20}). This implies either a relatively strong mass dependence in the luminosity at the onset of dust production, or a significant difference between the way that dust production works in globular clusters and our Galaxy \citep[cf.][]{MZ15a,MBG+19}. We expect this to be unrelated to the lower metallicity of globular clusters, since a lower metallicity would suggest a \emph{higher} threshold for sustaining a dust-driven wind. Instead, we suggest that this difference could be related to either RGB mass-loss efficiency \citep[e.g.][]{MZ15b,TML+20}, or the absence of third-dredge up in the lowest-mass stars \citep[cf.][]{Uttenthaler2019,Uttenthaler2024}.

\subsubsection{NESS in the context of other surveys}
\label{sec:deathstar}

Figure \ref{fig:deathstar} compares the entire NESS sample (781 AGB/RSG stars), to the LMC sample of \citet{Riebel12} and the DEATHSTAR (``Determining accurate mass-loss rates of thermally pulsing AGB stars'') project \citep{Ramstedt2020}. This informs us of how each survey samples the evolved-star distribution.

\citet{Riebel12} selected only stars brighter than RGB tip (plus some dust-producing stars up to 1\,mag below the RGB tip). This hard cutoff leads to significant incompleteness around the RGB tip once bolometric luminosities are computed, as some AGB stars will have been scattered below the RGB tip. The approximate completeness limit is expected to be just above the peak of the LMC luminosity function ($L \sim 4200$\,L$_\odot$). This severe selection effect hampers proper comparison of the LMC and Galactic luminosity functions. However, the LMC luminosity function is more strongly peaked than the NESS sample within the range of its completeness.

DEATHSTAR is essentially a meta-study of previously observed objects, and can therefore probe biases in literature sub-mm observations of evolved stars. We only have computed luminosities for 118 of the 201 stars that overlap with NESS, and it is this DEATHSTAR--NESS cross-matched list that forms the luminosity function in Figure \ref{fig:deathstar}. It comprises:
\begin{itemize}
    \item one of 48 Tier 4 stars,
    \item 41 of 301 Tier 3 stars,
    \item 73 of 220 Tier 2 stars,
    \item three of 97 Tier 1 stars,
    \item zero of 19 Tier 0 stars.
\end{itemize}
Assuming the overlapping 118 stars are broadly representative of the DEATHSTAR survey, existing literature preferentially misses both the most extreme AGB stars (which contribute most to the chemical enrichment of the Galaxy) and the lowest mass-loss rate stars (which are numerically the most common AGB stars). Instead, existing data concentrates on stars with intermediate mass-loss rates, which are not optically obscured, identifiable via long-period variability, and numerically common enough to be nearby (thus avoiding the heavy confusion and interstellar extinction in the Galactic plane).

NESS therefore crucially benefits our understanding of AGB stars by concentrating effort on both these lower-luminosity tiers filled with more typical AGB stars, while also trying to understand the dominant dust-producing sources in the Galaxy: the extreme, optically obscured AGB stars.

\subsubsection{Density of stars by tier}
\label{sec:lumfndensity}

\begin{figure}
\centering
\includegraphics[height=\linewidth,angle=-90]{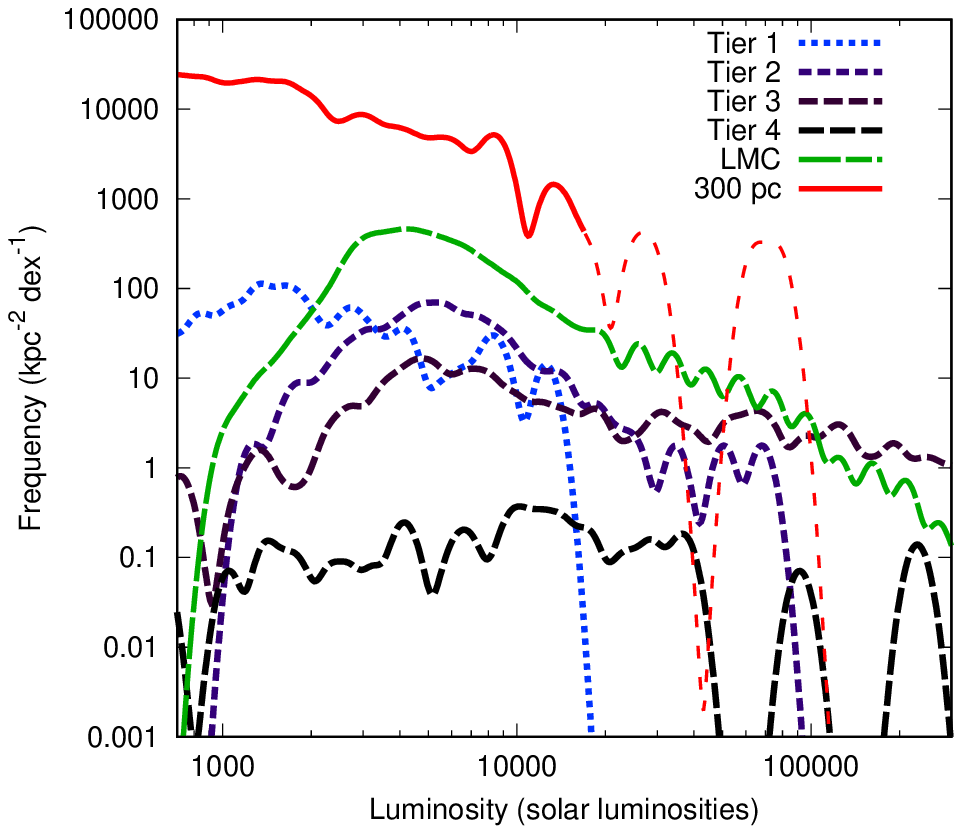}
\includegraphics[height=\linewidth,angle=-90]{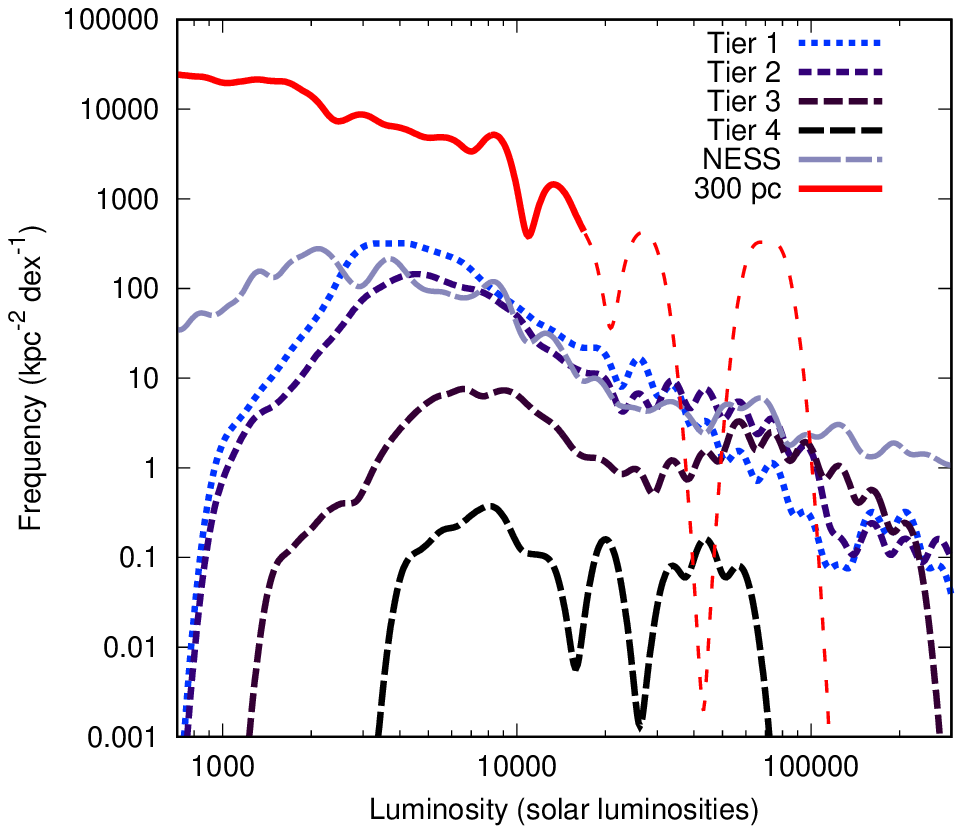}
\caption{Histograms, showing the luminosity density functions of the different NESS tiers (see Section \ref{sec:lumfndensity} for details). The top panel shows the NESS survey itself, broken down by tier. The corresponding luminosity functions for the \citet{Riebel12} LMC sample and the 300 pc sample are shown for comparison. The 300 pc sample becomes unreliable beyond $\sim$17\,000 L$_\odot$ due to small numbers of stars, indicated by the dashed line.  The bottom panel shows the same luminosity functions for the LMC, generated by mapping the NESS $\dot{D}$ tiering criteria onto the \citet{Riebel12} sample and assuming a 100 kpc$^2$ effective area for the LMC. The NESS and 300 pc samples are likewise shown for comparison. Tier 0 has too few data to show in NESS and no comparison in the LMC.}
\label{fig:lumfn2}
\end{figure}

Figure \ref{fig:lumfn2} compares the stellar density of our different samples and separates the restricted-data NESS sample into its respective tiers. Normalisation of the luminosity functions between the tiers is difficult, since it must assume that the stellar density is uniform across the different sampled radii. However, the larger tiers sample regions progressively further beyond the Galactic plane, where there are fewer stars (see Section \ref{sec:3D}). To approximate this behaviour, we have normalised the tiers to their stellar density per unit area of the Galactic plane. This normalisation will marginally under-estimate Tiers 0 and 1, as their spherical volumes only extend $\sim$1 scale height above the plane, however, it will also under-estimate Tiers 2, 3 and 4, due to the restrictions placed on including stars at $|b| < 1.5$\,deg (Table \ref{tab:NESS}). We therefore avoid comparisons between tiers, except to sum the stellar densities among different tiers to provide a luminosity function for the entire NESS survey: the amalgam of different scales means this function will only be approximate.

The NESS dust-production rate tiering system can be applied to the LMC data (no distance tiering is needed, as we can consider the galaxy as a whole), however, the Tier 0 stars is merged with Tier 1 since \citet{Riebel12} always provides a positive $\dot{D}$.

Even given the lack of completeness of the LMC sample below $\sim$4200\,L$_\odot$, With these caveats, we can still see that the extreme stars in the Magellanic Clouds are more concentrated at intermediate luminosities, while those in the NESS sample are scattered to higher and lower luminosities. The LMC has no Tier 4 (``extreme'') dust-producing stars below 4300\,L$_\odot$ nor above 56\,766\,L$_\odot$. This reflects the difficulty in establishing distances to these extreme stars in our Galaxy, resulting in inaccurate luminosities. Conversely, the absence of Tier 0, 1 and 2 stars above $L \sim 15\,000$ L$_\odot$ in the NESS sample is not reflected in the LMC, where Tier 1 stars still make up over half the sample up to $L \sim 30\,000$ L$_\odot$, and Tier 2 stars continue to dominate the sample up to $L \sim 100\,000$ L$_\odot$. The reasons lower $\dot{D}$ can be maintained in the LMC to higher luminosities is not clear but may be linked to the lower metallicity, star-formation histories and/or associated differences in the formation of carbon stars \citep[e.g.][]{Cioni2006} and their associated dust \citep[e.g.][]{Sloan2016}. Further exploration of this fact may prove useful in uncovering the effect of radiation pressure on dust and its role in driving a stellar wind \citep[cf.][]{GVM+16,MBG+19,MUZ+20}.

\subsubsection{Carbon-star luminosity functions}
\label{sec:lumfnc}

\begin{figure}
\centering
\includegraphics[height=\linewidth,angle=-90]{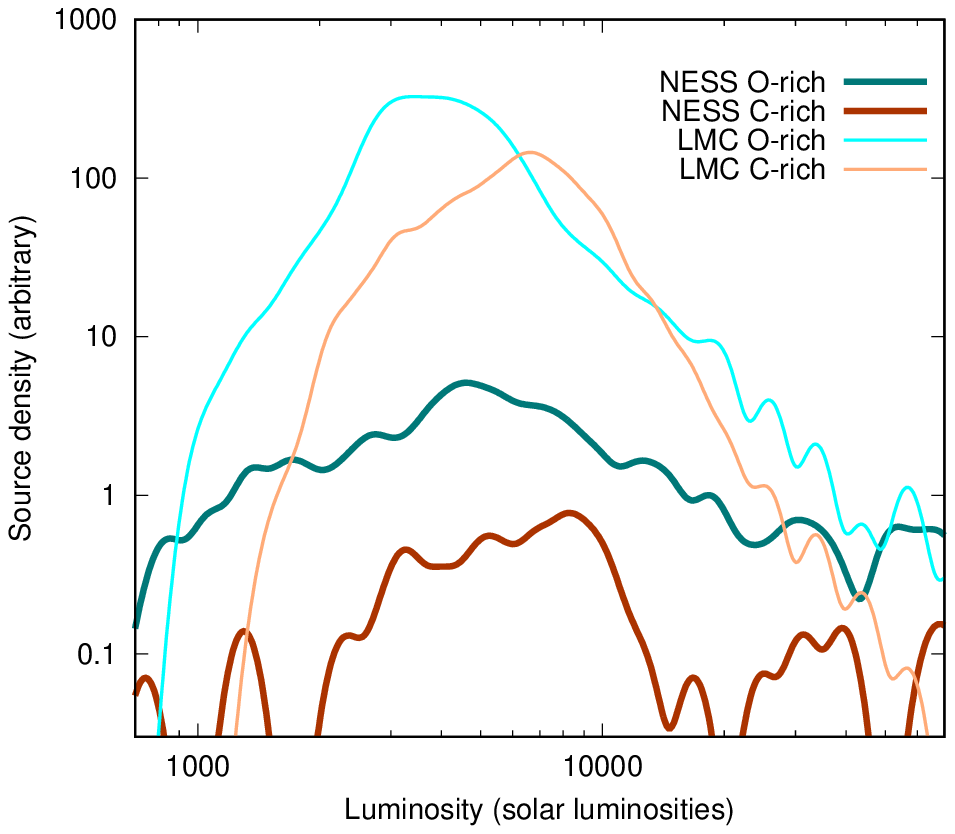}
\caption{Luminosity functions for C-rich and O-rich stars in the NESS criteria-matching sample and the LMC sample of \citet{Riebel12}. ``Extreme'' stars in the LMC, as defined by \citet{Riebel12}, have been merged into the C-rich sample. The absolute source densities for C-/O-rich stars are comparable within the NESS and LMC samples, but cannot be absolutely scaled between the galaxies.}
\label{fig:lumfnc}
\end{figure}

The ancillary data collected by {\sc PySSED} includes spectral types (see the {\tt ancillary.ness} file in the Supplementary Material for citations). K-type and M-type spectral classes were labelled as oxygen-rich, and C-type spectral classes labelled as carbon-rich. Other spectral classes were ignored (including classes of S-type stars). Of the 649 criteria-meeting NESS stars, 594 had optical spectral types.

The \emph{IRAS} Low-Resolution Spectrograph data \citep{Olnon1986} was used to separate obscured O-rich and C-rich stars, bringing the total number of stars with spectral classes to 635 out of the 649, of which 67 (11 per cent) are C-rich. A per-tier summary is given in Figure \ref{fig:venn}. Figure \ref{fig:lumfnc} shows the luminosity functions for O-rich and C-rich stars separately, alongside similar O-rich versus C-rich data for the LMC, generated from Table 3 of \citet{Riebel12}. Since the NESS C-/O-rich luminosity functions are from a tiered survey, they can only be compared against each other, and should not be taken as absolutely calibrated in shape or amplitude. Similarly, the LMC sample remains cut off at low luminosities.

Evolutionary models \citep[e.g.][]{Pastorelli2019} predict fewer carbon stars at higher metallicity, since stars need to generate more carbon to overcome a higher initial oxygen abundance. The lower mass boundary for carbon-star formation will therefore be higher in the Milky Way than in the LMC. The upper mass boundary is set by the onset of hot bottom burning, which is not expected to be strongly metallicity dependent. We therefore expect the Milky Way carbon-star luminosity function to begin at higher luminosities than the LMC and comprise fewer stars. A small subset of \emph{extrinsic} carbon stars will also exist in both galaxies due to pollution by carbon-rich binary companions.

The uncertain distances still cause problems with our Galactic luminosity functions, causing stars in both carbon- and oxygen-rich luminosity functions to sporadically appear at arbitrarily high and low luminosities. However, we note that the NESS carbon-star luminosity function peaks at a higher luminosity ($\sim$8300 L$_\odot$) than the LMC function ($\sim$6600 L$_\odot$), while the ratio of carbon:oxygen stars between 500 and 10\,000 L$_\odot$ is 32:165 ($\approx$1:5.2) in the NESS sample, whereas in the LMC the ratio is 1:0.9. The selection function of the NESS tiering system complicates an exact measurement, but we can approximate that the solar circle of the Milky Way contains $\sim$six times fewer carbon stars than the LMC.

\subsection{3D distribution of evolved stars}
\label{sec:3D}

\begin{figure*}
\centering
\includegraphics[height=0.45\linewidth,angle=-90]{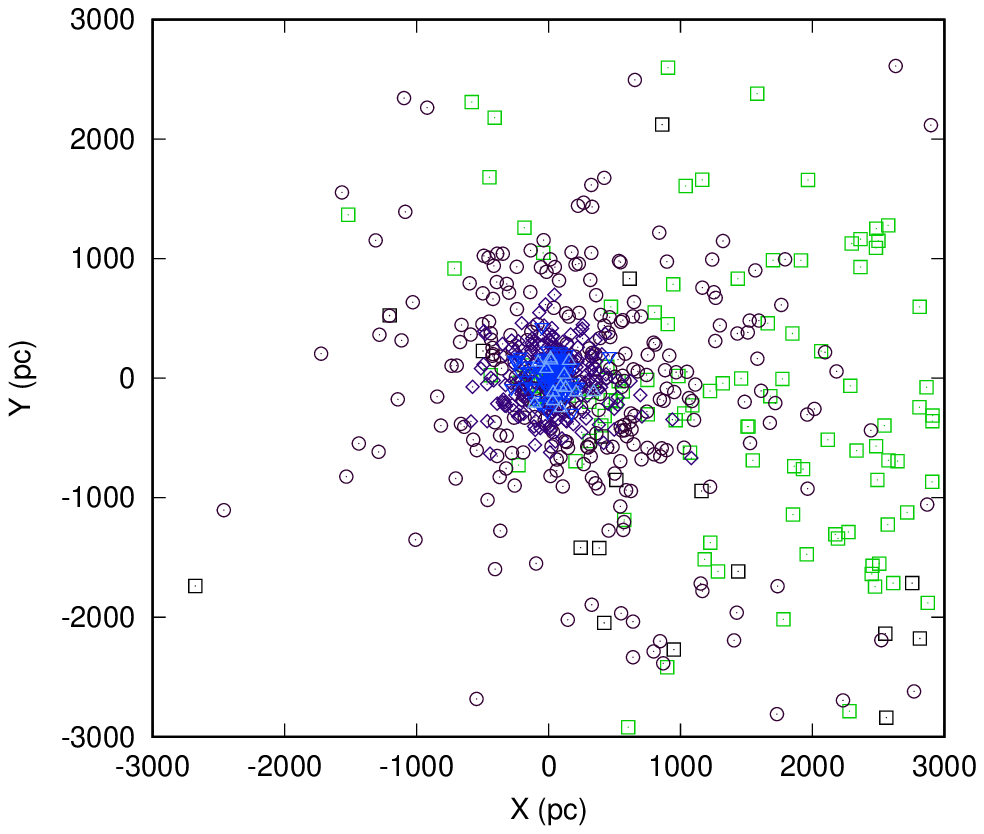}
\includegraphics[height=0.45\linewidth,angle=-90]{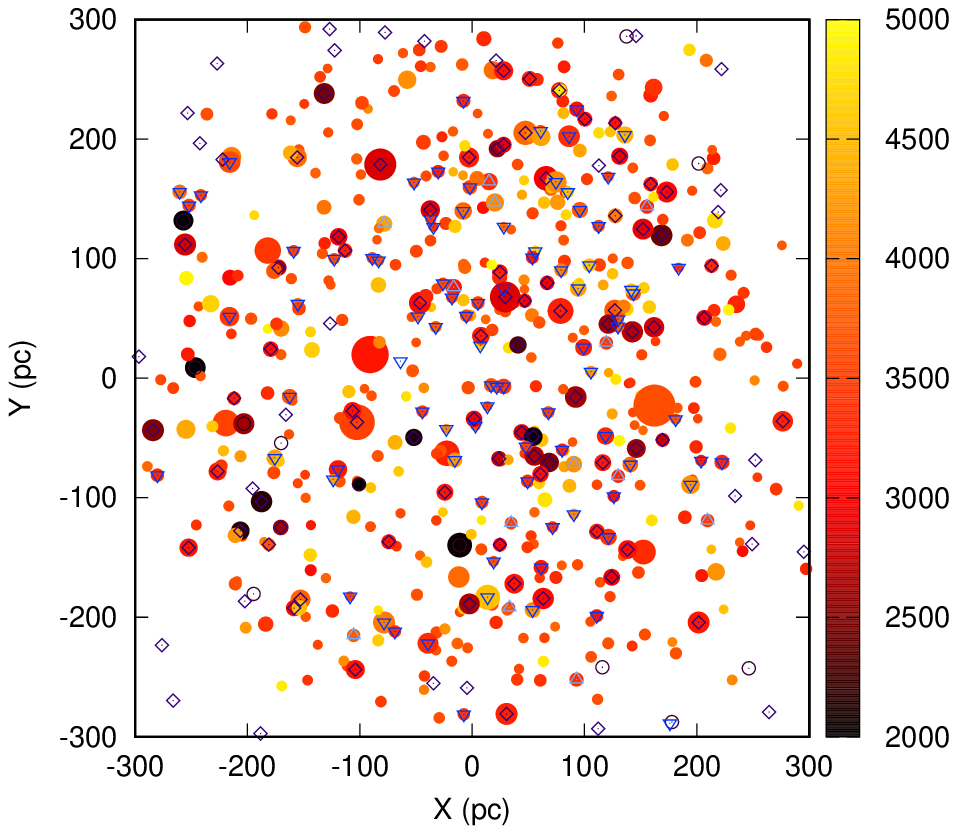}
\includegraphics[height=0.45\linewidth,angle=-90]{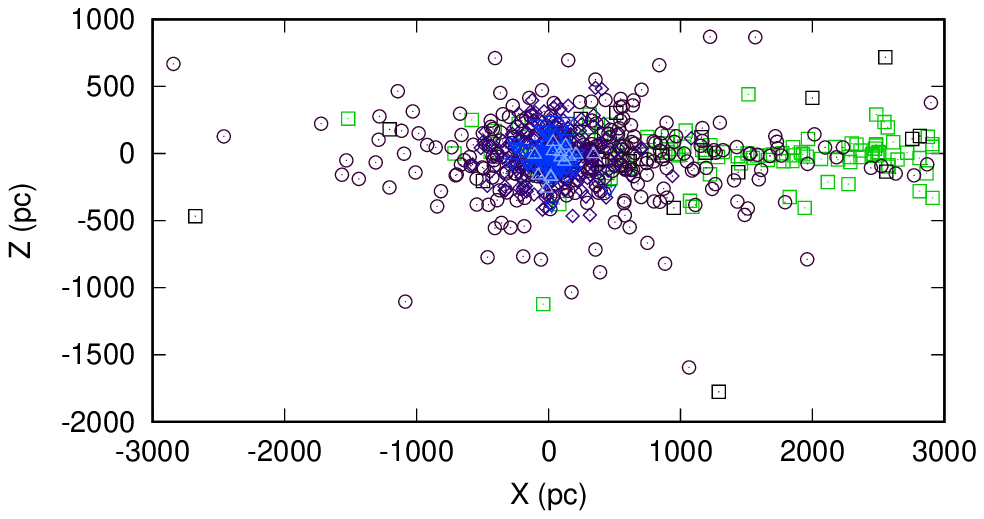}
\includegraphics[height=0.45\linewidth,angle=-90]{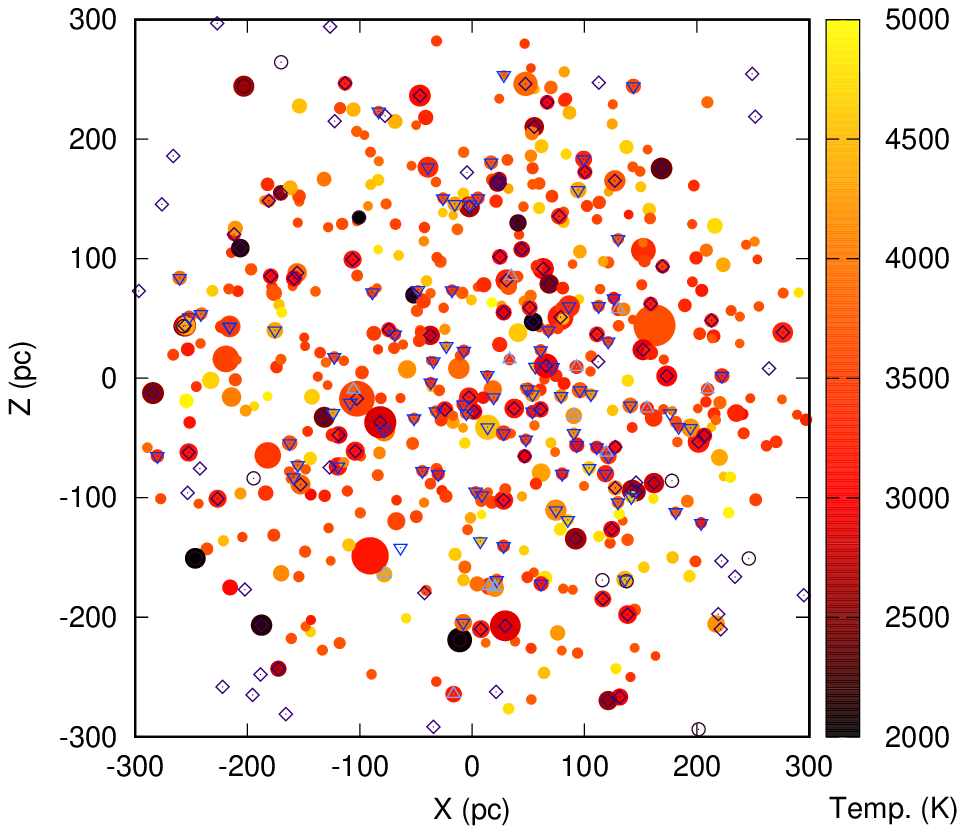}
\caption{\emph{Left panel:} Galactic XYZ co-ordinates of sources in the NESS criteria-meeting sample. The colours and point shapes represent different tiers, as in previous plots. Green points show other (non-rejected) NESS objects with parameters inconsistent with AGB/RSG stars ($T_{\rm eff} > 5000$\,K, $L < 700$ L$_\odot$ or $L > 200\,000$ L$_\odot$), or confirmed AGB/RSG stars with only period--luminosity- or luminosity-based distances. Most of these are in the ``extreme'' Tier 4 (see Figure \ref{fig:venn}). \emph{Right panel:} a zoom to within 300 pc of the Sun. The 300 pc sample is shown, colour-coded by temperature, with point size proportional to luminosity.}
\label{fig:XYZ}
\end{figure*}

\begin{figure}
\centering
\includegraphics[height=\linewidth,angle=-90]{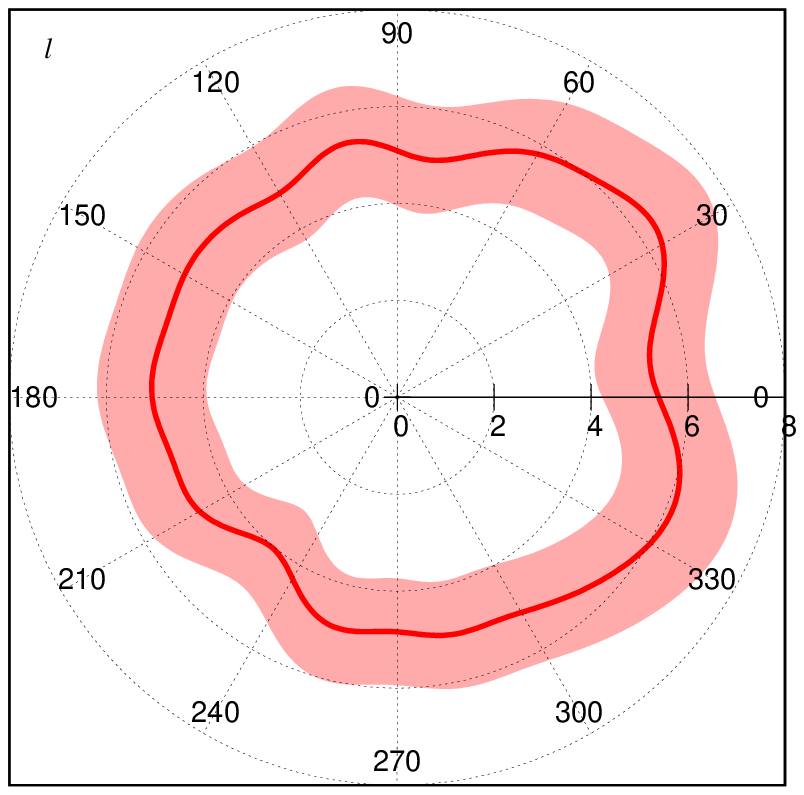}
\includegraphics[height=\linewidth,angle=-90]{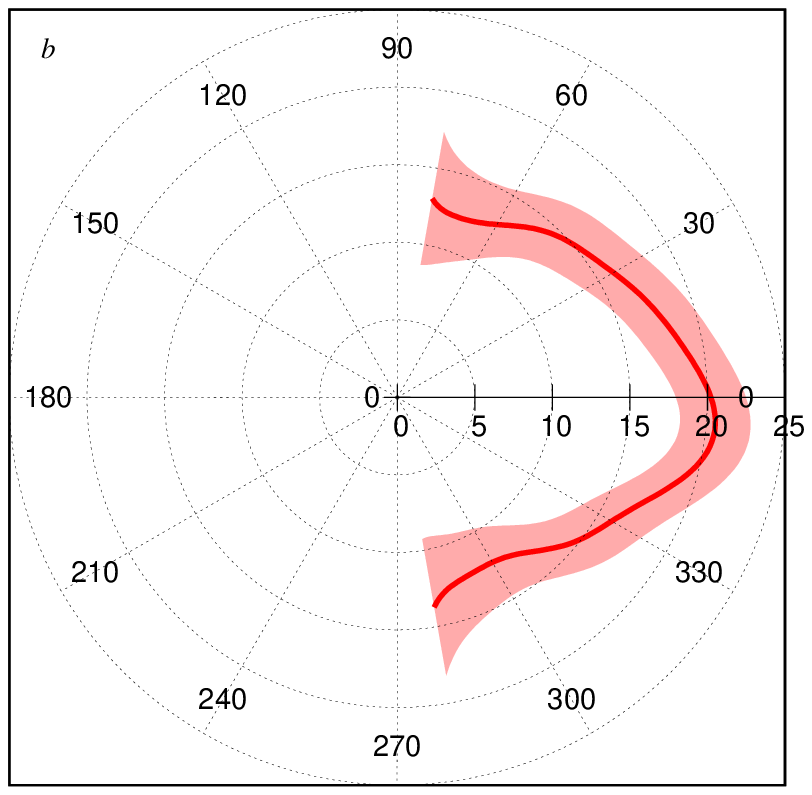}
\includegraphics[height=\linewidth,angle=-90]{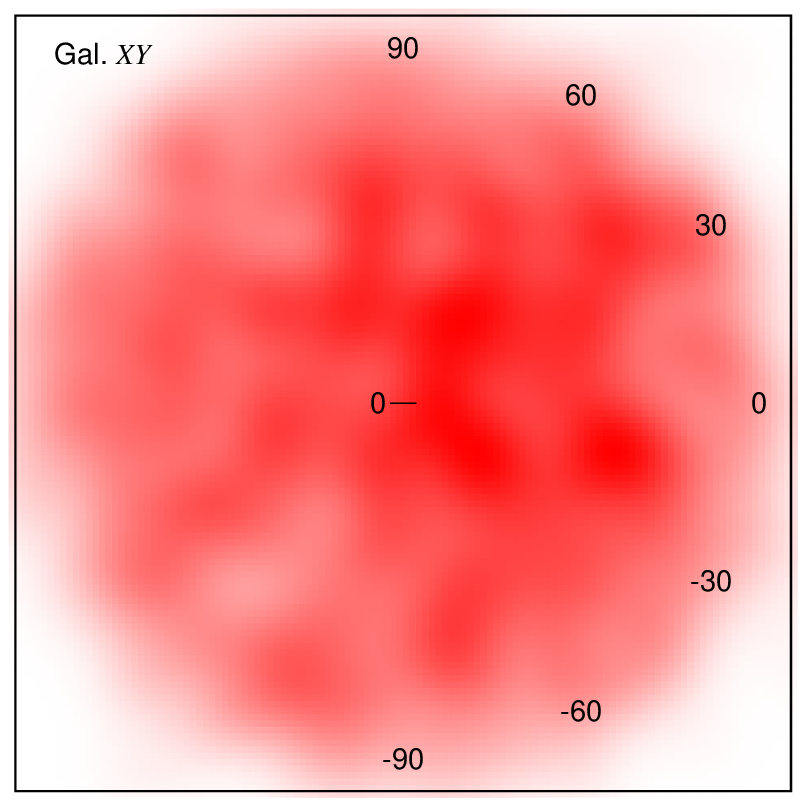}
\caption{Density plots showing number of stars per degree in Galactic longitude (\emph{top panel}) and Galactic latitude (\emph{middle panel}, a division by $\cos(\delta)$ is applied to account for area differences). A Gaussian smoothing factor of 10$^\circ$ has been applied to each plot, with 1$\sigma$ errors applied as the square root of the sum of Gaussians (coloured regions). \emph{Bottom panel:} a two-dimensional density plot of stars in the 300 pc sample, around the Galactic plane. A Gaussian smoothing factor of 25 pc is applied.}
\label{fig:radialhist}
\end{figure}

\begin{figure*}
\centering
\includegraphics[height=0.47\linewidth,angle=-90]{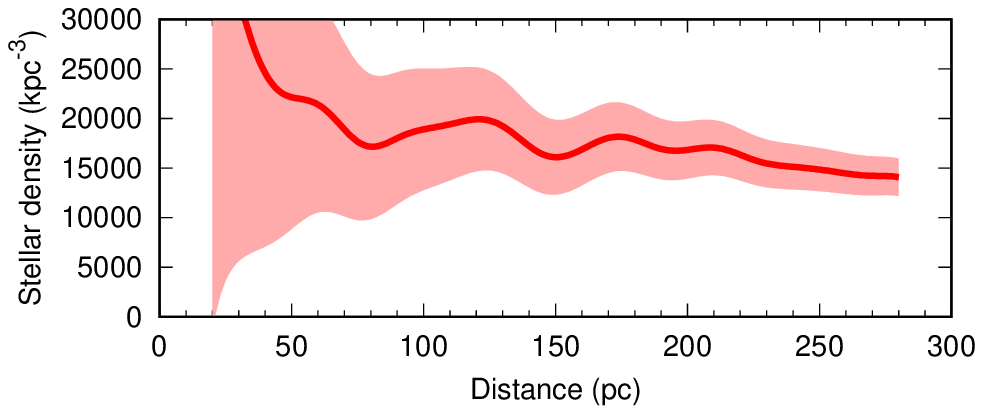}
\includegraphics[height=0.47\linewidth,angle=-90]{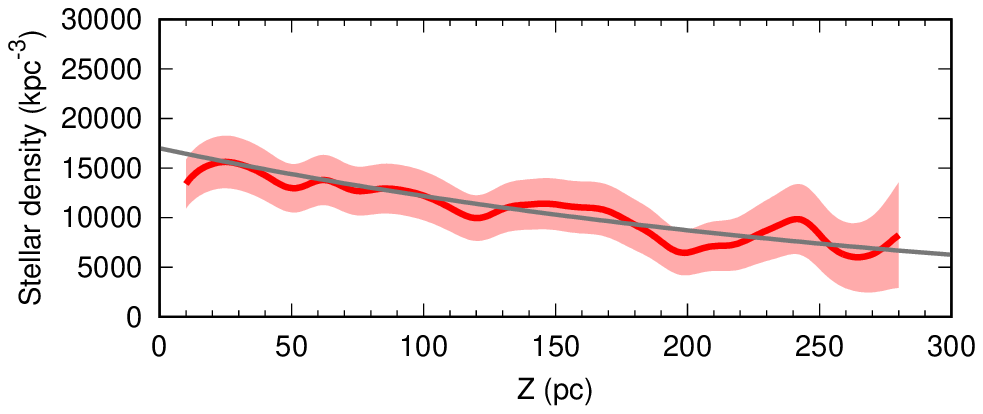}
\includegraphics[height=0.47\linewidth,angle=-90]{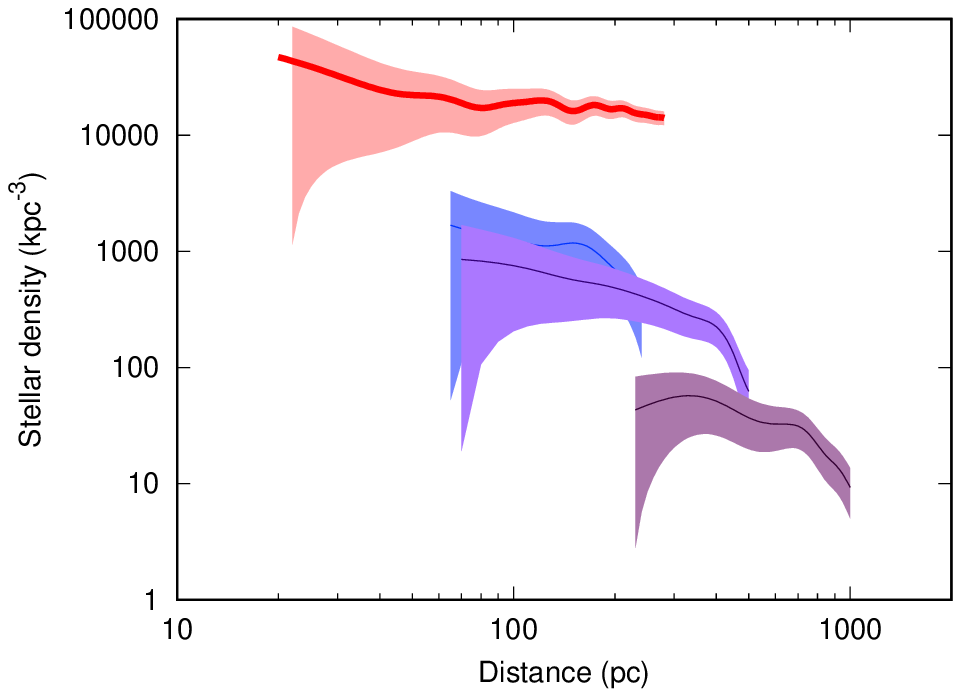}
\includegraphics[height=0.47\linewidth,angle=-90]{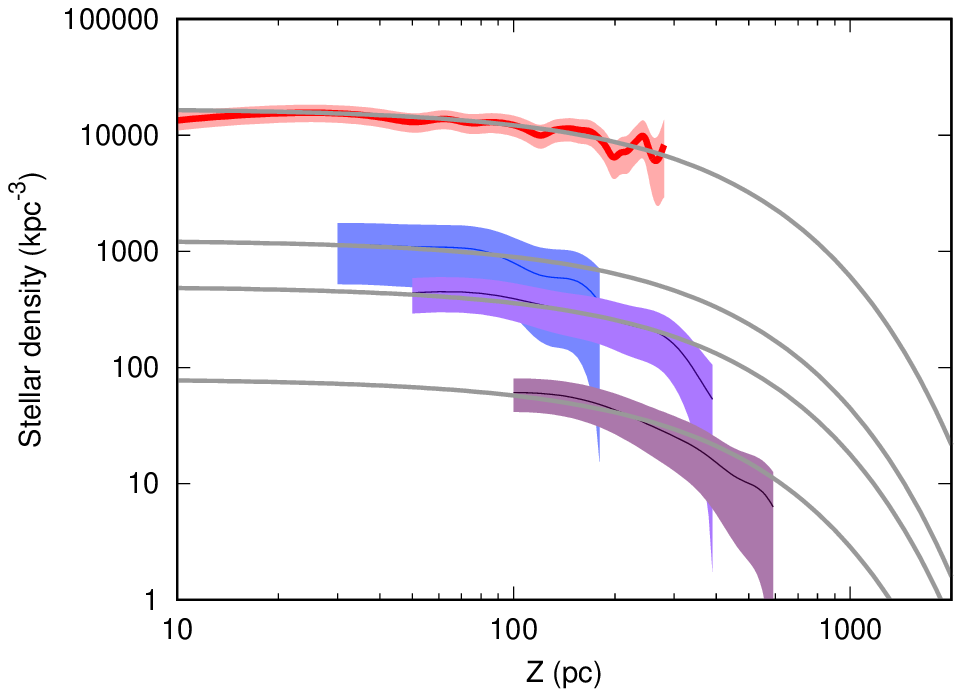}
\includegraphics[height=0.47\linewidth,angle=-90]{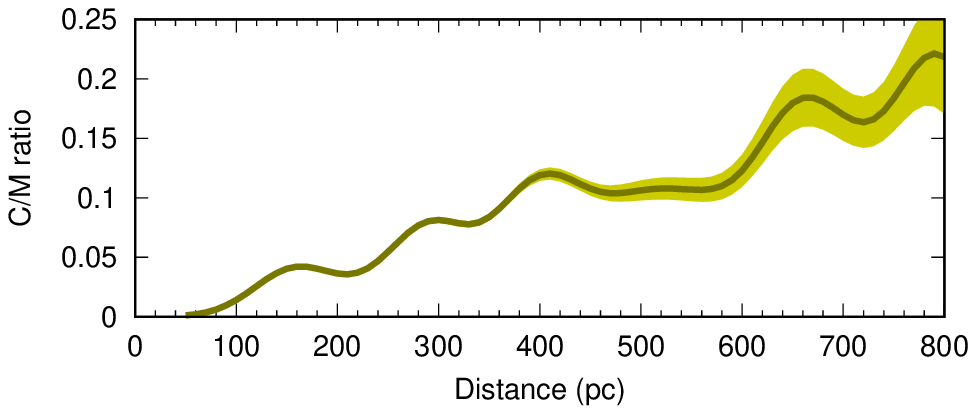}
\includegraphics[height=0.47\linewidth,angle=-90]{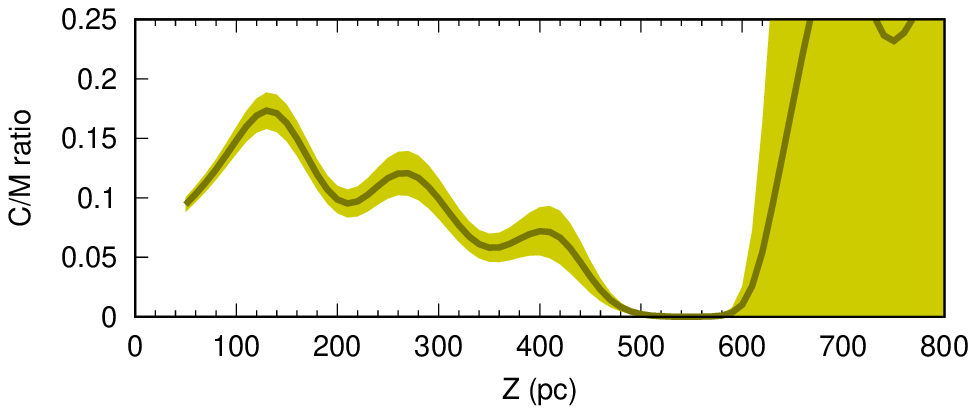}
\caption{Evolution of space density of evolved stars versus (left) distance from the Sun and (right) distance from the Galactic Plane. Coloured bands show the 1$\sigma$ confidence intervals. The top panels show the 300 pc sample; the middle panels include NESS Tiers 1, 2 and 3 (Tiers 0 and 4 have too few stars); the bottom panels show the ratio of C-rich to O-rich stars (C/M). For the right-hand panels, only stars within $\sqrt(X^2+Y^2) < 0.707\times$ the tier boundary are shown (2 kpc for the C/M ratio), to avoid problems near the Galactic Plane at larger distances, and the grey lines show tracks for a 300 pc scale height. Note a strong radial sampling bias exists in the C/M data due to the NESS tiering (see text).}
\label{fig:distz}
\end{figure*}

\subsubsection{300 pc sample}

Figure \ref{fig:XYZ} shows the 3D distribution of individual stars in both the NESS survey and the 300 pc sample. The statistical distribution of stars in the 300 pc sample can be seen in Figure \ref{fig:radialhist}.

These demonstrate that evolved stars within 300 pc of the Sun are comparatively homogeneous, with no identifiable patterns visible. Slight excesses of stars are visible toward Galactic longitudes $l \approx 30$ and $330$\,deg (Figure \ref{fig:radialhist}, top panel). These are individually statistically insignificant, but lead to a 3$\sigma$ excess\footnote{The statistical uncertainty, $\sigma$ is computed here by $\sqrt{N}/N$, which approximates Poisson uncertainties for large $N$.} of stars at positive $X$ ($l<90$, $l>270$\,deg). A Kolmogorov--Smirnoff (K--S) test against a uniform distribution shows a deficit in the direction $l = 108$\,deg, but with a $p$-value of only $p = 0.216$. If the $L>700$\,L$_\odot$ criterion is relaxed and fainter RGB/AGB stars are included, $p$ increases. 

Stellar density decreases by $\sim$30 per cent at high Galactic latitude (Figure \ref{fig:radialhist}, middle panel), a result of the $\approx$300\,pc scale height of the Galactic thin disc \citep{Juric2008}. This fall-off appears lop-sided, with more stars in the southern Galactic hemisphere: a K--S test against an arccos distribution shows a possible surplus of stars around $b = -13$\,deg ($p=0.145$). Removing the $L>700$\,L$_\odot$ criterion more confidently retrieves a surplus around $b = -18$\,deg ($p=0.017$). This could be explained by the Sun's position slightly above the Galactic plane. With a 17\,pc height above the plane \citep{Karim2017}, we would expect 52.6 per cent of stars within 300 pc to be at negative $Z$, whereas with a 34.2\,pc height above the plane \citep{Yin2024}, we would expect 54.9 per cent. We find 53.5 $\pm$ 1.6 per cent, which is consistent with both positions.

Figure \ref{fig:radialhist} (bottom panel) approximates the area density (stars kpc$^{-2}$) of evolved stars within the Galactic Plane. Variations on small scales ($\lesssim$25 pc) are subject to small-number statistics, while regions close to 300 pc boundary are affected by edge effects, since the sampled volume is a sphere. We can see a mostly homogeneous distribution of evolved stars around the Sun, with a few small concentrations and rarefactions consistent with statistical variation in the small numbers of stars. 

Figure \ref{fig:distz} shows the space density (stars kpc$^{-3}$) of evolved stars within 300 pc and its evolution with height above the Galactic plane (top-right panel) and distance from the Sun (top-left panel). 

Stellar densities at large Galactic scale height, $Z \sim 200$ pc, are approximately half the density of evolved stars in the Galactic mid-plane to $Z \sim 200$ pc, which is consistent with the established scale height of $\approx$300 pc (grey lines in the plot).

The vertical gradient of the Galactic disc means that space density of stars per shell transitions from $N \propto R^3$ to $N \propto R^2$ at large radii, with the transition becoming evident around the 300 pc scale height. We can observe this in the 300 pc sample as a very slight decrease seen in stellar density with radius, from $\sim$20\,000 evolved stars kpc$^{-3}$ near the Sun to $\sim$15\,000 evolved stars kpc$^{-3}$ beyond 200 pc from the Sun.

\subsubsection{NESS sample}

NESS Tiers 0 and 4 contain too few stars to meaningfully estimate their stellar density over distance or scale height, while Tier 0 is also only sampled at declinations of $\delta < -30^\circ$. The stellar density in Tiers 1, 2 and 3 drops towards the tier boundaries. This is partly due to a real decrease with Galactic scale height (over the portions following the grey curves in Figure \ref{fig:distz}, middle-right panel). It is also partly due to incompleteness near the tier boundaries, as distant stars are more likely to be missing distances (thus absent from the criteria-meeting dataset used in this plot) and as updated distances in \emph{Gaia} DR3 have smoothed the distribution near the tier boundaries.

Figure \ref{fig:XYZ} (left panels) shows an excess of NESS sources towards the Galactic centre (positive $X$). This surplus approximately corresponds to the known location of the Sagittarius--Carina arm. Broadening the sample to the unrestricted dataset (additional green points in Figure \ref{fig:XYZ}) shows a much more significant concentration in this direction. Some are likely correctly plotted, however this is also the region into which luminous stars from larger distances will be scattered (cf., the survey bias discussion in Section \ref{apx:bias}).

Comparisons of stellar types within the NESS tiers should remain largely valid regardless of completeness issues due to imprecise/unknown distances. Carbon stars are more common among higher-mass stars and in metal-poor populations, where enough nuclear fusion and dredge-up of fusion products occur to overcome natal atmospheric oxygen. The Galactic thick disc and halo are metal-poor, but lack high-mass stars, while the converse is true in the Galactic disc. Figure \ref{fig:distz} also shows how the ratio of C-rich to O-rich stars (C/M ratio\footnote{Distinct from the chemical C/O ratio.}) varies with distance and Galactic scale height. The C/M ratio artificially increases at large distance, as carbon stars (which tend to have higher mass-loss rates) are preferentially sampled by Tier 2 (300--600 pc) and Tier 3 (600--1200 pc). Despite this, a substantial \emph{decrease} in the number of carbon stars can be seen moving away from the Galactic plane, consistent with carbon stars coming almost entirely from the metal-rich but younger Galactic disc population in the Galactic plane.


\section{Conclusions} 
\label{sec:conc}

We have performed an automated literature search for photometric and ancillary data of the NESS catalogue of stars. We have vetted the photometric literature for concordance and assessed stellar types and evolutionary status for each star. We produce a reassessed set of distances, based on \emph{Gaia} DR3 and other measurements, and fitted each star's spectral energy distribution to assign photometric temperatures and luminosities. We identify:
\begin{itemize}
    \item The NESS survey contains 781 evolved stars, of which 685 have distances based mostly on parallactic data. Of these, 649 meet our criteria for evolved stars ($700 < L < 200\,000$ L$_\odot$, $T_{\rm eff} < 5000$\,K.
    \item Among these 649, there are 568 O-rich stars in the survey, 67 C-rich stars, two S-type stars, and 12 stars lacking a clear definition.
    \item There are 42 objects in the NESS survey that are not (highly) evolved stars, and 27 that are too evolved to meet our evolved-star criteria. Two additional objects have unclear status, but are probably not evolved stars. These objects were removed from the analysis.
    \item Removed objects are primarily from the ``extreme'' mass-losing tier of NESS sources. Since AGB stars in this tier dominate the dust-production rate in the Milky Way, the difficulty in separating evolved stars from false positives is a potential major source of uncertainty for both NESS and AGB research generally. A concerted all-sky search of indicators of AGB status (e.g., infrared stellar variability, masers) is recommended to better separate these two classes.
\end{itemize}

We compiled a comparison dataset of a complete sample of 1880 AGB and upper RGB stars within 300 pc of the Sun from \emph{Gaia} DR3. Of these, 507 meet the above temperature and luminosity criteria, and 178 overlap with the NESS sample. We have used these to assess the completeness of the NESS tiers 0 and 1, finding five sources potentially missing from each tier.

Methodologically, we have assessed the distance estimates to evolved stars from parallax, period--luminosity and average-bolometric-luminosity methods. At close distances, all methods are sufficiently accurate. All methods fail at large distances, as parallaxes become noisy and stars become extincted, and as samples become contaminated by RSG stars, which are both more luminous than average and tend to pulsate in overtone modes. Distances to dust-enshrouded stars still provide the most serious uncertainty in analysing their properties. A comprehensive survey of distances to evolved stars not covered by \emph{Gaia} is recommended, e.g., via very-long baseline interferometry.

We also assess different methods of temperature and luminosity estimation from SEDs. Stellar model atmospheres are strongly preferred for stars without appreciable dust production. Indicative temperatures from fitted blackbodies are often very imprecise for stellar-like SEDs. Estimating luminosity via trapezoidal integration of the SED is preferred when a star shows significant dust production. Comparing against spectroscopic surveys, we find:
\begin{itemize}
    \item Good agreement between non-\emph{Gaia} spectral temperatures and our SED fits for the 300 pc sample (median offset 24\,K), but with a large scatter (central 68 per cent confidence interval: --658 to 224\,K).
    \item The large scatter is considerably larger than the scatter generated during testing of {\sc PySSED} (--146 to 302 K; \citealt{McDonald2024}), which we attribute to: (a) the heterogeny of spectral methods used, including out-of-date methdologies, and (b) the intrinsic spectral variability of the stars, which is averaged out in our SED-fitted temperatures but not in individual spectra. We consider our SED-fitted temperatures to be more precise as a result.
    \item The \emph{Gaia} {\sc aspis} spectral temperatures for our sample are considerably higher and more scattered than both our SED-fitted temperatures (334\,K, 7 to 1031\,K) and the literature spectral temperatures (304\,K, 38 to 589\,K), suggesting that there is considerable room for improvement for \emph{Gaia} parameter estimation of brighter AGB stars.
\end{itemize}
For the NESS survey, we produce temperature estimates via a combination of these SED-fitting methods, relying mostly on stellar model atmospheres to fit stars in Tiers 0, 1 and 2, and blackbodies and trapezoidal integration to fit Tiers 3 and 4.

We compare the NESS survey sample against our sample of evolved stars from within 300 pc of the Sun and other surveys to measure statistics on evolved stars. Highlights include the following.
\begin{itemize}
    \item We present the luminosity function of evolved stars within 300 pc of the Sun. There is a notable absence of stars at luminosities above 10\,000 L$_\odot$ compared to (e.g.) the LMC.
    \item This is broadly reproduced by the known local SFH \citep{Alzate2021}, but at a lower luminosity than expected. This could be due an over-estimated star formation $\sim$1.0 Gyr ago, or represent imprecisions in how stellar evolution models treat mass loss around the transition between carbon stars and hot-bottom-burning stars.
    \item We use NESS to derive the luminosity function of dusty AGB stars within 300 pc of the Sun. Dust production occurs on either side of the RGB tip, but is concentrated mostly among stars in the region around 1300--5600 L$_\odot$. 
    \item The fraction of dusty stars increases with luminosity above the RGB tip (from $\sim$60 per cent at the RGB tip to $\sim$74 per cent at approximately a bolometric magnitude above it).
    \item Literature data on evolved stars, exemplified by the DEATHSTAR survey, shows historic under-observation of AGB and RSG stars with both the highest and lowest dust-production rates.
\end{itemize}

The uncertainty in the distances to the stars with the strongest dust production hampers our ability to reconstruct the 3D spatial distribution of evolved stars near the Sun. However, we identify that
\begin{itemize}
    \item the distribution of evolved stars within 300 pc of the Sun is largely homogeneous, excepting a decrease in density in the Galactic $Z$ direction, consistent with a thick-disc scale height of 300 pc, and a slight preference for stars in directions both at negative $b$ (consistent with the Sun being slightly above the Galactic plane) and towards the Galactic centre (consistent with an increase in stellar density at smaller Galactocentric radii);
    \item NESS Tiers 1, 2 and 3 have density gradients also consistent with a scale height of 300 pc, but are affected at their outer boundaries by incompleteness due to the updated distances of \emph{Gaia} DR3;
    \item the fraction of carbon stars within $\sim$600 pc of the Sun increases with distance in the NESS survey due to its tiered selection functions, but decreases with Galactic $Z$, indicating that the vast majority of carbon stars belong to the younger, metal-rich Galactic thin disc, not the older but metal-poor Galactic halo.
\end{itemize}


\section*{Acknowledgements}

The James Clerk Maxwell Telescope is operated by the East Asian Observatory on behalf of The National Astronomical Observatory of Japan; Academia Sinica Institute of Astronomy and Astrophysics; the Korea Astronomy and Space Science Institute; Center for Astronomical Mega-Science (as well as the National Key R\&D Program of China with No. 2017YFA0402700). Additional funding support is provided by the Science and Technology Facilities Council of the United Kingdom and participating universities in the United Kingdom and Canada. Program ID: M17BL002.

This research has been financially supported by the Ministry of Science and Technology of Taiwan under grant numbers MOST104-2628-M-001-004-MY3 and MOST107-2119-M-001-031-MY3, and by Academia Sinica under grant number AS-IA-106-M03.

IM and AAZ acknowledge support from the UK Science and Technology Facilities Council under grants ST/L00768/1 and ST/P000649/1, and from the EU 'EXPLORE' program funded under Horizon 2020 grant 101004214, and the OSCARS project, which has received funding from the European Commission’s Horizon Europe Research and Innovation programme under grant agreement No. 101129751.
SS acknowledges support from the UNAM PAPIIT programs IA104820, IA104822, and IA104824.
PS and HI were supported by Daiwa Anglo-Japan Foundation and the Great Britain Sasakawa Foundation. 
OCJ has received support from an STFC Webb fellowship.
JH thanks the support of NSFC project 11873086. This work is sponsored (in part) by the Chinese Academy of Sciences (CAS), through a grant to the CAS South America Center for Astronomy (CASSACA) in Santiago, Chile.
JPM acknowledges support by the National Science and Technology Council of Taiwan under grant NSTC 112-2112-M-001-032-MY3.
TD is supported in part by the Australian Research Council through a Discovery Early Career Researcher Award (DE230100183). This research is supported in part by the Australian Research Council Centre of Excellence for All Sky Astrophysics in 3 Dimensions (ASTRO 3D), through project number CE170100013.
JC is supported by a Discovery Grant from the Natural Sciences and Engineering Research Council (NSERC).
HK acknowledges the support by the National Research Foundation of Korea (NRF) grant (No. 2021R1A2C1008928) and the Korea Astronomy and Space Science Institute (KASI) grant (Project No. 2023-1-840-00), both funded by the Korean government (MSIT).

This research has made use of the VizieR catalogue access tool, CDS, Strasbourg, France (DOI : 10.26093/cds/vizier). The original description of the VizieR service was published in A\&AS 143, 23. This research has made use of ``Aladin sky atlas'' developed at CDS, Strasbourg Observatory, France. This research has made use of the SIMBAD database, operated at CDS, Strasbourg, France. This research has made use of the Astrophysics Data System, funded by National Aeronautics and Space Administration (NASA) under Cooperative Agreement 80NSSC21M00561. This research has made use of the NASA/IPAC Infrared Science Archive, which is funded by the National Aeronautics and Space Administration and operated by the California Institute of Technology.


\section*{Data Availability}

The {\sc PySSED} v1.1 code is available from \url{https://github.com/iain-mcdonald/PySSED}. Input files for {\sc PySSED}, output files from {\sc PySSED}, and code to generate all datafiles and plots are included in the Supplementary Material (see Appendix \ref{apx:data}), as are machine-readable versions of tables as indicated in the main text. We recommend that readers consult the notes in the \LaTeX{}\ source of the paper on arXiv if they wish to reproduce specific numbers or plots.




\bibliographystyle{mnras}
\bibliography{references.bib,biblio.bib}


\section*{Author affiliations}

$^{1}$Jodrell Bank Centre for Astrophysics, School of Physics and Astronomy, University of Manchester, M13 9PL, Manchester, UK \\
$^{2}$Instituto de Radioastronom\'ia y Astrof\'isica, UNAM. Apdo. Postal 72-3 (Xangari), Morelia, Michoac\'an 58089, Michoac\'{a}n, M\'{e}xico\\
$^{3}$Centre for Astrophysics Research, University of Hertfordshire, Hatfield, UK\\
$^{4}$UK Astronomy Technology Centre, Royal Observatory, Blackford Hill, Edinburgh, EH9 3HJ, UK \\
$^{5}$Institute of Astronomy, KU Leuven, Celestijnenlaan 200D bus 2401, 3001 Leuven, Belgium\\
$^{6}$School of Physics and Astronomy, Monash University, Clayton, 3800, Victoria, Australia\\
$^{7}$ARC Centre of Excellence for All Sky Astrophysics in 3 Dimensions (ASTRO 3D), Clayton, 3800, Victoria, Australia\\
$^{8}$Yunnan Observatories, Chinese Academy of Sciences, 396 Yangfangwang, Guandu District, Kunming, 650216, People's Republic of China \\
$^{9}$Chinese Academy of Sciences South America Center for Astronomy, National Astronomical Observatories, CAS, Beijing 100101, China \\
$^{10}$Departamento de Astronom\'ia, Universidad de Chile, Casilla 36-D, Santiago, Chile\\
$^{11}$Institute of Astronomy and Astrophysics, Academia Sinica, 11F of AS/NTU Astronomy-Mathematics Building, No. 1, Sec. 4, Roosevelt Rd, Taipei 106319, Taiwan\\
$^{12}$Lennard-Jones Laboratories, Keele University, ST5 5BG, UK. \\
$^{13}$Department of Physics and Astronomy, University College London, Gower St, London WC1E 6BT, UK\\
$^{14}$Institut de Ci\`encies de l'Espai (ICE, CSIC), Can Magrans, s/n, E-08193 Cerdanyola del Vall\`es, Barcelona, Spain\\
$^{15}$ICREA, Pg.~Llu\'{\i}s Companys 23, E-08010 Barcelona, Spain\\
$^{16}$Institut d'Estudis Espacials de Catalunya (IEEC), E-08860 Castelldefels, Barcelona, Spain\\
$^{17}$School of Physics \& Astronomy, Cardiff University, The Parade, Cardiff CF24 3AA, UK\\
$^{18}$Max Planck Institute for Astronomy (MPIA), K\"onigstuhl 17, D-69117 Heidelberg, Germany\\
$^{19}$Center for Computational Astrophysics, Flatiron Institute, 162 5th Ave, New York, NY 10010, USA\\
$^{20}$ Center for Cosmology and Particle Physics, New York University, 726 Broadway, New York, NY 10003, USA \\
$^{21}$Department of Physics and Astronomy, University of Western Ontario, London, ON, N6A 3K7, Canada\\
$^{22}$Institute for Earth and Space Exploration, University of Western Ontario, London, ON, N6A 3K7, Canada\\
$^{23}$SETI Institute, 189 Bernardo Avenue, Suite 100, Mountain View, CA 94043, USA\\
$^{24}$Korea Astronomy and Space Science Institute (KASI) 776, Daedeok-daero, Yuseong-gu, Daejeon 34055, Republic of Korea \\
$^{25}$Institute for Scientific Research, Boston College, 140 Commonwealth Avenue, Chestnut Hill, MA 02467, USA \\
$^{26}$AURA for the European Space Agency, Space Telescope Science Institute, 3700 San Martin Drive, Baltimore, MD 21218, USA \\
$^{27}$Department of Physics and Astronomy, Kagoshima University, 1-21-35 Korimoto, Kagoshima, Japan \\
$^{28}$Amanogawa Galaxy Astronomy Research Center (AGARC), Graduate School of Science and Engineering, Kagoshima University, 1-21-35 Korimoto, Kagoshima 890-0065, Japan \\ 
$^{29}$Department of Physical Sciences, The Open University, Walton Hall, Milton Keynes MK7 6AA, UK \\
$^{39}$Center for General Education, Institute for Comprehensive Education,  Kagoshima University, 1-21-30 Korimoto, Kagoshima 890-0065, Japan \\ 
$^{31}$East Asian Observatory (JCMT), 660 N. A`ohoku Place, Hilo, Hawai`i, USA, 96720\\
$^{32}$Valencian International University, Master in Astronomy and Astrophysics, Pintor Sorolla nº 21, Valencia, Valencia 46002, Spain 
$^{33}$NASA Goddard Space Flight Center, 8800 Greenbelt Road, Greenbelt, MD 20771, USA \\ 
$^{34}$National Science Foundation, 2415 Eisenhower Avenue, Alexandria, Virginia 22314, USA \\ 



\appendix

\section{Methods}
\label{apx:more}

\subsection{Matching NESS sources from \emph{IRAS} to \emph{Gaia} DR3} 
\label{sec:xmatch:togaia}

The large \emph{IRAS} beam\footnote{The \emph{IRAS} beam is non-circular and wavelength-dependent, varying between $1^\prime \times 5^\prime$ at 12 $\mu$m to $4^\prime \times 5^\prime$ at 100 $\mu$m. A smaller, synthesised beam, generated from multiple passes of the satellite, provides astrometric accuracy for uncrowded sources on the scale of 2--16$^{\prime\prime}$. See \url{https://lambda.gsfc.nasa.gov/product/iras/docs/exp.sup/toc.html}.} provides adequate astrometric precision for the pointing of the NESS sub-mm survey (i.e., $\lesssim$15$^{\prime\prime}$), but insufficient precision and source separation to match to the observed object(s) in optical surveys. To obtain an accurate list of cross-matches to other surveys, we must first identify higher-precision astrometric cross-detections, beginning from the \emph{IRAS} detection and working towards higher-resolution surveys and towards optical wavelengths.

Since most of our stars are bright and isolated point sources, the nearest cross-match in other catalogues is usually the correct one. However, in some cases, proper motion, optical obscuration of the target star by circumstellar or interstellar dust, or nearby blended objects can cause confusion. Hence, we require verification to obtain a set of high-quality cross-matches across a broad wavelength range.

\subsubsection{General approach} 

\begin{figure}
\centering
\includegraphics[width=\linewidth]{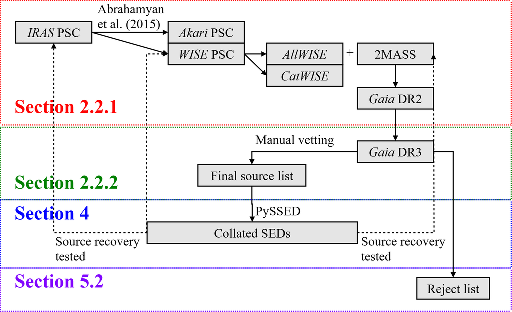}
\caption{Simplified diagram of the steps taken to cross-match \emph{IRAS} PSC objects to other catalogues. Coloured boxes denote the Section numbers in which they are discussed. Note that the \emph{AllWISE}$\rightarrow$2MASS conversion is taken from the \emph{AllWISE} catalogue. Other steps use cone searches unless otherwise stated.}
\label{fig:reduce}
\end{figure}

To cross-match the \emph{IRAS} PSC to higher-resolution and shorter-wavelength surveys, we gradually step down to smaller resolutions/wavelengths as summarised in Figure \ref{fig:reduce}. Cross-matching then proceeds in a semi-automated fashion, but with manual confirmation of the output. Cross-matches were later checked (Section \ref{sec:reduce}) using proper-motion-corrected coordinates of the final \emph{Gaia} DR3 cross-match, to ensure internal consistency.

In the first step, objects were mapped from the \emph{IRAS} PSC\footnote{The sample also includes $\lambda$ Vel, a source in the \emph{IRAS} point-source reject catalogue. See \citet{Scicluna22} for details.} to the \emph{Akari} IRC and FIS PSCs and the \emph{Wide-field Infrared Survey Explorer} \citep[\emph{WISE};][]{Wright2010} PSC, using cross-matches from \citet{Abrahamyan15}. Based on these matches, a ``best coordinate'' pair was assigned to each NESS target from, in order of preference\footnote{This follows the order of resolution and, for Rayleigh--Jeans energy distributions, the order of photometric depth, thus preserves the greatest astrometric accuracy.}, the \emph{WISE}, \emph{Akari} IRC and \emph{Akari} FIS catalogues.

In the second step, \emph{WISE} sources were updated to the later \emph{AllWISE} data release \citep{Cutri2013}. Three sets of 60$^{\prime\prime}$ cone searches were then performed, one each for the IRAS position, the best coordinate pair, and the \emph{WISE} position, if available. A cone search with a 6$^{\prime\prime}$ radius was also used to match each resulting \emph{AllWISE} position with the astrometrically similar but photometrically different \emph{unWISE} \citep{Schlafly2019} and \emph{catWISE} \citep{MEF+21} catalogues. \emph{UnWISE} provides more realistic flux estimates in the $W1$ and $W2$ bands for saturated sources, so we use it in preference to \emph{AllWISE} for sources with $W1$ or $W2 < 5$ mag. Data from \emph{catWISE} \citep{MEF+21} are used in preference to \emph{unWISE} or \emph{allWISE} for fainter sources ($W1$ and $W2 > 5$ mag). This is based on the more consistent match of the \emph{catWISE} photometry with the flux expected from our final models. It should be noted, however, that both \emph{unWISE} and \emph{catWISE}, on average, overestimate the flux for fainter sources compared to the stellar models. This is expected due to Malmquist bias, but could be in part intrinsic, as fainter sources tend to be the more-extreme stars in higher NESS tiers.

In most cases, the \emph{AllWISE} cross-match is straightforward: the closest \emph{AllWISE} match to the three positions agrees in 833 out of 852 of cases. However, saturation and high proper motion, sometimes decomposes the \emph{AllWISE} match into two or three detections. These were rectified manually by selecting the most-representative detection, reverting to the original \emph{WISE} photometry where necessary. If no match was found within 60$^{\prime\prime}$, or if the source was rejected during manual inspection as being implausibly far away and/or had the wrong magnitude, the \emph{IRAS} source position was retained.

In the third step, the \emph{WISE}/\emph{AllWISE} detections were mapped to the Two-Micron All-Sky Survey (2MASS; \citealt{Cutri2003,Skrutskie2006}) catalogues using the \emph{AllWISE} catalogue's cross-identifiers. Some heavily saturated sources lacked \emph{WISE} or \emph{AllWISE} detections and were manually mapped from the \emph{IRAS} co-ordinates using {\sc aladin}\footnote{\url{https://aladin.cds.unistra.fr/AladinLite}}.

Finally, 2MASS detections were mapped onto \emph{Gaia} DR2 \citep{GaiaDR2}, using its internal database of cross-matches\footnote{This step was performed before the release of \emph{Gaia} DR3.}. For sources that had \emph{AllWISE} matches but no corresponding 2MASS match, we retained the \emph{AllWISE} astrometry and mapped to \emph{Gaia} directly.

In some cases, multiple possible counterparts existed; in others, proper motion had moved the star so that it was no longer the closest cross-matching object, while a small number of \emph{IRAS} detections did not correspond to a point-like object in higher resolution surveys. Consequently, the SEDs of all potential \emph{Gaia}--2MASS--\emph{Akari}--\emph{WISE} cross-matches were manually inspected, alongside survey images accessed through the {\sc aladin} service. A small fraction of sources had cross-matches that were visibly wrong from either their SEDs or on visual imagery, and these were manually updated to the correct match where possible. Several potentially problematic cases were also identified where sources exhibited blending with other points or diffuse objects in the field. With these steps, we have an accurate position and, for most sources, a proper motion for each star.

\subsubsection{Conversion from \emph{Gaia} DR2 to DR3 and treatment of problem cases} 

On the release of \emph{Gaia} DR3, a cross-match between the \emph{Gaia} DR2 and DR3 positions was performed using a cross-matching radius of 1$^{\prime\prime}$. The majority (652) of sources had a direct DR2-to-DR3 cross-match. A few stars had multiple matches, or a significant difference in magnitude ($|\Delta G| > 1.5$ mag). These cases were checked individually to ensure the magnitude, colour, and sky position (via comparison with imaging surveys using {\sc aladin}) matched the AGB star. This was also done for 39 bright, high-proper-motion AGB stars where the correct cross-match did not lie within 1$^{\prime\prime}$. A small number of sources with 2MASS cross-identifiers but no \emph{Gaia} DR2 match obtained a \emph{Gaia} DR3 match. Two sources (\emph{IRAS} 21417+0938 = \emph{Gaia} DR2 1765433632573306496 and \emph{IRAS} 09251$-$0826 = \emph{Gaia} DR2 5741512800984781824) did not have a \emph{Gaia} DR3 counterpart. We retain the \emph{Hipparcos} identifiers for R Dor and L$_2$ Pup, as the \emph{Gaia} DR3 cross-matches do not contain proper motion information.

Some 32 sources did not translate directly from the \emph{Gaia} DR2 to DR3 catalogues, and had to have their optical counterparts manually extracted. The majority of these 32 stars are close to the saturation boundary and did not have any \emph{Gaia} DR3 counterpart.

Especially in the Galactic plane, optical source confusion and high infrared backgrounds both contribute to cross-matching uncertainty. If later modelling (Section \ref{sec:reduce}) did not correctly recover any mid-infrared photometry longward of 10\,$\mu$m, the source co-ordinates were examined and realigned where appropriate to a different cross-identifier (e.g., an OH maser source).

The final cross-matches and fitting results (Appendix \ref{apx:data}) do not contain the 71 rejected sources discussed in Section \ref{sec:rejects}, as counterparts were not always sought if objects were identified as contaminants. Of the listed 781, the \emph{Gaia} DR3 counterpart was the preferred co-ordinate solution for 683. While some of the remaining sources have 2MASS counterparts, these are often not automatically resolvable by {\sc simbad}, hence the primary data source for 83 of the remaining 98 is the \emph{position} of the 2MASS source (except \emph{IRAS}\,18257$-$1000, 18460$-$0254 and 21318+5631, where the \emph{WISE} co-ordinates were used). For 14 sources, the \emph{Hipparcos} astrometry \citep{vanLeeuwen07} was used instead, including proper motions. Finally, for \emph{IRAS}\,21417+0938, we retain the source \emph{Gaia} DR2 1195189725172268288, as there is no corresponding DR3 counterpart.

\subsection{Constructing the 300\,pc sample}
\label{apx:300pc}

\begin{figure*}
\centering
\includegraphics[width=0.8\linewidth]{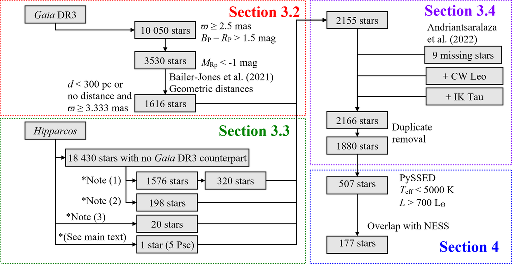}
\caption{Simplified diagram of the steps taken to create a catalogue of evolved stars within 300 pc of the Sun. Coloured boxes denote the Section numbers in which they are discussed. Notes: (1) \emph{Hipparcos} stars not listed in \citet{McDonald17} (1576 stars), restricted to 320 stars by selection of \emph{Hipparcos} $\varpi \geq 3.333$\,mas, $B_{\rm T}-R_{\rm T} > 1$\,mag and $M_{R_T} < -1$\,mag; (2) \emph{Hipparcos} stars with $\varpi \geq 3.333$\,mas, $T_{\rm eff}<5500$\,K and $L>350$\,L$_\odot$ in \citet{McDonald17}; (3) \emph{Gaia} stars with $B_{\rm P}-R_{\rm P} > 1.5$\,mag, $M_{R_P} < -1$\,mag, with no \emph{Gaia} parallax, but with a \emph{Hipparcos} parallax of $\varpi \geq 3.333$\,mas.}
\label{fig:reduce2}
\end{figure*}

\subsubsection{Generating a 300 pc comparison sample}

The NESS Overview paper demonstrated that \emph{Gaia} DR3 parallaxes \citep{GaiaDR3} are substantially more accurate than both pre-existing parallaxes and luminosity-based distances, at least for stars within a few hundred pc of the Sun. Hence, we can now use \emph{Gaia} DR3 to define a volume-complete set of evolved stars within a few hundred pc of the Sun based on parallax data alone, defining a cutoff here of 300 pc to match the NESS survey's Tiers 0 and 1. From this, we can re-evaluate the completeness of these tiers and better tie the rarer objects in the upper tiers of NESS to the properties of local stars. Unfortunately, the astrometric noise of the optically faint, self-obscured and highly variable AGB stars typical of NESS's upper tiers, and contamination from other types of object with near-zero but noisy parallaxes, means that 300 pc marks the approximate limit where a volume-complete sample can be drawn without encountering an overwhelming number of edge cases and problems in robustly identifying a complete set of optically obscured stars missing from \emph{Gaia}.

\subsubsection{\emph{Gaia} DR3 giant stars within 300 pc}

The steps we use to create this catalogue are outlined in Figure \ref{fig:reduce2}. We first remind the reader that an authoritative catalogue of evolved stars at this distance cannot be performed with current technology (see discussion in Section \ref{sec:xmatch:distances:paper}).

To construct our catalogue, we begin by querying the \emph{Gaia} DR3 catalogue for stars with parallaxes of $\varpi \geq 2.5$ mas (see below for discussion), colours $B_P - R_P > 1.5$ mag and $R_P < 10$ mag. These limits respectively select most stars with $T \lesssim 5700$ K and $L \gtrsim 300$ L$_\odot$ at 300 pc, except the most obscured or heavily extincted AGB stars. From this dataset of 10\,030 stars, we use simple inversion of the parallax to assign an approximate distance, allowing us to further select those stars with absolute magnitudes $M_{Rp} < -1$ mag. This selects only luminous stars (the RGB tip is $M_{Rp} \approx -2.7$ mag), leaving 3530 stars, of which 1589 have parallaxes of $\varpi \geq 3.333$ mas and are thus likely to be within 300 pc.

To convert \emph{Gaia} DR3 parallax to distance, we use the geometric distances listed in \citet{BJRF+21}, which has the added advantage of dealing with asymmetric errors and the parallax zero-point uncertainty of \emph{Gaia} DR3 (Appendix \ref{apx:dist:zpt}). Only 3207 of the 3530 stars have distances in \citet{BJRF+21}. For the remainder, we retain distances based on inversion of the parallax (1/$\varpi$). Generally, the geometric and parallax distances agree to within 1--2 per cent, though there are a handful of larger outliers. Our previously chosen limit of $\varpi \geq 2.5$ mas allows us to identify 27 sources that \citet{BJRF+21} place within our 300 pc radius that 1/$\varpi$ does not. Adding these to our sample leaves a total of 1616 \emph{Gaia} DR3 sources likely to lie within 300 pc.

\subsubsection{Completing the catalogue}

There are 18\,430 \emph{Hipparcos} stars without a \emph{Gaia} DR3 counterpart\footnote{\url{http://cdn.gea.esac.esa.int/Gaia/gedr3/cross_match/hipparcos2_best_neighbour/}}. Mostly this is due to saturation, though some are duplicates that have not been successfully cross-matched due to their proper-motion anomaly. We extract two groups of stars from these. First, 1576 stars lack counterparts in \citet{MZW17} because their spectral energy distributions (SEDs) were too poorly fit to publish: from these we extract the 320 which have $1/\varpi_{\rm Hip} < 300$ pc, $B_T - V_T > 1$ mag\footnote{This approximates the $B_P - R_P > 1.5$ mag limit applied to \emph{Gaia} and should conservatively retain all evolved stars. The same Lutz--Kelker corrections could not be applied to the \emph{Hipparcos} data, as the stars are not in the \citet{BJRF+21} sample, however stars bright enough not to be in \emph{Gaia} are largely restricted to stars with good astrometry within 300 pc.} and $M_{\rm RT} < -1$ mag. Second, there are 198 stars that meet the criteria $1/\varpi_{\rm Hip} < 300$ pc, and have published parameters in \citet{MZW17} of $T_{\rm eff} < 5500 K$ and $L > 350$ L$_\odot$: these broader $T_{\rm eff}$ and $L$ criteria allow us to check whether additional photometric and distance data move edge cases in or out of the $T_{\rm eff} < 5000 K$ and $L > 700$ L$_\odot$ criteria used in this work.

An additional 20 \emph{Hipparcos} stars have \emph{Gaia} DR3 counterparts, $B_P - R_P > 1.5$ mag and $M_{Rp} < -1$ mag in \emph{Gaia}, have $\varpi_{\rm Hip} > 3.33$ mas in \emph{Hipparcos}, but lack \emph{Gaia} parallax estimates. A final object, HIP\,114273 (5 Psc) has its $R_P$ magnitude and parallax in two different \emph{Gaia} DR3 sources (though was ultimately found to be too hot for our study). The addition of these \emph{Hipparcos} stars brings the total number of sources to 2155. 

To this list, we add the carbon star CW Leo (IRC\,+10216; $95 \pm 15$\,pc; \citealt{Sozzetti2017}) and the OH/IR star IK Tau (260 pc; \emph{Gaia} DR3), which appear in \emph{Gaia} but are too obscured to meet the $R_P < 10$ mag target. We anticipate that these are the only sufficiently obscured sources within 300 pc, otherwise they would have been identified by NESS and other surveys. A further nine NESS sources (R Aqr, S Dra, T Ari, U Her, W Ori, X TrA, Y CVn, Y Lyn and $\chi$ Cyg) have distances in \citet{BJRF+21} that are $>$300 pc, but distances in \citet{ARVDB22} that are $<$300 pc (see also Section \ref{sec:reduce}), hence are pre-emptively added to the sample. This leaves 2166 \emph{Gaia} DR3 sources in total.

Duplications among the \emph{Gaia--Hipparcos} stars were identified as any two AGB stars with the same {\sc simbad} co-ordinates. Removing these duplicates leaves a clean list of 1880 sources.
Four \emph{Gaia} sources are not identified by {\sc simbad} and were replaced by the corresponding {\sc simbad} primary identifiers: HD\,174569, $\pi$ Pup, $\zeta$ Ara and $\chi$ Cyg (also mentioned above).

These 1880 sources represent a list of stars that could potentially match our evolved-star criteria. However, most of these are less-evolved, lower-luminosity RGB and AGB stars that will ultimately not meet our temperature and luminosity criteria, but which need their SEDs fitted before that can properly be determined. These steps are performed in Section \ref{sec:reduce}.

\section{Data sources}
\label{apx:sources}

\begin{table*}
    \caption{Point-source catalogues and their short-hand in this paper.}
    \label{tab:photosources}
    \centering
    \begin{tabular}{@{}l@{}l@{\,}l@{}c@{}c@{}c@{}c@{\,}c@{}l@{}}
        \hline\hline
        \multicolumn{1}{c}{Short-hand}    & \multicolumn{1}{c}{Full name}   & \multicolumn{1}{c}{VizieR} & \multicolumn{1}{c}{Epoch}  & \multicolumn{1}{c}{\clap{$r_{\rm beam}$}} & \multicolumn{1}{c}{\clap{$r_{\rm match}$}}  & \multicolumn{1}{c}{Filters}  & \multicolumn{1}{c}{\clap{Wavelength}}  & \multicolumn{1}{c}{Reference} \\
        \multicolumn{1}{c}{name}    & \multicolumn{1}{c}{\ }   & \multicolumn{1}{c}{table}  & \multicolumn{1}{c}{(yr)}  & \multicolumn{1}{c}{($^{\prime\prime}$)} & \multicolumn{1}{c}{($^{\prime\prime}$)}  & \multicolumn{1}{c}{\ }  & \multicolumn{1}{c}{range}  & \multicolumn{1}{c}{\ } \\
        \hline
        \emph{XMM}        &  X-ray Multi-Mirror Mission	& 	II/378/xmmom6s & 2005.0 & 1.7 & 1.0 & $UV_{W2}\,UV_{W1}\,UBV$ & UV & \citet{PBT+12}\\
        \                   & \rlap{...  -- Newton Optical Monitor (XMM-OM)  Serendipitous Source Survey Catalogue} &  &  &  &  &  & \\
        \emph{GALEX}        &  (Revised) all-sky survey of	& 	II/335/galex\_ais & 2008.5 & 2.8 & 1.5 & $FUV\,NUV$ & UV & \citet{BST17}\\
        \                   & \rlap{... Galaxy Evolution Explorer sources} &  &  &  &  &  & \\
        \emph{Gaia} DR3    & \emph{Gaia} Data Release 3  & I/355/gaiadr3    & 2016.0 & 0.5 & 0.4 & $B_{\rm P}\,G\,R_{\rm P}$    & Optical & \citet{GaiaDR3}\\
        \emph{Hipparcos}    & \emph{Hipparcos}: The New Reduction & I/311/hip2	& 1991.25 & 1.0 & 1.0 & $H_{\rm P}$  & Optical & \citet{vanLeeuwen07}\\
        \emph{Tycho}	    & Tycho                             & I/239/tyc\_main	& 1991.25 & 1.0 & 1.0 & $B_{\rm T}\,V_{\rm T}$ & Optical & \citet{Perryman97} \\
        Pan-STARRS          & Panoramic Survey Telescope \& & II/349/ps1 & 2011.9 & 0.5 & 0.5 & $grizy$ & Optical & \citet{CMM16} \\
        \                   & \rlap{... Rapid Response System Data Release 1} &  &  &  &  &  & \\
        APASS       	    & American Association of  & II/336/apass9    & 2015.0	& 3.1	& 1.3   & $BVgri$ & Optical & \citet{HLTW15} \\
        \              	    & \rlap{... Variable Star Observers (AAVSO) Photometric All-Sky Survey} & \   & \	& \	& \   & \ & \ & \ \\
	    SDSS16          	    & Sloan Digital Sky Survey & II/376/sdss16	    & 2008.0	& 1.4	& 0.5   & $u^{\prime}g^{\prime}r^{\prime}i^{\prime}z^{\prime}$ & Optical & \citet{AAMA+09} \\
        \              	    & (SDSS) Data Release 16 & \   & \	& \	& \   & \ & \ & \ \\
        IGAPS   & The merged IPHAS and UVEX 	& V/165	& 2010.5	& 2.3	& 1.5 & $U_{\rm GRO}\,{\rm H}\alpha$ & Optical & \citet{Monguio2020}\\
        \              	    & \rlap{... optical surveys of the northern Galactic plane} & \   & \	& \	& \   & \ & \ & \ \\
	    CMC15         	    & Carlsberg Meridian Catalog 15 & I/327/cmc15	    & 2005.25	& 1.5	& 0.5   & $r^\prime$ & Optical & $^1$ \\
        Morel78             & \                             & II/7A/catalog & 2000.0 & 5.0 & 5.0 & \llap{$U$}$BVRIJHKLM$\rlap{$N$} & Opt/IR & \citet{MM78} \\
        Johnson1975	& \     & II/84/catalog 	& 1975.0	& 60	& 30    & $^7$   & Optical   &   \citet{Johnson1975} \\
        Mermilliod1989	& \ & II/164/mean	& 1989.0	& 60	& 30    & $^8$   & Optical   &   \citet{Mermilliod1989} \\
        Mermilliod1991	& \ & II/168/ubvmeans	& 1991.0	& 60	& 30    & $UBV$   & Optical   &   Online only \\
        Rufener1988	& \     & II/169/main	& 1988.0	& 60	& 30    & $^9$   & Optical   &   Online only \\
        Ducati2002	& \     & II/237/colors	& 2002.0	& 60	& 30    & $BVRIJHKLMN$   & Opt/IR   &   Online only \\
        INTEGRAL	& The first International     & J/A+A/548/A79/	& 2000.0	& 2.0	& 3.0  & $V$ & Optical  & \citet{AlfonsoGrazon2012} \\
        \              	    & \rlap{...  Gamma-Ray Astrophysics Laboratory catalogue of optically variable sources} & \   & \	& \	& \   & \ & \ & \ \\
        Hackstein2015	& Bochum Galactic Disk Survey & J/AN/336/590/varsum	& 2000.0	& 2.0	& 3.0 &   $ri$    & Optical & \citet{Hackstein2015} \\
        DES	& The Dark Energy Survey DR2 & II/371/des\_dr2	& 2017.0	& 0.5	& 0.5   & $grizY$    & Optical  & \citet{DESDR2} \\
        VPHAS	& The Very Large Telescope & II/341/vphasp	& 2014.0	& 0.7	& 0.7  & $ugri\,H\alpha$  &  Optical  & \citet{Drew14} \\
        \              	    & \rlap{...  (VLT) Survey Telescope (VST) Photometric H$\alpha$ Survey of the Southern Galactic Plane and Bulge Data Release 2} & \   & \	& \	& \   & \ & \ & \ \\
        TASS	& The Amateur Sky Survey & II/271A/patch2	& 2005.0	& 1.0	& 1.0  & $VI$    & Optical  & \citet{DRSC06} \\
        
        2MASS	            & Two-Micron All-Sky Survey & II/246/out	    & 1999.5	& 1.0	& 1.0   & $JHK_{\rm s}$ & Near-IR & \citet{Cutri2003}\\
	    DENIS	            & Deep Near-Infrared Survey & B/denis/denis	& 1998.5	& 3.0	& 1.0   & $IJK_{\rm s}$ & Near-IR & $^2$\\
        \              	    & \rlap{... of the Southern Sky} & \   & \	& \	& \   & \ & \ & \ \\
	    VVV	                & Visible and Infrared Survey & II/376/vvv4	    & 2016.0	& 0.3	& 1.0   & $ZYJHK_{\rm s}$ & Near-IR & \citet{MLE+10}\\
        \              	    & \rlap{... Telescope for Astronomy (VISTA) Variables in the Via Lactea Data Release 4} & \   & \	& \	& \   & \ & \ & \ \\
	    VHS	                & VISTA Hemispheric Survey DR5  & II/367/vhs\_dr5	& 2016.0	& 0.3	& 1.0   & $YJHK_{\rm s}$ & Near-IR & \citet{MBG+13}\\
        Whitelock2008	& \ & J/MNRAS/386/313/	& 2000.0	& 2.0	& 3.0  & $JHKL$ & Near-IR & \citet{WFvL08} \\
        UKIDSS	&  United Kingdom Infrared & II/316	& 2007.0	& 1.0	& 1.0   & $JHK$   & Near-IR & \citet{Lucas2008} \\
        \              	    & \rlap{...  Telescope (UKIRT) Deep Sky Survey Data Release 6} & \   & \	& \	& \   & \ & \ & \ \\
	    All\emph{WISE}	    & $^3$ & II/328/allwise	& 2010.25	& 6.0	& 2.0   & [3.4]\,[4.6]\,[11.3]\,[22] & Mid-IR & \citet{Cutri2013}\\
	    cat\emph{WISE}	    & $^3$ & II/365/catwise	& 2010.25	& 6.0	& 2.0   & [3.4]\,[4.6]\,[11.3]\,[22] & Mid-IR & \citet{MEF+21}\\
        un\emph{WISE}	    & $^3$ & II/363/unwise	& 2010.25	& 6.0	& 2.0   & [3.4]\,[4.6]\,[11.3]\,[22] & Mid-IR & \citet{Lang14}\\
        GLIMPSE	& The \emph{Spitzer Space Telescope} & II/293	& 2006.0	& 2.0	& 2.0   & [3.6]\,[4.5]\,[5.8]\,[8]    & Mid-IR & $^{10}$ \\
        \              	    & \rlap{...  Galactic Legacy Infrared Mid-Plane Survey Extraordinaire} & \   & \	& \	& \   & \ & \ & \ \\
        SSTSL2	& & The \emph{Spitzer} source list	& 2006.0	& 2.0	& 2.0    & [3.6]\,[4.5]\,[5.8]\,[8]    & Mid-IR & $^{11}$ \\
	    \emph{MSX}	        & \emph{Mid-course Space Experiment} & V/114/msx6\_main	& 1996.5	& 20.0	& 2.0   & $A\,B_1\,B_2$ & Mid-IR & \citet{Egan2003}$^4$\\
	    \emph{IRAS}	        & \emph{Infrared Astronomical Satellite} & II/125/main	    & 1985.0	& 180	& 15  & [12]\,[25]\,[60]\,[100] & Far-IR & \citet{HW88} \\
	    \emph{DIRBE}	    & \emph{Cosmic Background Explorer} & J/ApJS/154/673/DIRBE	& 1991.9	& 2520	& 10  & [1.25]--[240]$^5$ & IR & \citet{SPB04} \\
        \              	    & \rlap{...  (\emph{COBE}) Diffuse Infrared Background Experiment} & \   & \	& \	& \   & \ & \ & \  \\
	    \emph{Akari} IRC    & \emph{Akari} Infrared Catalogue & II/297/irc	    & 2008.0	& 5.5	& 1.0   & [9]\,[18] & Mid-IR & \citet{IOK+10} \\
        MIPSGAL	& The \emph{Spitzer Space Telescope} & J/AJ/149/64	& 2006.0	& 5.9	& 2.0  & [24]    & Far-IR  & \citet{MIPSGAL} \\
        \              	    & \rlap{... Multiband Imaging Photometer (MIPS) Galactic Plane Survey} & \   & \	& \	& \   & \ & \ & \ \\
	    \emph{Akari} FIS    &  \emph{Akari} Far-Infrared Surveyor & II/298/fis	    & 2008.0	& 26.5	& 5.0   & [65]\,[90]\,[140]\,[160] & Far-IR & \citet{KBB+07}\\
	    \emph{Herschel}	    & \emph{Herschel} Photodetector Array & VIII/106/hppsc070& 2011.5	& 7.0	& 3.0   & [70] & Far-IR & $^5$\\
	    \ 	                & ...Camera \& Spectrometer & VIII/106/hppsc100& 2011.5	& 7.0	& 3.0   & [100] & Far-IR &\\
	    \ 	                & ... (PACS) & VIII/106/hppsc160& 2011.5	& 7.0	& 3.0   & [160] & Far-IR &\\
        \emph{Planck}  &  	Second \emph{Planck} Catalogue & J/A+A/594/A26/	& 2011.5	& $^{12}$	& 10.0 & 0.35--10 mm & Far-IR  & \citet{PlanckPSC} \\
        \hline
        \multicolumn{9}{p{\textwidth}}{$^1$\url{http://svo2.cab.inta-csic.es/vocats/cmc15/docs/CMC15_Documentation.pdf} 
        $^2$\url{http://cds.u-strasbg.fr/denis.html} 
        $^3$All\emph{WISE}, cat\emph{WISE} and un\emph{WISE} are different reductions of \emph{Wide-field Infrared Survey Explorer} data. 
        $^4$\url{https://irsa.ipac.caltech.edu/Missions/msx.html} 
        $^5$ \emph{DIRBE} filters are [1.25], [2.2], [3.5], [4.9], [12], [25], [60], [100], [140] and [250].
        $^6$\url{https://doi.org/10.5270/esa-rw7rbo7}.
        $^7$\citet{Johnson1975} photometry on the 13-colour system (0.33--1.10 $\mu$m).
        $^8$\citet{Mermilliod1989} uses the DDO photometric system (0.35--0.48 $\mu$m).
        $^9$Rufener's catalogue uses the Geneva photometric system ($UBVG$).
        $^{10}$\url{http://irsa.ipac.caltech.edu/data/SPITZER/GLIMPSE}.
        $^{11}$\url{https://irsa.ipac.caltech.edu/data/SPITZER/Enhanced/SEIP/overview.html}.
        $^{12}$\emph{Planck} has a frequency-dependent beam size.}\\
        \hline
    \end{tabular}
\end{table*}

\begin{table}
    \caption{Sources of distance used in this paper.}
    \label{tab:distsources}
    \centering
    \begin{tabular}{@{}l@{\ }l@{\ }l@{}c@{}}
        \hline\hline
        \multicolumn{1}{c}{Reference}    & \multicolumn{1}{c}{VizieR}  & \multicolumn{1}{c}{Type} & \multicolumn{1}{c}{\llap{Prior}ity$^1$} \\
        \hline
\citet{ARVDB22}	& J/A+A/667/A74	  & Parallax distance	& 8 \\
\citet{BJRF+21}	    & I/352/gedr3dis    	  & Parallax distance	 & 9 \\
\emph{Hipparcos}	& I/311/hip2	          & Parallax        	 & 9 \\
\emph{Gaia} DR3	    & I/355/gaiadr3	          & Parallax	         & 10 \\
\citet{MRZ+21}	    & J/A+A/646/A74	          & Kinematic            & 11 \\
\citet{SLK+18}      & II/364/tableb1	      & Parallax	         & 11 \\
\citet{ARS+06}	    & V/136/tycall   	      & PM + colour          & 12 \\
\citet{ZRF+21}	    & J/A+A/650/A112	  & Parallax + kinematic & 12 \\
\citet{FVA07}	    & J/A+A/469/1221	  & Parallax	         & 12 \\
\citet{QAC+20}	    & J/A+A/638/A76	          & Photogeometric	     & 13 \\ 
\citet{RMF+10}$^\dag$  & ---	                  & Period--luminosity	 & $^\ast$ \\
\citet{Scicluna22}	& ---                     & Luminosity distance  & $^\ast$ \\
    \hline
    \multicolumn{4}{p{0.95\columnwidth}}{$^\ast$Smaller numbers indicate preferential use. \citet{RMF+10} and \citet{Scicluna22} given priority 9 in the final catalogue, but are excluded from some parts of the discussion (see Section \ref{sec:xmatch:distances:paper}). $^\dag$Sources of pulsation periods are listed in Table \ref{tab:persources}. Table \ref{tab:persources} lists sources of pulsation periods.}\\
    \hline
    \end{tabular}
\end{table}

\begin{table}
    \caption{Sources of pulsation period used in this paper.}
    \label{tab:persources}
    \centering
    \begin{tabular}{@{}l@{\ }l@{\ }l@{}c@{}}
        \hline\hline
        \multicolumn{1}{c}{Reference}    & \multicolumn{1}{c}{VizieR}  & \multicolumn{1}{c}{Regime} & \multicolumn{1}{c}{Priority$^1$} \\
        \hline
\citet{KKMS+02} & J/A+A/384/925/catalog & Optical & 9 \\
\citet{WWVG04} & J/AJ/128/2965/table4 & Optical & 10 \\
\citet{TMW+05} & J/AJ/130/776/table1 & Optical & 9 \\
\citet{WHP+06} & B/vsx/vsx & Optical & 9\\
\citet{TBK+09} & J/MNRAS/400/1945/table3 & Optical & 10--13$^\ast$ \\
\citet{PSK+10} & J/ApJS/190/203/var & Infrared & 10$^\dag$ \\
\citet{VCQFM+16} & J/ApJS/227/6/table1 & Optical & 10 \\
\citet{SKD+17} & B/gcvs/gcvs\_cat & Optical & 9 \\
\citet{BTS+18} & J/A+A/617/A32/tablea1 & Optical & 9\\
\citet{HTD+18} & J/AJ/156/241 & Optical & 10 \\
\citet{JKS+18} & II/366/catv2021 & Optical & 9 \\
\citet{ORS+18} & J/AJ/155/39/Variables & Optical & 13 \\
\citet{AMP+20} & J/ApJS/247/44/table2 & Optical & 9\\
    \hline
    \multicolumn{4}{p{0.95\columnwidth}}{$^\ast$The dominant period is given the highest priority; others are merely recorded. $^\dag$Multiple periods are given at different wavelengths: adopting the same priority provides a sigma-clipped average.}\\
    \hline
    \end{tabular}
\end{table}

\begin{table}
    \caption{Sources of temperature used in this paper.}
    \label{tab:tempsources}
    \centering
    \begin{tabular}{@{}l@{\ }l@{\ }l@{\ }c@{}}
        \hline\hline
        \multicolumn{1}{c}{Shorthand}    & \multicolumn{1}{c}{Reference}    & \multicolumn{1}{c}{VizieR}  & \multicolumn{1}{c}{Priority} \\
        \hline
APOGEE DR16         & \citet{APOGEEDR16}    & III/284/allstars    & 8\\
RAVE DR5            & \citet{RAVEDR5}    & III/279/rave\_dr5    & 8\\
Xiang2019        & \citet{Xiang2019}    & J/ApJS/245/34       & 8\\
Queiroz2020         & \citet{Quieroz2020}    & J/A+A/638/A76       & 8\\
\emph{Gaia} Apsis   & \citet{GaiaApsis}    & I/355/paramp        & 9\\
PASTEL              & \citet{PASTEL}    & B/pastel/pastel     & 9\\
LAMOST DR5          & \citet{LAMOST}    & V/164/stellar5      & 9\\
    \hline
    \end{tabular}
\end{table}

The default catalogues of {\sc PySSED} were queried, including all photometric data and \emph{Gaia} ancillary data and incorporating their bad-data rejection criteria. Additional data catalogues were sourced from VizieR near the start of this work (July 2021). To construct this list, VizieR was queried for the NESS list of targets, and the number of results per catalogue was identified from the list of 2783 catalogues returned. All catalogues containing more than 50 NESS sources were inspected individually, to which a small number of manually selected catalogues that included 20 or more NESS sources and probed important stellar parameters were also added. Superseded and outdated catalogues were then removed from the list, and each catalogue was inspected for data columns that were both considered relevant and could be manipulated into the {\sc PySSED} interface\footnote{Very few tables could not be parsed into a format interpretable by {\sc PySSED}, however photometric data listed in colour format was first downloaded and converted to magnitudes, then fed into {\sc PySSED} as a file.}, resulting in a list of 333 individual VizieR queries that {\sc PySSED} makes for each star. Catalogues were updated to their post-2021 versions as work progressed. 

For all catalogues, the cross-matching radius for optical data was assigned to approximate the 95 per cent confidence bounds for the astrometric precision of each catalogue. The cross-matching radii for mid-IR photometry (3.4--100\,$\mu$m) in Table \ref{tab:photosources} was increased (and notably increased from the {\sc PySSED} default settings) to acknowledge the brightness of our objects at these wavelengths and the corresponding decrease in the likelihood of a bad cross-match for such extremely bright sources.

Tables \ref{tab:photosources}, \ref{tab:distsources}, \ref{tab:persources} and \ref{tab:tempsources} respectively list sources of photometry, distances, pulsation periods and spectral temperature used in this work. Priority to data sources is chosen such that sources that typically have better accuracy due to higher resolution or signal-to-noise are given the higher priority. Full data sources for all parameters can be found in the input files {\tt catalogues.ness} and {\tt ancillary.ness} in the Supplementary Material, with criteria for rejecting bad data in the {\tt rejects.ness} file.

\section{Comparison of fitting and distance-estimation methods}
\label{apx:dist}

\subsection{Model-atmosphere versus blackbody fits}
\label{apx:dist:lumtypes}

\begin{figure}
\centering
\includegraphics[height=\linewidth,angle=-90]{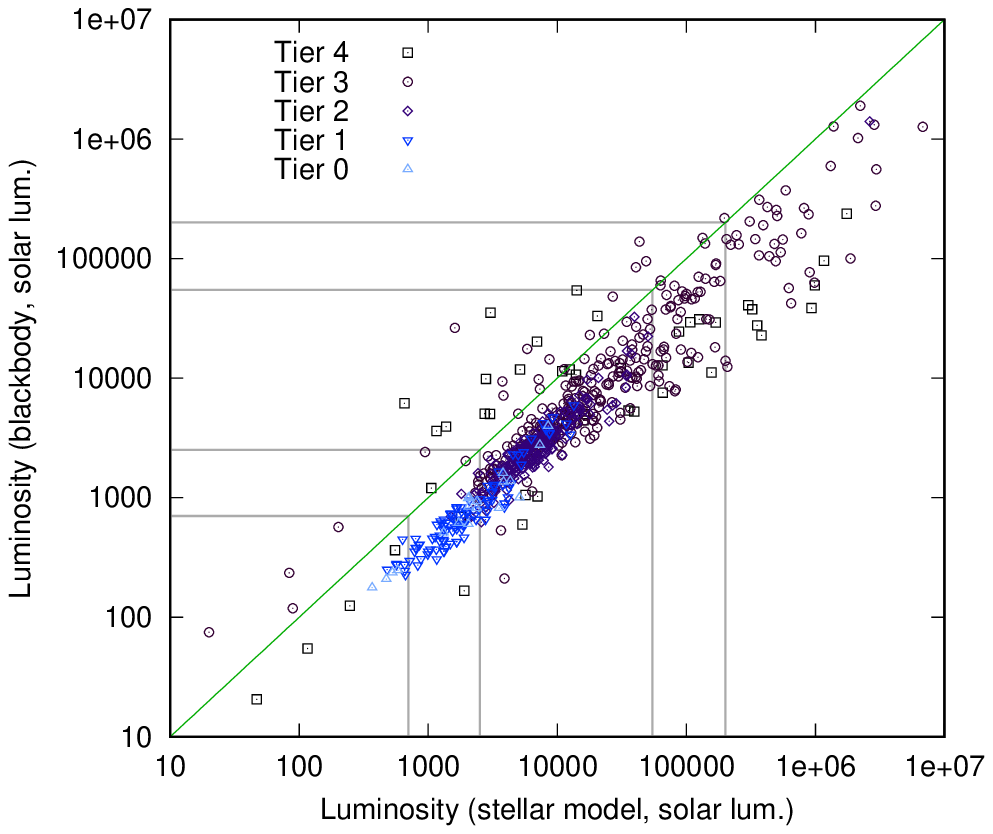}
\includegraphics[height=\linewidth,angle=-90]{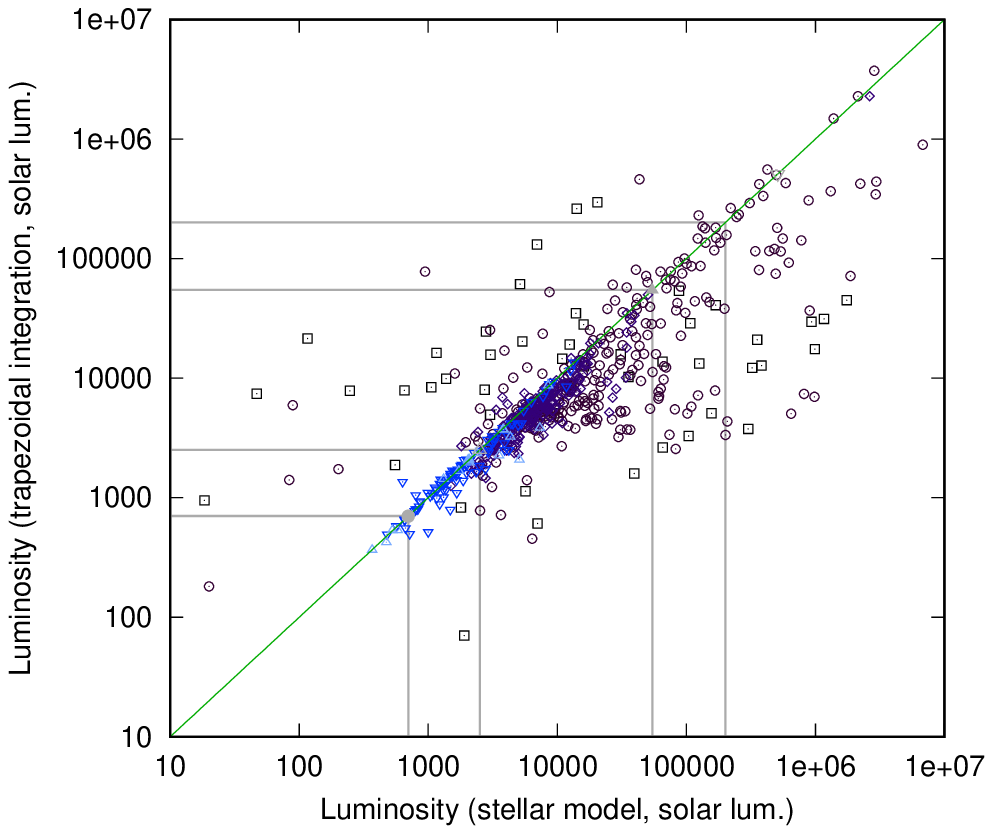}
\includegraphics[height=\linewidth,angle=-90]{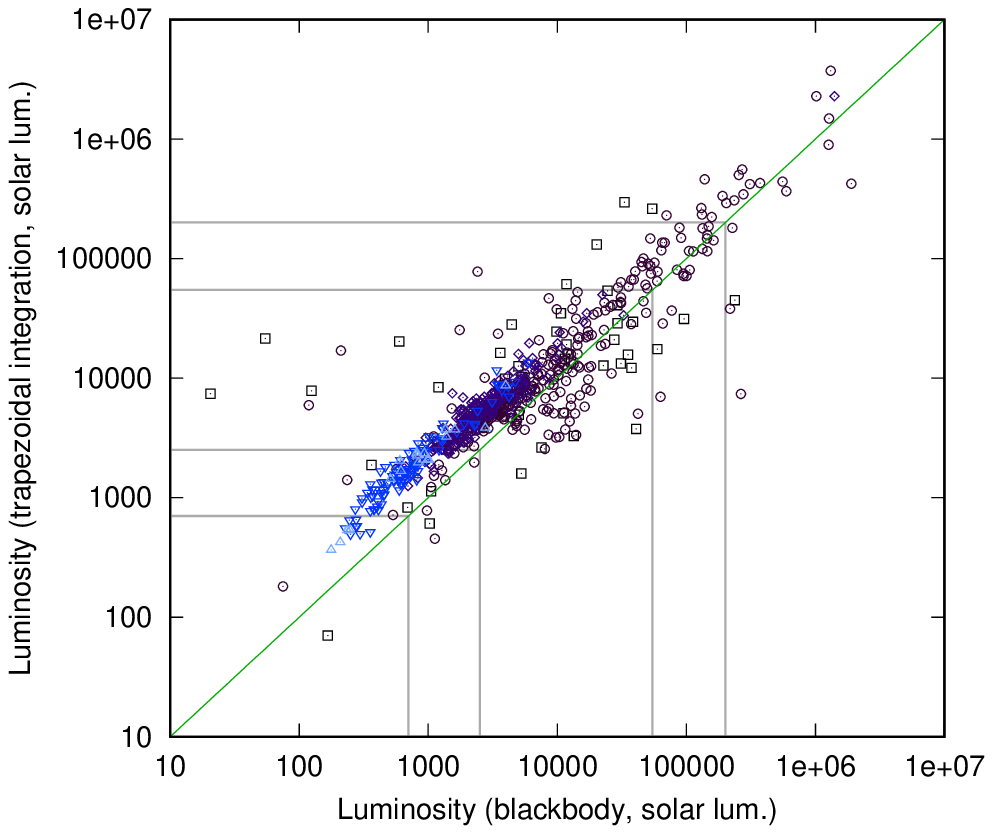}
\caption{Comparison of luminosities from fitting SED models and blackbodies with {\sc PySSED}, and simple trapezoidal integration of the SED. Coloured points denote the NESS tiers. The green line denotes 1:1 parity, while the grey lines denote luminosity boundaries at (in increasing luminosity) our 700\,L$_\odot$ cutoff, the RGB tip, the classical AGB limit and the approximate observed luminosity limit for RSG stars.}
\label{fig:lum_fit_trap_bb}
\end{figure}

Limitations in the ability of available broadband photometry to describe an SED, and in the availability of accurate distances to our objects, has consequent limitations in our abilities to extrapolate properties from complex SEDs of these dusty stars. We also note that our luminosities reflect spherical extractions from the flux received in our line of sight, and may not truly represent the luminosity of stars with spherically asymmetric surfaces or envelopes.

Figure \ref{fig:lum_fit_trap_bb} compares the luminosities derived from fitting simple (non-dusty) stellar atmosphere models to our stars, versus fitting blackbodies and fitting luminosities derived from trapezoidal integration of the SED without outlier rejection  (the trapezoid luminosities have no fitting parameters). Distance uncertainties shift stars along the parity line; differences in fit quality scatter stars from the diagonal parity line. The blackbody fit has several differences from the stellar model fit:
\begin{itemize}
    \item Long-wavelength photometry between $20 < \lambda \leq 1000\,\mu$m is included in the blackbody fit.
    \item More weighting is given to points far from the SED peak ({\tt WeightedTSigma} = 2 instead of 1).
    \item A starting temperature of 500 K instead of 3000 K is used.
    \item To fit optically thick sources, the lower temperature limit is relaxed from 1000\,K (with a 1000\,K softening parameter) to 100\,K (and 10\,K).
    \item The \emph{Gaia} GSP-Phot spectroscopic temperature is no longer used as either a prior ({\tt UsePriorsOnTspec}) or a starting point ({\tt UseGaiaModelStart}) for fitting.
\end{itemize}

For the less-extreme NESS tiers 0, 1 and 2, the model-derived and trapezoid-integrated luminosities match each other closely, demonstrating the close agreement of the stellar models with the data and the well-sampled SEDs. Most of the ``high'' Tier 3 stars are similarly well fit. However, a fraction of the Tier 3 and most of the ``extreme'' Tier 4 stars do not show such good agreement. This is expected, since the SEDs of ``extreme'' stars are dominated by their dust, not the underlying starlight.

The luminosity for less-extreme sources is significantly under-estimated by the blackbody fit compared to the other two methods. Typically, these (mostly oxygen-rich) stars are still warm enough that most of their bolometric flux is emitted at $\lesssim$1\,$\mu$m, meaning the deep optical TiO bands distort the spectrum significantly from a blackbody, forcing additional output in the near-IR, and lowering the blackbody-fitted temperature significantly but without strongly affecting the fitted radius.

For the more-extreme Tier 3 and 4 stars, there is better agreement between the trapezoid-integrated and blackbody-derived luminosities than with the model-derived luminosities, as the stars depart significantly from dustless stellar atmosphere models. However, there is considerable disagreement between all three methods in a few cases. These sources generally have double-peaked SEDs, where the infrared dust excess dominates, but where there is still a strong optical component. Such sources can be binary stars or chance super-positions of sources, but often this double-peaked SED is indicative of a post-AGB object \citep[e.g.][]{RKJ+15}.

\begin{figure*}
\centering
\includegraphics[height=0.47\linewidth,angle=-90]{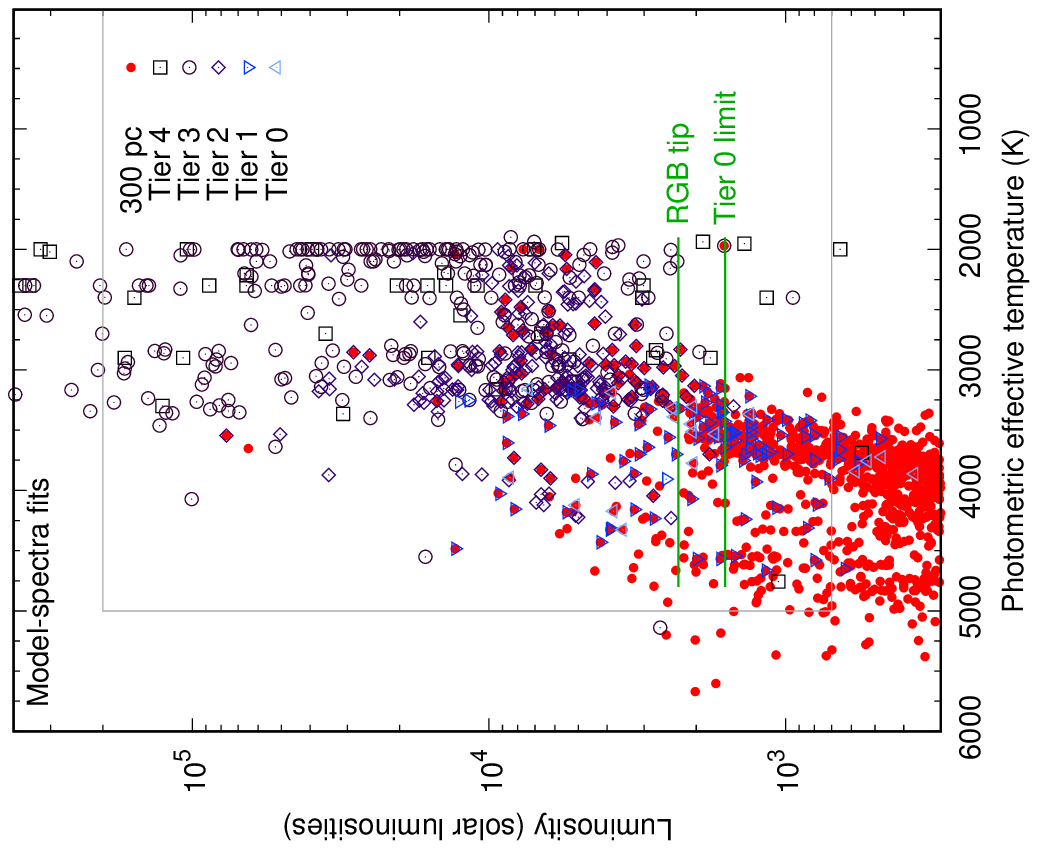}
\includegraphics[height=0.47\linewidth,angle=-90]{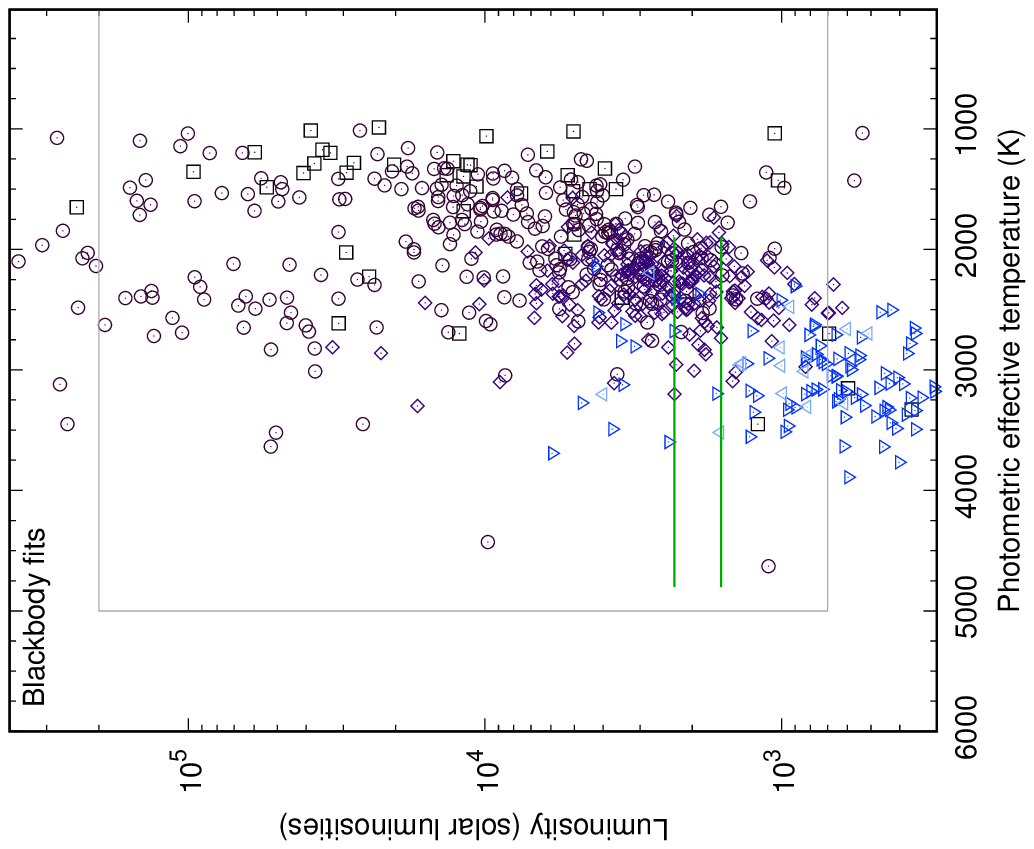}
\caption{H--R diagrams of the NESS sample, fitted with (left) a stellar model atmosphere and (right) a blackbody. The grey boundary shows the temperature (5000 K) and luminosity (700 / 200\,000 L$_\odot$) bounds denoting evolved stars for the purposes of this work. The short green lines at 1600 and 2300 L$_\odot$ respectively mark the nominal luminosity cutoff for Tier 0 and the approximate location of the RGB tip.}
\label{fig:bb-model-hrd}
\end{figure*}

\begin{table}
    \centering
    \caption{Difference between average properties of model-atmosphere and blackbody fits by tier.}
    \label{tab:fitaverages}
    \begin{tabular}{@{}c@{\ }l@{\ }c@{}c@{}c@{}c@{}c@{}c@{}@{}}
    \hline
    Tier& Descriptor & $<\!T_{\rm model}\!>$ & $<\!L_{\rm model}\!>$ & $<\!T_{\rm bb}\!>$ & $<\!L_{\rm bb}\!>$ & $<\!\frac{T_{\rm bb}}{T_{\rm model}}\!>$ & $<\!\frac{L_{\rm bb}}{L_{\rm model}}\!>$\\
    \ & \ & (K) & (L$_\odot$) & (K) & (L$_\odot$) & \ & \ \\
    \hline 
      0 & Very low     & 3640 & 3081 & 2983 & 1099 & 0.82 & 0.39\\
      1 & Low          & 3691 & 2706 & 3052 & 1023 & 0.83 & 0.39\\
      2 & \rlap{Intermediate} & 2970 & 20\,363 & 2317 & 9291 & 0.79 & 0.39\\
      3 & High         & 2597 & 205\,227 & 1980 & 64\,554 & 0.77 & 0.46\\
      4 & Extreme      & 2746 & 142\,822 & 1933 & 26\,572 & 0.68 & 0.96\\
    -- & {\it All}    & {\it 2910} & {\it 107\,349} & {\it 2263} & {\it 33\,434} & {\it 0.78} & {\it 0.46}\\
    \hline
    \end{tabular}
\end{table}

Figure \ref{fig:bb-model-hrd} shows a comparison of the H--R diagrams generated by fitting both model-atmosphere spectra and blackbodies to the SEDs of the NESS sample stars, showing both the luminosities from Figure \ref{fig:lum_fit_trap_bb} and the corresponding temperatures. Only those in the restricted dataset are shown; trapezoidal integration results lack temperatures, so cannot be shown.

The blackbody fits can clearly be seen to fit stars as cooler and fainter, as well as allowing fits below the 2000\,K limit of the stellar atmosphere models. The difference in the average properties is shown in Table \ref{tab:fitaverages}. In general, temperatures are $\sim$20 per cent lower and luminosities $\sim$60 per cent lower for the blackbody fits than for the models. The exception to this is the ``extreme'' Tier 4 (and a few sources in the ``high'' Tier 3), where the model fits are limited by the available model grid to those above 2000\,K, so the corresponding temperature difference is larger and models often fail to properly fit the SED at all. For the lower tiers, the properties of the AGB stars retrieved by the stellar atmosphere models more accurately represent those expected for AGB stars (i.e., stars are generally above the $\sim$2500\,L$_\odot$ RGB-tip luminosity), so we anticipate that these models are accurate for Tiers 0, 1 and 2, and most stars in Tier 3. For some stars in Tier 3 and most stars in Tier 4, we anticipate that the blackbody fits are more accurate.

\subsection{Comparison of distance estimates}
\label{apx:dist:comp}

\begin{figure*}
\centering
\includegraphics[height=0.33\linewidth,angle=-90,trim={0 0mm 0 6mm},clip]{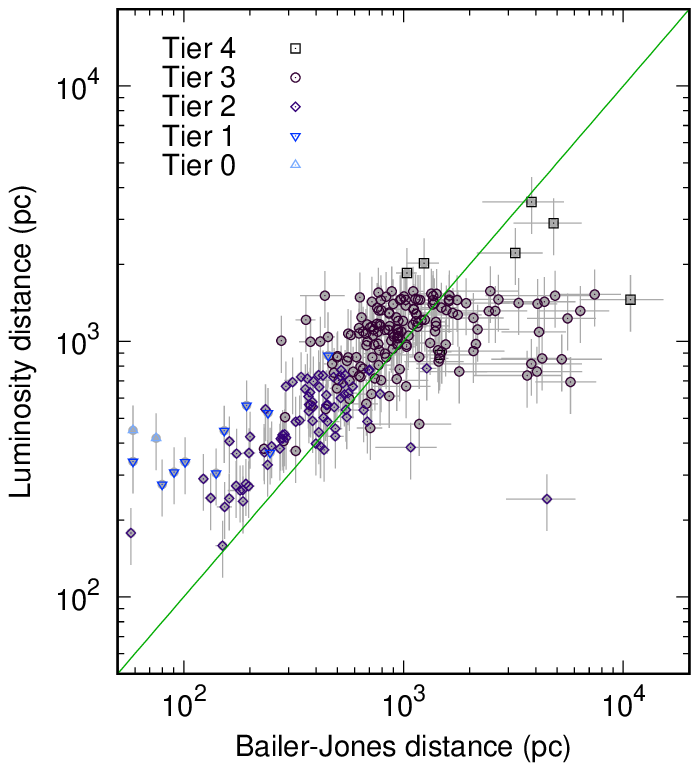}
\includegraphics[height=0.33\linewidth,angle=-90,trim={0 0mm 0 6mm},clip]{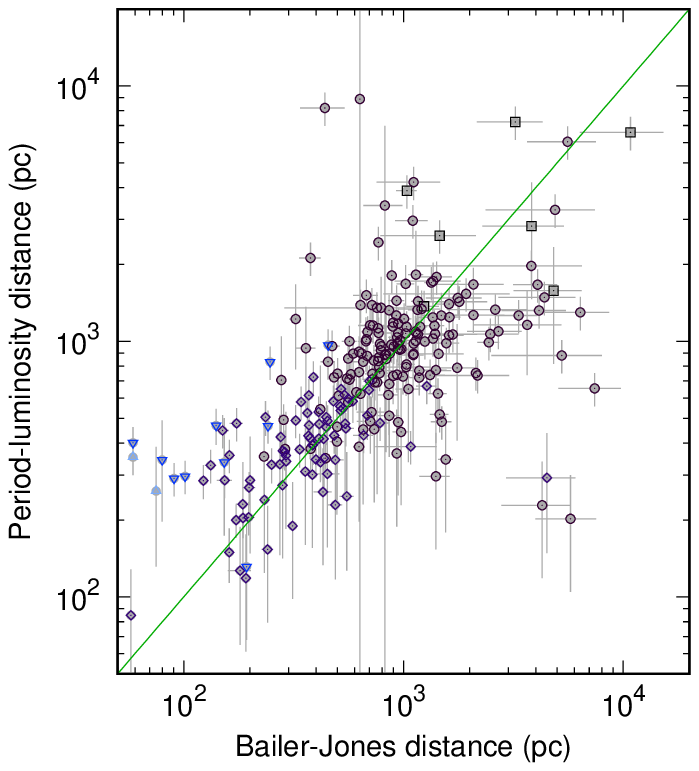}
\includegraphics[height=0.33\linewidth,angle=-90,trim={0 0mm 0 6mm},clip]{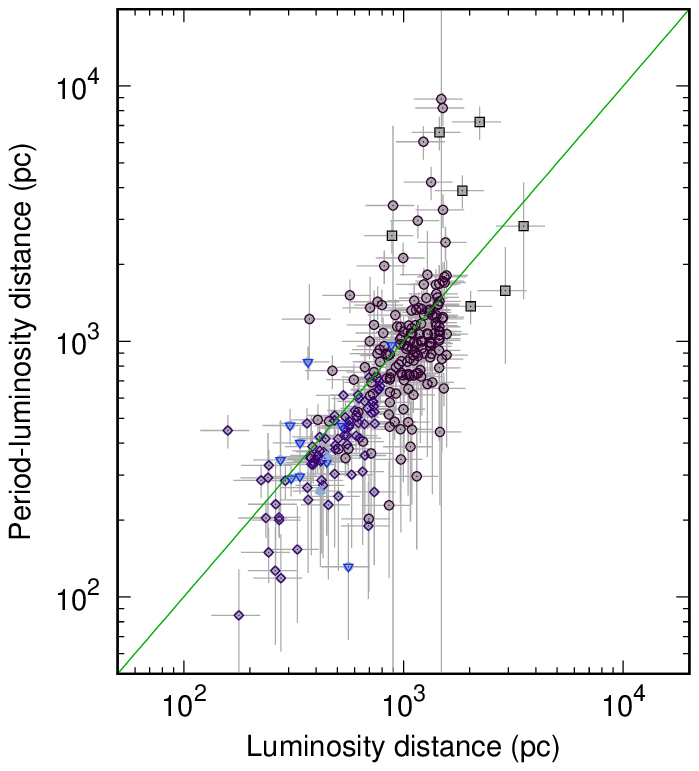}
\caption{Comparison of distance estimates by different methods: parallax distances from \citet{BJRF+21}, period--luminosity distances and the original NESS luminosity distances. Errors are indicative. Diagonal lines show 1:1 correspondence of estimated distance.}
\label{fig:distcompare}
\end{figure*}

\begin{figure}
\centering
\includegraphics[height=\linewidth,angle=-90,trim={0 0mm 0 6mm},clip]{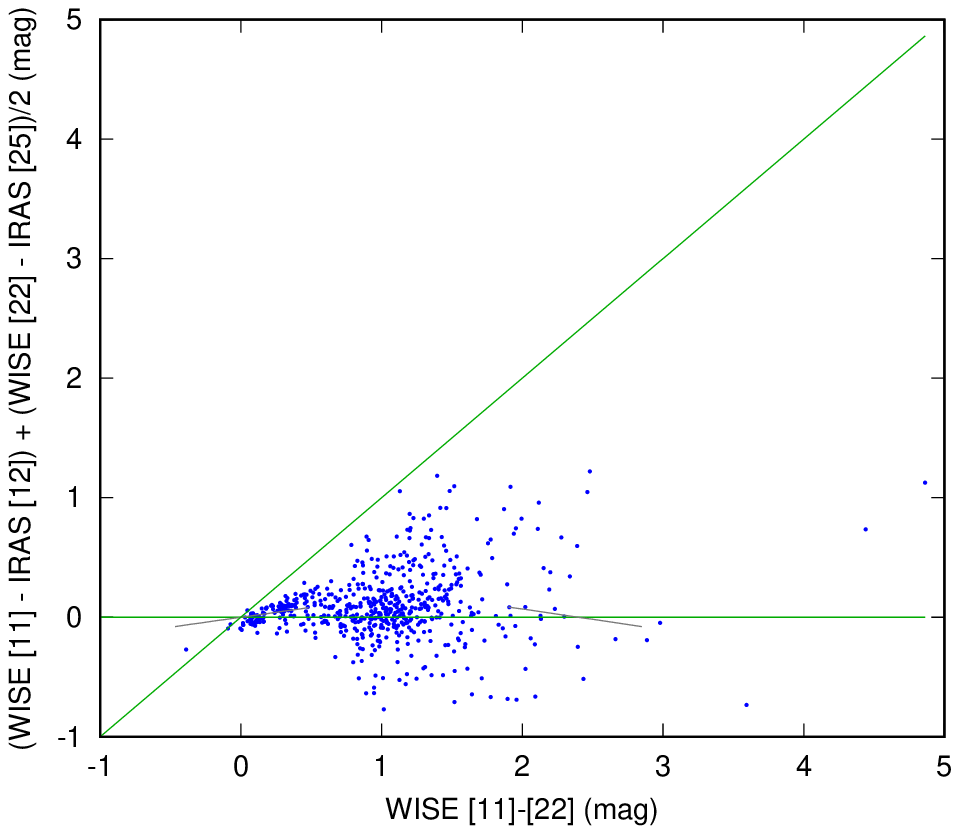}
\caption{\emph{WISE} versus \emph{IRAS} colour index for the NESS sample, showing the differences between neighbouring filters with different telescopes ([11]--[12] and [22]--[25]) versus the more distantly spaced \emph{WISE} [11]--[22] filters. Magnitudes in all filters use a Vega zeropoint. Most stars are expected to lie close to the zero line shown in green: a pure $F \propto \lambda^{-2}$ (Rayleigh--Jeans) spectrum would closely follow the dashed grey line. A unity line has also been shown for comparison.}
\label{fig:wiseiras}
\end{figure}

\begin{figure}
\centering
\includegraphics[height=\linewidth,angle=-90,trim={0 0mm 0 6mm},clip]{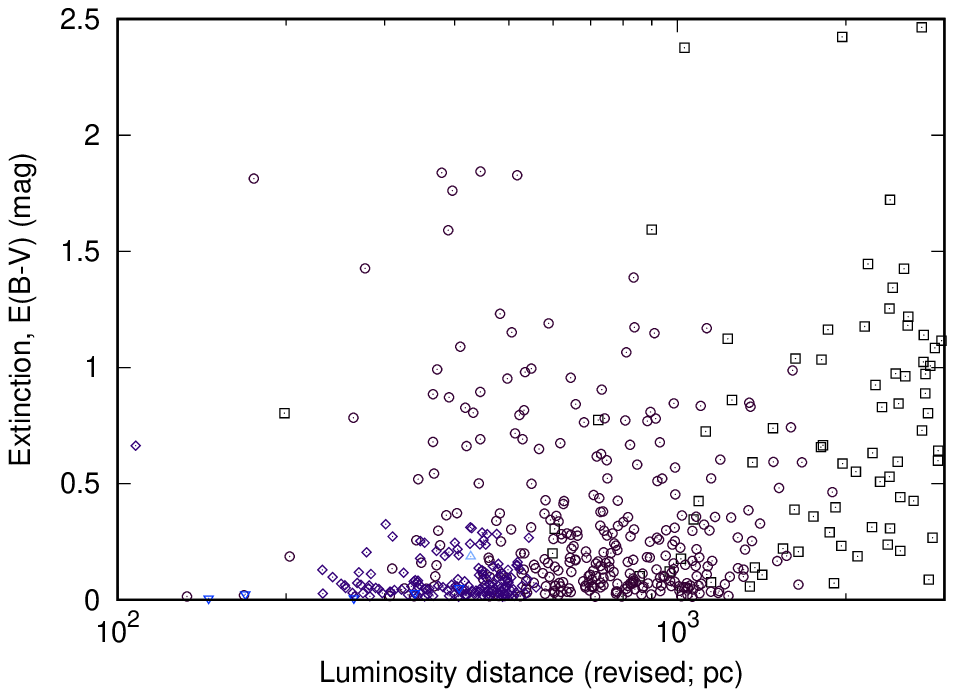}
\caption{Distribution of extinctions assumed for NESS targets. The colour scale is as in Figure \ref{fig:distcompare}.}
\label{fig:ebv}
\end{figure}

\begin{figure}
\centering
\includegraphics[height=\linewidth,angle=-90,trim={0 0mm 0 2mm},clip]{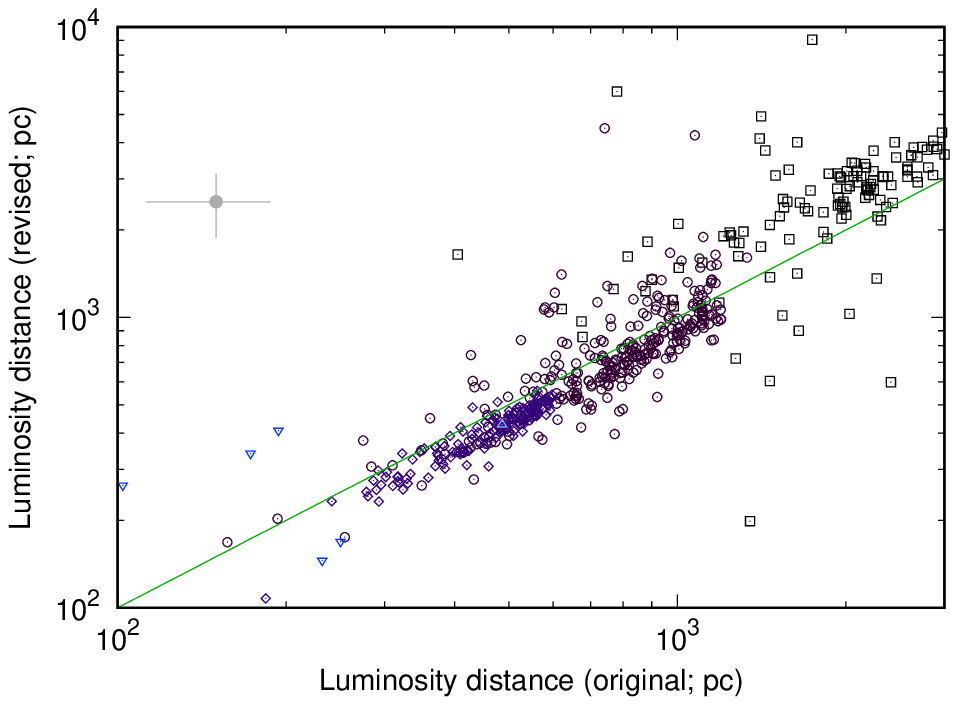}
\caption{Comparison of the original NESS luminosity distances and revised luminosity distances calculated with updated photometry and dereddening. Representative error bars for individual sources are shown. The colour scale is as in Figure \ref{fig:distcompare}.}
\label{fig:newlumdist}
\end{figure}

In this section, we compare the three main methods used to find distances to stars in the NESS sample: \emph{Gaia} parallaxes from \cite{BJRF+21}, period--luminosity relations, and the original NESS luminosity-based distances \citep{Scicluna22}. Figure \ref{fig:distcompare} compares each pair in turn.

\subsubsection{Parallax distance versus luminosity distance}

Comparing parallax versus luminosity distances contrasts our most accurate measure of distance at short distances against the distance measure used to define the NESS survey. Between parallax distances of 200 and 2000 pc, there is a good correlation, with the luminosity distance over-estimating the parallax distance by a median factor of 1.27 (with a 68 per cent interval of 0.89--1.71). The range of $\sim\pm$32 per cent is consistent with statements made in \citet{Scicluna22} noting an expected $\sim$25 per cent uncertainty in stellar distance using this method, but the increase of the average distance by 27 per cent is notable.

Below 200 pc, we expect parallax measurements to be accurate, as these stars tend to be from the lower NESS tiers, thus warmer and weaker pulsators not subject to astrometric noise: in general, their \emph{Hipparcos} and \emph{Gaia} parallaxes match each other closely. However, Tiers 0 and 1 may include upper RGB stars as well as low-luminosity AGB stars, and Figure \ref{fig:distcompare} shows that luminosity distances are over-estimated compared to parallax distances.

Similarly, beyond 2000 pc, the luminosity distance under-estimates compared to the parallax distance. While parallax distances may be underestimated due to astrometric noise affecting the parallax solution, sources at these distances are mostly very late-type (hence very luminous) OH/IR stars or known supergiants. This demonstrates a bias towards these stars in the NESS survey.

\subsubsection{Parallax distance versus period--luminosity distance}

Comparing the parallax distance to the distance from the $P$--$L$ relation, again in the range 200--2000 pc, the median ratio of parallax to period distances is 0.94 (0.63--1.42), making the $P$--$L$ relation a more accurate measure than the luminosity distance but, with its $\sim\pm$50 per cent scatter, a less precise one. Again, the closer stars from lower tiers have much larger distances based on $P$-$L$ relations than from parallaxes. In part, this could because variables are being recognised on sequence $D$ (the long-secondary period sequence) rather than the fundamental mode $C$.

Some stars with parallax distances beyond 2000 pc show very scattered distances, showing a general breakdown of the $P$-$L$ relationship near periods of $\sim$700 days, as circumstellar dust affects the validity of the $P$-$L$ relation, possibly compounded by reduced accuracy of the \citet{BJRF+21} Galactic model. However, there are also a small number of stars that cluster at a factor of 2--3 below the parity line: these are massive stars pulsating in the first overtone, whose distances are underestimated as a result.

\subsubsection{Old versus new luminosity-based distances}
\label{apx:dist:comp:oldnew}

The luminosity-based distances used to select the Tier 2, 3 and 4 NESS sources in \citet{Scicluna22} were based on trapezoidal integration of the 2MASS $JHK_{\rm s}$, and \emph{IRAS} [12] and [25] fluxes. Each star was assumed to have a luminosity of 6200\,L$_\odot$ (the LMC median, as used in \citet{Scicluna22}), which was used to scale the distance to the object by $F \propto d^{-2}$. We can now update these luminosity distances, using the same method but making five improvements.

Firstly, we add many more photometric surveys, extending the SED into the optical and further into the infrared. This allows a more accurate SED to be constructed, particularly for warm sources with an optical SED peak. These additions could increase or decrease the integrated flux, so could also increase or decrease the projected distance.

Secondly, this paper corrects the SEDs for interstellar extinction (Figure \ref{fig:ebv} shows the $E(B-V)$ distribution). This increases the optical integrated flux, thus decreasing the projected distance while increasing fitted temperature. For a typical NESS target\footnote{We can use the intermediate-tier T Sge as one with typical properties.}, with a median extinction of $E(B-V) = 0.17$ mag, the luminosity increases by $\approx$14 per cent, thus the distance decreases by $\approx$7 per cent. The effect will be stronger for stars with higher extinction and warmer temperatures.

Thirdly, we can improve on the procedure by assuming a $F \propto \lambda^{-4}$ Wien tail and a $F \propto \lambda^3$ Rayleigh--Jeans tail beyond either end of the observed photometry, and we update the wavelengths of the filters to the effective wavelengths from the SVO catalogue.

Fourthly, we update our median luminosity of choice from $6200^{+2800}_{-3900}$\,L$_\odot$, which represents the median luminosity and 68 per cent confidence interval of the sampled LMC stars, to $5363^{+9613}_{-2910}$\,L$_\odot$, which represents the median luminosity and 68 per cent interval of the NESS stars which have known distances (see main text).

Finally, we colour-correct the \emph{IRAS} photometry: \citet{Scicluna22} used the \emph{IRAS} photometric fluxes in their raw catalogue form, which assumes $F_\nu \propto \nu^{-1}$. However, most of our sources (even the extreme sources) are better represented\footnote{A spectrum of $F_\nu \propto \nu^2$ represent a (colour-corrected) \emph{IRAS} or \emph{WISE} colour of zero in the Vega system. A spectrum of $F_\nu \propto \nu^{-1}$ represents an \emph{IRAS} colour of [12]--[25] $\approx$ 2.4 mag, or a \emph{WISE} colour of [11]--[22] $\approx$ 2.1 mag. While some stars do reach these colours (Figure \ref{fig:wiseiras}), there are relatively few of them.} beyond $\lambda \sim 9\,\mu$m by a $F_\nu \propto \nu^2$ Rayleigh--Jeans law (as assumed by \emph{WISE}\footnote{\url{https://wise2.ipac.caltech.edu/docs/release/allwise/faq.html}}). In this work, colour-corrections have been applied to \emph{MSX}\footnote{\url{https://irsa.ipac.caltech.edu/data/MSX/docs/MSX_psc_es.pdf}}, \emph{IRAS}\footnote{\url{https://lambda.gsfc.nasa.gov/product/iras/iras_colorcorr.html}}, \emph{DIRBE}\footnote{\url{https://lambda.gsfc.nasa.gov/data/cobe/dirbe/ancil/colcorr/DIRBE_COLOR_CORRECTION_TABLES.ASC}} and \emph{Akari}\footnote{\url{https://data.darts.isas.jaxa.jp/pub/akari/AKARI_Documents/AKARI-IRC/DataUserManual/IRC_DUM_v2.2_20160706.pdf}} based on available colour corrections for a 5000 K blackbody\footnote{The temperature of the blackbody does not significantly affect the colour corrections, provided the observations are on the Rayleigh--Jeans tail of the SED: a 5000 K temperature is applied for typical stars in PySSED, but suffices for our $\sim$2000--4000 K evolved stars at these wavelengths.}. For \emph{IRAS}, this colour correction amounts to a factor of 1.4 decrease in flux, resulting in a increase in projected distance of up to $\sim$18 per cent (since $\sqrt{1.4} \approx 1.18$), though the actual amount of increase will depend on the contribution of the \emph{IRAS} flux to the overall SED.

Figure \ref{fig:wiseiras} shows the revised \emph{WISE} and \emph{IRAS} colours with this colour correction in place: the colour on the vertical axis, which represents the colour excess between the \emph{WISE} and \emph{IRAS} magnitudes, has (for \emph{WISE} [11]--[22] = 0 mag) decreased from $\sim$0.4 mag to close to zero, as expected for the pure Rayleigh--Jeans tail of dustless stars. For stars of increasing [11]--[22] colour, the mineralogy of dust around the star scatters the stars from the zero line, though to generally slightly positive colours. With the revised colour correction, the median colours of the NESS sources and their 68 per cent confidence intervals are:
\begin{itemize}
    \item $[11]-[12] = 0.13\ (-0.14 - 0.47$) mag,
    \item $[22]-[25] = 0.01\ (-0.13 - 0.34$) mag,
    \item $[11]-[22] = 0.98\ (0.35 - 1.48$) mag, and
    \item $[12]-[25] = 0.85\ (0.21 - 1.32$) mag.
\end{itemize}
Much of the remaining difference in [11]--[12] is likely to be intrinsic to the stars, since the [12] and [11] filters differ in their coverage of the 10\,$\mu$m silicate emission feature. It should also be noted that all objects are above the \emph{WISE} [11] nominal saturation limit of +4 mag, while most stars also above the [22] limit of +0 mag, beyond which the \emph{AllWISE} catalogue is at risk of not being fully calibrated.

The overall effect of these four improvements to the luminosity distance estimate can be seen in Figure \ref{fig:newlumdist}. Extinction correction and updated photometry causes scatter in the relationship, particularly at large distances, but a general trend can be seen. A small offset in photometry at $\sim$300 pc becomes, at maximum, at 30 per cent decrease in distance for stars at $\sim$800 pc. This trend then reverses as distant objects become more extreme and concentrated in the Galactic Plane. Here, confusion affecting the \emph{IRAS} photometry (and ultimately Malmquist bias) can decrease an object's flux in higher-resolution surveys, which increases its luminosity distance.

\subsection{Revised tiering}
\label{apx:dist:newtier}

\begin{figure}
\centering
\includegraphics[height=\linewidth,angle=-90]{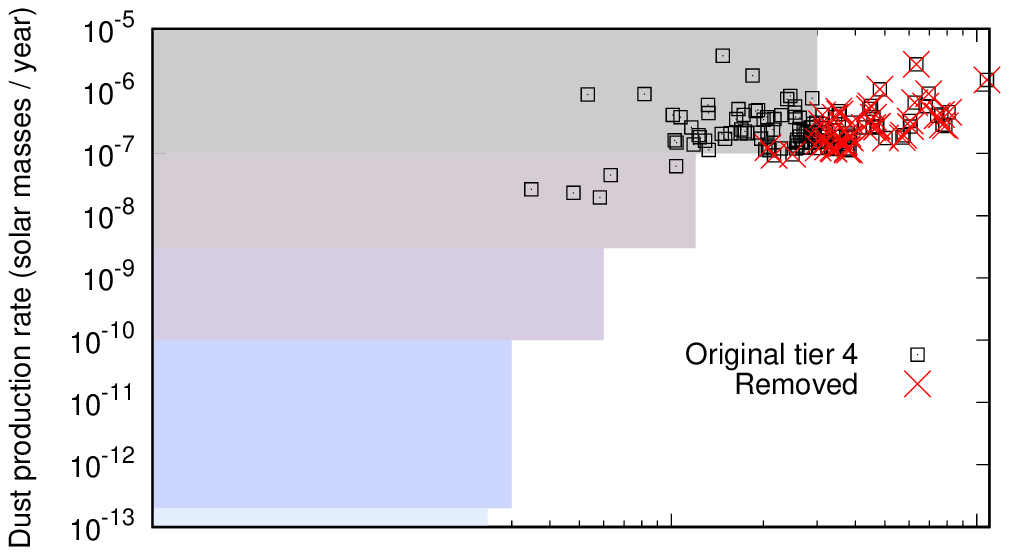}
\includegraphics[height=\linewidth,angle=-90]{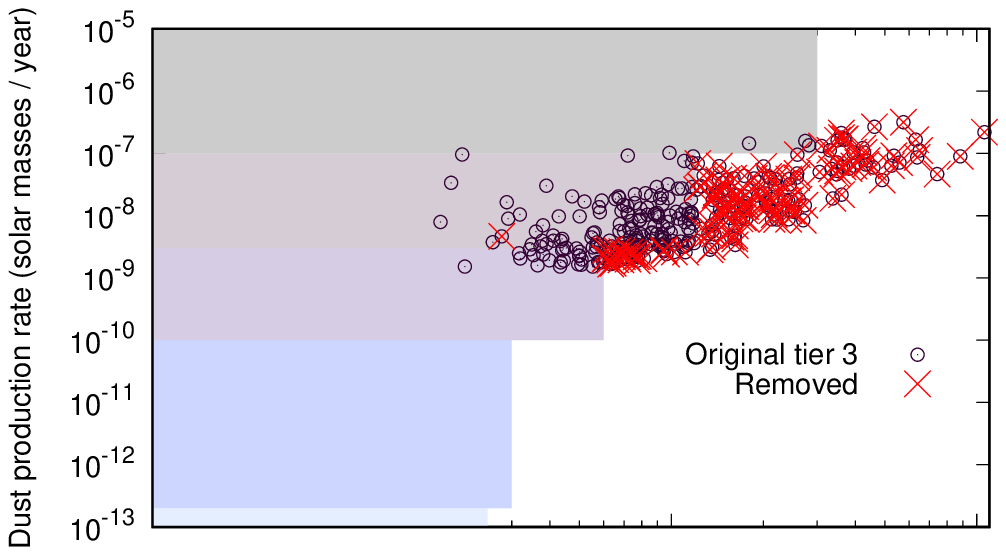}
\includegraphics[height=\linewidth,angle=-90]{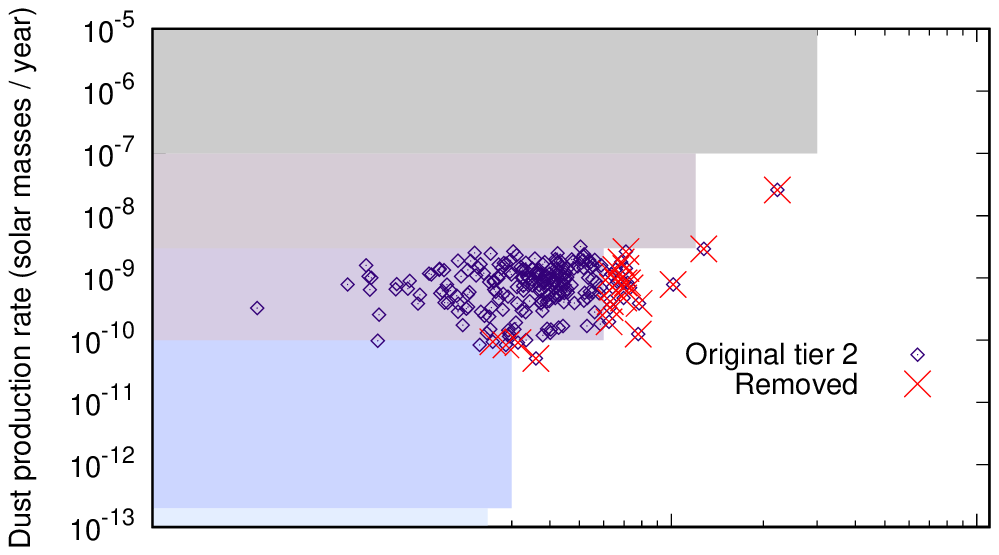}
\includegraphics[height=\linewidth,angle=-90]{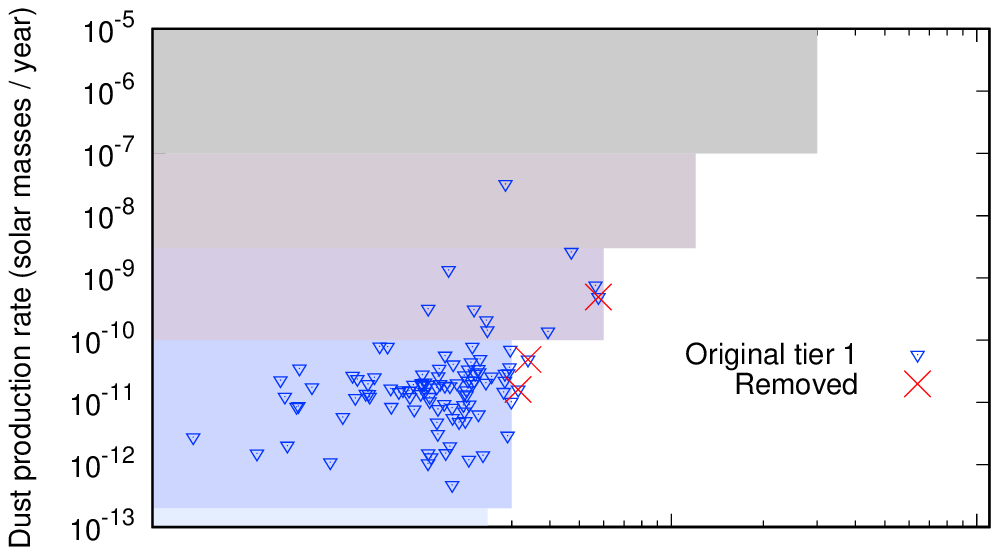}
\includegraphics[height=\linewidth,angle=-90]{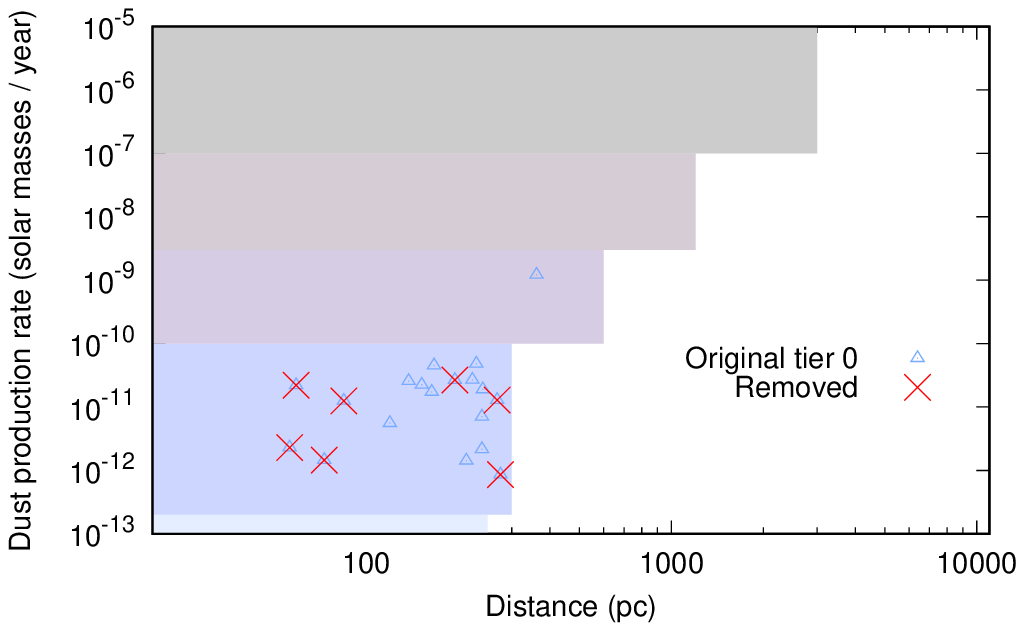}
\caption{Revised tiering from distance and DPR changes by tier. Filled boxes show tier boundaries. Note that the boundary for tier 0 is a luminosity-based boundary, rather than a $\dot{D}$ boundary.}
\label{fig:dDdottiers}
\end{figure}

Figure \ref{fig:dDdottiers} shows, on a tier by tier basis, how revisions to the distance estimates affect the dust-production rate ($\dot{D}$) estimates for the NESS survey, and therefore how the stars in various tiers should be either redistributed throughout the existing NESS tiers or removed from the survey altogether.

\section{Rejected sources}
\label{apx:reject}

\subsection{Objects that are not evolved stars}
\label{apx:reject:not}

With a few notable exceptions, rejects are generally young stellar objects (YSOs), which have similar observable properties to heavily embedded evolved stars. In some cases, it is difficult to determine whether objects are evolved stars (RSGs or AGB stars) or YSOs. A general philosophy adopted here is that objects forming within the last $\sim$10$^7$ yr are likely to be too massive to evolve into RSGs (instead undergoing supernova as blue or yellow supergiants), and that star-forming clouds in the immediate vicinity of the star should have dispersed by this time. Consequently, AGB stars should be physically separated from star-forming regions (or at worst superimposed on them), and should not be hot enough to generate their own H\,{\sc ii} regions.

The following list provides the sources that we manually reject as not being evolved stars, plus a note or discussion on the reason they were rejected. These objects are not explored further in this paper.
\begin{itemize}
\item \emph{IRAS}\,05362-0626: part of the Orion Nebula.
\item \emph{IRAS}\,05389-6908 and 05401-6940: parts of the Tarantula Nebula surrounding 30 Doradus.
\item \emph{IRAS}\,05377+3548: associated with a pair of embedded, star-forming clusters within the wider H\,{\sc ii} region Sh 2-235, surrounding the O9.5V star BD+35\,1201.
\item \emph{IRAS}\,05388-0147: associated with a region of the Flame Nebula (NGC 2024).
\item \emph{IRAS}\,06050-0623: associated with the B1 star BD--06 1415 and the nebula that surrounds it. Part of the wider Orion Molecular Cloud.
\item \emph{IRAS}\,06491-0654: classified in {\sc simbad} as a Herbig Ae/Be star with spectral type A1Ib/II.
\item \emph{IRAS}\,07422+2808 ($\beta$ Gem): although classified as a K0III giant, this star was later found to be sufficiently far down the red giant branch (i.e., sufficiently less evolved) to warrant its exclusion.
\item \emph{IRAS}\,09572-5636, 09576-5644 and 09578-5649; \emph{IRAS}\,11254-6244, 11260-6241 and 11266-6249: parts of associated infrared nebulae spanning Vela and Carina, identified in numerous literature sources as likely star-formation sites. The latter three are knots in the wider nebula RAFGL 4132. \emph{IRAS}\,11254-6244 blended with the star TYC 8976-3711-1.
\item \emph{IRAS}\,10431-5925: $\eta$ Car. While this interacting binary could be considered an evolved star under some classifications, it is too hot to include in our criteria here.
\item \emph{IRAS}\,11202-5305 (HD\,98922): classified in {\sc simbad} as a Herbig Be star with spectral type B9Ve.
\item \emph{IRAS}\,13416-6243: classified in {\sc simbad} as a post-AGB star, this is spectroscopically determined to be a G1 supergiant by \citet{Hu1993}.
\item \emph{IRAS}\,14050-6056 and 16434-4545: these do not appear to be associable with any mid-infrared (\emph{WISE} or \emph{Akari}) source.
\item \emph{IRAS}\,14359-6037 ($\alpha$ Cen): mistakenly included due to its high proper motion.
\item \emph{IRAS}\,15141-5625: a blend between the star 2MASS\,15180114-5637360 and the probable young stellar object 2MASS\,15175464-5636357 (G322.0970+00.7105). Lies within the extended structure of the molecular cloud GAL 322.2+00.6, and treated as contamination from the infrared-bright nebula.
\item \emph{IRAS}\,16124-5110: embedded source near an H\,{\sc ii} region; at Galactic latitude $b = -0\fdeg42$.
\item \emph{IRAS}\,16545-4012: embedded source near infrared nebulae; at Galactic latitude $b = +1\fdeg58$.
\item \emph{IRAS}\,16555-4237 (V921 Sco): classified in {\sc simbad} as a Be star with spectral type B0IVe.
\item \emph{IRAS}\,16557-4002: recognised as an H\,{\sc ii} region; at Galactic latitude $b = +1\fdeg51$.
\item \emph{IRAS}\,17326-3324 (HD\,159378): a yellow supergiant, classified by {\sc simbad} as G3Ia spectral type, i.e., too early a spectral type for inclusion.
\item \emph{IRAS}\,17423-2855: Sgr A$^\ast$.
\item \emph{IRAS}\,17441-2822: an H\,{\sc ii} region close to the Galactic Centre.
\item \emph{IRAS}\,17590-2337 (WR\,104): a WC9 + B2V binary with substantial dust production \citep{Soulain2023}.
\item \emph{IRAS}\,18008-2425 (SV Sgr): a K-type FU Ori variable within NGC 6530, the young open cluster associated with the Lagoon Nebula (Messier 8).
\item \emph{IRAS}\,18072-1954: \emph{Spitzer} GLIMPSE and \emph{WISE} imagery shows this to be a star (2MASS\,18101404-1954084) creating an H\,{\sc ii} region inside a dark cloud; at Galactic latitude $b = -1\fdeg32$.
\item \emph{IRAS}\,18155-1206: a knot in a diffuse infrared-bright nebula; at Galactic latitude $b = +1\fdeg70$.
\item \emph{IRAS}\,18288-0207: within the H\,{\sc ii} region W40.
\item \emph{IRAS}\,18585-3701: this source represents a young star cluster (the Coronet Cluster) and associated nebulosity, NGC\,6729. Optically, it mostly represents a blend of the Herbig Ae/Be star R CrA and the F-type star T CrA; the infrared is dominated by the surrounding nebulosity.
\item \emph{IRAS}\,18595+0107: within the H\,{\sc ii} region W48.
\item \emph{IRAS}\,19117+1107: associated with a pair of infrared nebulae with known methanol maser detections; at Galactic latitude $b = +0\fdeg13$.
\emph{IRAS}\,19597+3327A: Source appears to be extended in optical images. \citep{Samal2010} identify it as a massive protostar (their source IRAS-B) within the larger Sharpless 2-100 star-forming region.
\item \emph{IRAS}\,20002+3322: appears associated with a knot in the nebula W\,58.
\item \emph{IRAS}\,20081+2720: part of a nebula; at Galactic latitude $b = -3\fdeg18$.
\item \emph{IRAS}\,20101+3806: part of a nebula with no obvious stellar counterpart.
\emph{IRAS}\,22133+5837 (V653 Cep): two-component fit plus anomalously bright UVEX $U_{\rm GRO}$-band observation. Fitted SED is not consistent with a low-luminosity ($\sim$9\,L$_\odot$) source at the stated distance (5.2 $\pm$ 3.4 kpc). May be related to the nearby (3$\farcm$7) star-forming region containing \emph{IRAS}\,22134+5834.
\item \emph{IRAS}\,22540+6146 (2MASX J22560350+6202554), 22544+6141 and 22548+6147: embedded young stellar objects within the wider star-forming region Cepheus A.
\end{itemize}

\subsection{Objects with unclear classifications}
\label{apx:reject:unknown}

Three objects have unclear classifications. We reject the following two. The third is \emph{IRAS}\,17205-3418, mentioned in the main text.
\begin{itemize}
    \item \emph{IRAS}\,13428-6232 (PM 2-14): this is a complex source at low Galactic latitude ($b = -0\fdeg59$), superimposed on a wider star-forming region, which includes a reflection nebula surrounding V766 Cen and open cluster NGC 5281. It comprises a bipolar outflow with an obscuring torus and has been observed by both \emph{Herschel} \citep{Groenewegen2011,RamosMedina18}, the \emph{Hubble Space Telescope} \citep{Siodmiak08}, and \emph{ISO} (TDT 60600505), the latter showing a rising but featureless spectrum. \citet{Suarez06} identifies this as a post-AGB star with a proto-planetary nebula, and we adopt that designation here. The object is therefore too evolved for this analysis and is rejected.
    \item \emph{IRAS}\,16437-4510: another complex source at low Galactic latitude ($b = -0\fdeg07$). This source is a known OH-IR star \citep{teLintelHekkert91}. It appears to have a counterpart in OH\,340.042-0.092, but this is 1$\farcm$3 from the \emph{IRAS} position \citep{Sevenster97}. \emph{WISE} imagery suggests the \emph{IRAS} detection is a blend of two stars: OH\,340.042-0.092 in the west and an eastern IR-blue star (AllWISE J164719.89-451615.6). The OH-IR star itself is moderately blended further in \emph{WISE} with the red star 2MASS 16473293-4516496. We adopt this object as an AGB/RSG star, hence part of our primary study.
\end{itemize}

\subsection{Objects that are highly evolved}
\label{apx:reject:pne}

In addition, several objects were discovered that are highly evolved objects (post-AGB stars, proto-planetary nebulae, and planetary nebulae themselves). These objects are evolved stars, but have either completed their AGB evolution. They are no longer actively losing mass from their surfaces, though their remaining circumstellar matter has yet to be ejected. Consequently, they do not contribute to the AGB properties and dust budgets examined in this paper, but are listed separately as they can be included in some of the remits of the NESS survey. These sources also have photometry extracted assuming they are point sources, hence the properties extracted in the catalogue accompanying this paper may not be valid if they host extended nebulae. Sources have been checked for extended nebulae, spectral type, or other literature confirmation before removal. These sources are not counted among the evolved stars in this work.

Also in this list are a number of yellow hypergiants. These supernova progenitors are considered too hot for the present study, and the methods used here are not particularly suitable for determining their properties.

\begin{itemize}
    \item \emph{IRAS}\,04395+3601 (V353 Aur; RAFGL\,618; Westbrook Nebula). A bipolar nebula of several arcseconds extent is resolved in PanSTARRS images.
    \item \emph{IRAS}\,05251-1244 (IC\,418; Spirograph Nebula). A well-known elliptical planetary nebula. Central star has spectral type O7fp.
    \item \emph{IRAS}\,06176-1036 (HD 44179; Red Rectangle). A well-known bipolar nebula surrounding a post-AGB star. Spectral type B9Ib/II.
    \item \emph{IRAS}\,08011-3627 (AR Pup). A post-AGB star with an edge-on circumbinary disc that obscures the post-AGB component completely \citep{Ertel2019}.
    \item \emph{IRAS}\,09256-6324 (IW Car). A post-AGB star with a circumbinary hourglass nebula \citep[e.g.][]{Bujarrabal2017}.
    \item \emph{IRAS}\,10197-5750 (MR\,22). A complex bipolar nebula of $\sim$8$^{\prime\prime}$ arcseconds in extent, resolved in PanSTARRS images.
    \item \emph{IRAS}\,13428-6232 (PM 2-14): see above, Section \ref{apx:reject:unknown}.
    \item \emph{IRAS}\,14562-5406 (WRAY 15-1269). A carbon-rich planetary nebula with spectral type [WC11] \citep[e.g.][]{Parthasarathy12}.
    \item \emph{IRAS}\,15445-5449 (OH\,326.5-0.4). A post-AGB star with a small bipolar nebula, visible in the infrared \citep{Lagadec2011}.
    \item \emph{IRAS}\,15452-5459. The nebula is resolved in images and a fast CO outflow has been found \citep{Cerrigone2012}.
    \item \emph{IRAS}\,16133-5151 (Menzel 3; Ant Nebula). A bipolar nebula easily visible on optical imagery.
    \item \emph{IRAS}\,16594-4656 (SS\,293). A small nebula, partially resolved in DES images, with a central star of spectral type Ae \citep{Suarez06}.
    \item \emph{IRAS}\,17103-3702 (NGC 6302, Bug Nebula). A well-known bipolar nebula.
    \item \emph{IRAS}\,17150-3224 (Cotton Candy Nebula). A bipolar nebula inside a spherical halo.
    \item \emph{IRAS}\,17163-3907 (Fried Egg Nebula). A yellow hypergiant exhibiting strong mass loss, e.g., \citep{Wallstrom2017}.
    \item \emph{IRAS}\,17347-3139. A small, bipolar planetary nebula, partly obscured by an overlying star \citep[e.g.][]{Tafoya2009}.
    \item \emph{IRAS}\,17427-3010 (M\,1-26). A small, complex planetary nebula, resolved in \emph{Hubble Space Telescope (HST)} images (programme GO6563)\footnote{See \url{https://faculty.washington.edu/balick/pPNe/} for all \emph{HST} images}.
    \item \emph{IRAS}\,17251-3505 (H\,1-13). An elliptical planetary nebula with a bright torus is resolved in DES images.
    \item \emph{IRAS}\,18450-0148 (W43a). Identified as a proto-planetary nebula \citep{Chong15} or post-common-envelope system \citep{Khouri2022}. This object is invisible in optical and near-infrared images.
    \item \emph{IRAS}\,18458-0213. Classified as a planetary nebula by \citet{Urquhart09}; later confirmed by \citet{Irabor18}. Not visible at optical wavelengths.
    \item \emph{IRAS}\,19244+1115 (IRC+10420). A yellow hypergiant with spectral type A \citep{Koumpia2022}, thus too hot for our criteria.
    \item \emph{IRAS}\,19327+3024 (HD\,184738, Campbell's hydrogen star \citep{Campbell1893}). A [WC] star at the centre of a small planetary nebula.
    \item \emph{IRAS}\,19374+2359. Listed on {\sc simbad} as proto-planetary nebula with spectral type B. \emph{HST} images show a small planetary nebula with complex morphology (programme GO6364).
    \item \emph{IRAS}\,20028+3910. Listed on {\sc simbad} as a proto-planetary nebula with spectral type F. \emph{HST} images show a small, probably bipolar nebula with possible jets \citep[programme GO8210;][]{Hrivnak2001}.
    \item \emph{IRAS}\,20547+0247 (U Equ): a rapidly warming post-AGB star \citep{Kaminski2024}. This previously M-type star showed TiO and VO bands in both absorption {\it and emission}, but now shows a spectral type of $\sim$F6.
    \item \emph{IRAS}\,21282+5050. \emph{HST} images show a small, multipolar planetary nebula (programme GO9463). {\sc Simbad} lists the central star as having spectral type 	O7(f)/[WC11].
    \item \emph{IRAS}\,23541+7031 (M2-56). \emph{HST} images resolve a small bipolar nebula with a larger, much fainter, diffuse structure (programme GO9463).
\end{itemize}

\subsection{Objects not meeting our temperature and luminosity criteria}
\label{apx:reject:criteria}

These objects tend to be more-extreme sources that have erroneous distances, therefore are scattered to luminosities that are too low or too high. The exceptions include some (very) low mass-loss rate sources whose updated distances in \emph{Gaia} DR3 place them closer to the Earth than the \emph{Hipparcos}/\emph{Tycho--Gaia} solution distances, therefore reducing their luminosities below the 700 L$_\odot$ limit.


\subsubsection{Hot sources ($T > 5000$\,K)}

Tier 4 (``extreme'' mass-loss rate) sources:
\begin{itemize}
\item \emph{IRAS}\,16383-4626 (OH 338.5-00.2; also sub-luminous): SED shows hot and cold components. Unclear in optical imagery whether the two components are physically associated.
\end{itemize}

\subsubsection{Under-luminous sources ($T < 5000$\,K, $L < 700$\,L$_\odot$)}

Tier 0 (``very low'' mass-loss rate) sources:
\begin{itemize}
\item \emph{IRAS}\,01261-4334 ($\gamma$ Phe): spectroscopic binary, classified as K4--M0III by various authors.
\item \emph{IRAS}\,07276-4311 ($\sigma$ Pup): classified K5 or M0.
\item \emph{IRAS}\,10193+4145 ($\mu$ UMa): spectroscopic binary, K5 or M0. Parallax substantially higher in \emph{Gaia} (17.80 $\pm$ 0.39 mas) than \emph{Hipparcos} (14.16 $\pm$ 0.54 mas), reducing inferred luminosity.
\item \emph{IRAS}\,15186-3604 ($\phi_1$ Lup): high proper motion, classified K4 or K5.
\end{itemize}

\noindent
Tier 1 (``low'' mass-loss rate) sources:
\begin{itemize}
\item \emph{IRAS}\,03479-7423 ($\gamma$ Hyi)
\item \emph{IRAS}\,03557-1339 ($\gamma$ Eri)
\item \emph{IRAS}\,05217-3943 (SW Col)
\item \emph{IRAS}\,05271-0107 (31 Ori): K4 spectral type.
\item \emph{IRAS}\,16117-0334 ($\delta$ Oph)
\item \emph{IRAS}\,18142-3646 ($\eta$ Sgr)
\item \emph{IRAS}\,19320-5307 (HD\,184192)
\end{itemize}

\noindent
Tier 3 (``high'' mass-loss rate) sources:
\begin{itemize}
\item \emph{IRAS}\,00193-4033 (BE Phe)
\item \emph{IRAS}\,05405+3240 (RAFGL 809, carbon star)
\item \emph{IRAS}\,20570+2714 (RAFGL 2686, carbon star)
\end{itemize}

\noindent
Tier 4 (``extreme'' mass-loss rate) sources:
\begin{itemize}
\item \emph{IRAS}\,16280-4154: poor fit to extremely red source. Crowded field. Distance of 348 $\pm$ 113 pc from \citet{BJRF+21}, based on a \emph{Gaia} parallax of 3.75 $\pm$ 0.66 mas, may be an under-estimate if this source is truly an OH/IR star.
\item \emph{IRAS}\,17121-3915: SED is not well-represented by a blackbody or stellar model. Large (44 per cent) distance uncertainty.
\item \emph{IRAS}\,17128-3748 (V1013 Sco): SED shows hot and cool components. The {\sc PySSED} fit applies to the hotter component.
\item \emph{IRAS}\,19178-2620 (RAFGL 2370): poor fit to extremely red source. Parallax appears reasonable.
\end{itemize}

\subsubsection{Over-luminous sources ($T < 5000$\,K, $L > 200\,000$\,L$_\odot$)}

Tier 2 (``intermediate'' mass-loss rate) sources:
\begin{itemize}
\item \emph{IRAS}\,21419+5832 ($\mu$ Cep): red supergiant, but with a luminosity over-estimated by a factor of $\sim$10. This is likely due to an over-estimated distance (2223 pc), which is based on the weighted average of a distance from \citet{BJRF+21} (4496 $\pm$ 1567 pc) and \emph{Hipparcos} (1818 $\pm$ 661 pc). Note that the \emph{Gaia} parallax (0.12 $\pm$ 0.26 mas) is consistent with zero and marginally inconsistent with the much larger \emph{Hipparcos} parallax (0.55 $\pm$ 0.20 mas).
\end{itemize}

\noindent
Tier 3 (``high'' mass-loss rate) sources:
\begin{itemize}
\item \emph{IRAS}\,02192+5821 (S Per)
\item \emph{IRAS}\,11145-6534 (V832 Car, carbon star)
\item \emph{IRAS}\,11179-6458 (V538 Car, red supergiant)
\item \emph{IRAS}\,12233-5920 (EN Cru)
\item \emph{IRAS}\,13436-6220 (HD 119796, yellow supergiant)
\item \emph{IRAS}\,15576-5400 (HD 143183, red supergiant)
\item \emph{IRAS}\,16340-4634 (OH\,337.9+00.3)
\item \emph{IRAS}\,17104-3146 (IRC\,-30285)
\item \emph{IRAS}\,17163-3835
\item \emph{IRAS}\,17328-3327 (red supergiant)
\item \emph{IRAS}\,17393-3004 (IRC\,-30316, OH/IR star): parallax is very uncertain.
\item \emph{IRAS}\,17485-2209 (IRC\,-20394)
\item \emph{IRAS}\,18050-2213 (VX Sgr)
\item \emph{IRAS}\,19007-3826 (RAFGL 5553)
\end{itemize}

\noindent
Tier 4 (``extreme'' mass-loss rate) sources:
\begin{itemize}
\item \emph{IRAS}\,17327-3319: parallax consistent with zero. Significant uncertainty in \citet{BJRF+21} distance.
\item \emph{IRAS}\,19422+3506 (RAFGL 2445): parallax negative with 4.2 $\sigma$ significance. Significant uncertainty in \citet{BJRF+21} distance.
\end{itemize}


\section{AGB stars that have complex requirements}
\label{apx:reject:complex}

Several NESS sources are retained in this analysis, but are noted as having complex requirements for data extraction and/or subsequent analysis. The following sources have no distances (other than those based on GRAMS and/or $P$--$L$ relations) so are present in only the unrestricted dataset:
\begin{itemize}
    \item \emph{IRAS}\,16055--4621: this object is in a crowded field, offset from its \emph{IRAS} position by 20$^{\prime\prime}$. The \emph{IRAS} Low-Resolution Spectrograph (LRS) spectrum indicates it is an O-rich AGB star. No parallax or proper motion is available but the star is expected to be a luminous OH/IR star at $\sim$1\,kpc distance.
    \item \emph{IRAS}\,16440--4518 (OH\,339.974--0.192): AllWISE resolves this into two sources, both of which are likely associated with the same point source. The region exhibits variable extinction, and there is a nearby bright optical star, requiring careful photometric extraction.
    \item \emph{IRAS}\,17205--3418 has two nearby (10$^{\prime\prime}$) blends with AllWISE J172349.32--342103.0 and 2MASS 17235091--3421064. It is listed as a variable star at 8 $\mu$m and an AGB candidate in the GLIMPSE survey (G352.9382+00.9606) by \citet{Robitaille08}. However, it also lies within the ATLASGAL infrared dark clump AGAL G352.9413+0.9606 \citep{Csengeri14} and on a similar line-of-sight to known YSOs. The object is very faint in 2MASS and optically hidden behind a reflection nebula. \emph{NEOWISE}-R (the \emph{WISE} satellite warm mission reactivation; \citealt{Mainzer2014}) observations exist within 5$^{\prime\prime}$ of the \emph{IRAS} and \emph{AllWISE} positions. Visual inspection of the light curves does not reveal a strong time dependence, and the range of variation is relatively small ($\sim$0.3 mag).
    \item \emph{IRAS}\,17411--3154 (RAFGL\,5379; OH\,357.311--1.337): this object is clearly visible in \emph{WISE} and \emph{Spitzer} imagery, but the near-infrared and optical counterpart of this object is hidden in the very dense star field, which lies in the Galactic Bulge. A 2MASS source is offset from the \emph{WISE} and \emph{Spitzer} positions by 4$\farcs$5, but appears to represent a different star. The \emph{ISO}/SWS spectrum (TDT 84300128) indicates an oxygen-rich AGB star. This source also has data from the APEX ATLASGAL survey \citep{Schuller09} and several observations by the \emph{Herschel Space Observatory} \citep[e.g.][]{RamosMedina2018}. Its position was manually extracted at 17$^h$44$^m$23$^s.$92192 --31$^\circ$55$^\prime$39$\farcs$5125.
    \item \emph{IRAS}\,18009--2407: an obvious error in cross-identification was found in the \citet{Abrahamyan15} catalogue, which mistakenly links to \emph{IRAS}\,07240--2532 instead. This source was manually matched via the 2MASS source linked in {\sc simbad} (2MASS\,J18040106--2407083).
\end{itemize}
The following sources are also present in the restricted dataset:
\begin{itemize}
    \item \emph{IRAS}\,06027--1628 (17 Lep, SS Lep): a symbiotic binary. The M-type star (1200\,L$_\odot$, 3250\,K) is within our selection criteria, but is transferring mass to a bright (1900\,L$_\odot$, 9000\,K) A-type companion \citep{Verhoelst2007}.
    \item \emph{IRAS}\,17328--3327 (CD--33 12241): a red supergiant star in the cluster Trumpler 27, requiring careful photometric extraction.
    \item \emph{$\lambda$ Vel}: this object was included specially because it met the criteria for tier 0, but it does not exist in the \emph{IRAS} point source catalogue due to its comparative mid-IR faintness. It has a detection in the \emph{IRAS} reject catalogue (\emph{IRAS}\,R09061--4313).
\end{itemize}

\section{Possible changes to NESS tiers as a result of revised distances}
\label{apx:newtier}

\subsection{Completion of NESS Tier 0}

The 300 pc sample reveals five candidate additions to NESS Tier 0 (volume limit 250 pc), which were missed due the way in which the sample was built to cover data gaps in both \citet{MZB12} and \citet{MZW17}:
\begin{itemize}
\item $\eta_2$ Dor (\emph{IRAS}\,06111--6534) had no valid fit in \citet{MZB12} but is properly fitted here.
\item V913 Cen (\emph{IRAS}\,11352--6037) and GM Lup (\emph{IRAS}\,15014--4040) have a luminosity in \citet{MZW17} of $L < 1600$\,L$_\odot$ but now have a greater luminosity due to revised distances and photometry;
\item NO Aps (\emph{IRAS}\,17220--8049) had an assigned distance in \citet{MZW17} of $d > 250$ pc, while BQ Tuc (\emph{IRAS}\,00515--6308) had $d > 250$\,pc in \citet{MZB12}, but both now have distances in \citet{BJRF+21} below 250\,pc.
\end{itemize}

\subsection{Completion of NESS Tier 1}

There are also 327 candidate additions to NESS Tier 1, based on the unrestricted dataset. Whether a source should be included in Tier 1 also depends on its dust production. Of the 327 sources, only five have infrared excess as defined by a presence in Table 3 of \citet{MZW17}: HD\,112278, RR CrB, V2113 Oph, V568 Lyr and V1070 Cyg. The luminosities of these sources in \citet{MZW17} are below the RGB tip, so were considered too faint to include, and their infrared excess was not identified from \emph{IRAS} photometry alone.

\subsection{Possible exclusions from NESS due to refined distances}

The new distances (from the unrestricted catalogue) also remove some objects from the categories used to define the NESS tiers, as they are now at a distance greater than the tier boundary:
\begin{itemize}
\item Tier 0: \emph{IRAS}\,00254--1156, 11098--5809 and 16520--4501, of which the latter would move up to Tier 2.
\item Tier 1: \emph{IRAS}\,00084--1851, 16469--3412, 16520--4501, 19098+6601 and 20141--2128, which would move up to Tier 2, and \emph{IRAS}\,05254+6301 and 12319--6728, which would not.
\item Tier 2: 19 sources to be removed (no sources would be moved to Tier 3).
\item Tier 3: 120 sources to be removed, of which three sources (\emph{IRAS}\,08124--4133, 09429--2148 and 17328--3327) move into Tier 4.
\item Tier 4: 47 sources to be removed.
\end{itemize}

Accounting for changes to $\dot{D}$ removes the following sources (the adjustments are not applicable to tiers 0 and 1):
\begin{itemize}
    \item Tier 2: six sources (\emph{IRAS}\,05028+0106, 11164--5754, 17123+1426, 17553+4521, 18157+1757 and 21399+3516) to be removed. Of these, \emph{IRAS}\,17553+4521 would be retained in Tier 1.
    \item Tier 3: 52 sources to be removed, of which 25 sources are moved down into lower tiers.
    \item Tier 4: seven sources (\emph{IRAS}\,03149+3244, 10481--6930, 13517--6515, 14119--6453, 16280--4154, 19178--2620 and 19396+1637) to be removed, of which all but 10481--6930 and 14119--6453 are moved down into Tier 3.
\end{itemize}


\section{Using the digitised information}
\label{apx:data}

\subsection{Overview}

A ZIP file containing digital Supplementary Material is provided with this paper, which will recreate the files used to prepare this paper. To recreate the {\sc PySSED} output files, you will need:
\begin{itemize}
    \item {\sc PySSED} version 1.1 from \url{https://github.com/iain-mcdonald/PySSED}.
    \item A {\sc Python} 3 installation with the {\sc PySSED} pre-requisites (see Manual or run {\tt pyssed.py} with no arguments).
    \item Items in the {\tt inputs/} folder of the ZIP file should be placed in the {\tt src/} directory.
    \item Items in the {\tt inputdata/} folder should be placed in a {\tt data/} directory.
    \item The directories {\tt 300pc-*} and {\tt ness-*} contain the full {\sc PySSED} outputs. The second line of the contained {\tt output.dat} files indicates the input command required for {\sc PySSED}.
\end{itemize}

To recreate the analysis output files and plots from the ZIP file, you will need to run {\tt POSTPROC-public.bash}. Plots also require an installation of {\sc Python} 3 and {\sc Gnuplot} (v.\ 5 or higher). This also recreates the following Supplementary Tables in tab-separated form.

\subsection{Supplementary Tables}

The tab-separated versions of the Supplementary Tables can also be downloaded from the journal site. With the exception of the list in Table G2, each file contains a header prompt describing the column contents. Full descriptions of each table column are listed below.

\renewcommand{\labelenumi}{(\arabic{enumi})}
\begin{itemize}
    \item {\it Table G1:} Cross-identifiers for NESS sources.
        \begin{enumerate}
            \item \emph{IRAS} identifier,
            \item Right Ascension (degrees),
            \item Declination (degrees),
            \item {\sc simbad} source identifier.
        \end{enumerate}
    \item {\it Table G2:} List of input stars to the 300 pc sample.
    \item {\it Table G3:} Distances to NESS stars from different estimation methods.
        \begin{enumerate}
            \item \emph{IRAS} identifier,
            \item Parallactic distance (pc),
            \item Parallactic distance error (pc),
            \item Bolometric-luminosity distance (pc),
            \item Bolometric-luminosity distance error (pc),
            \item Period--luminosity distance (pc),
            \item Period--luminosity distance error (pc),
            \item NESS tier.
        \end{enumerate}
    \item {\it Table G4:} Comparison of parameter estimation from model atmosphere versus blackbody versus trapezoidal integration with goodness of fit for NESS sources.
        \begin{enumerate}
            \item \emph{IRAS} identifier,
            \item {\sc simbad} source identifier,
            \item $T_{\rm eff}$ (stellar atmosphere model, K),
            \item $L$  (stellar atmosphere model, L$_\odot$),
            \item $T_{\rm eff}$ (blackbody, K),
            \item $L$  (blackbody, L$_\odot$),
            \item $L$  (trapezoidal integration, L$_\odot$),
            \item Adopted distance (pc),
            \item NESS tier,
            \item $\dot{D}$ (M$_\odot$\,yr$^{-1}$)
            \item $GOF$ (stellar atmosphere model),
            \item $GOF$ (blackbody),
            \item Selection (model or blackbody).
        \end{enumerate}
    \item {\it Table G5:} Table of final parameters for the 300 pc sample.
        \begin{enumerate}
            \item {\sc simbad} source identifier,
            \item Right Ascension (degrees),
            \item Declination (degrees),
            \item $T_{\rm eff}$ (K)
            \item $L$  (L$_\odot$),
            \item Adopted distance (pc).
        \end{enumerate}
    \item {\it Table G6:} Table of final parameters for NESS sources.
        \begin{enumerate}
            \item \emph{IRAS} identifier,
            \item {\sc simbad} source identifier,
            \item $T_{\rm eff}$ (K)
            \item $L$  (L$_\odot$),
            \item Adopted distance (pc),
            \item NESS tier,
            \item $\dot{D}$ (M$_\odot$\,yr$^{-1}$)
            \item Selection (model or blackbody).
        \end{enumerate}
    \item {\it Table G7:} Revised tiers and DPRs for NESS sources.
        \begin{enumerate}
            \item \emph{IRAS} identifier,
            \item {\sc simbad} source identifier,
            \item $T_{\rm eff}$ (K)
            \item $L$  (L$_\odot$),
            \item original NESS tier,
            \item $\dot{D}$ (M$_\odot$\,yr$^{-1}$)
            \item original distance from \citet{Scicluna22} (pc),
            \item revised NESS tier,
            \item revised $\dot{D}$ (M$_\odot$\,yr$^{-1}$),
            \item revised distance (pc).
        \end{enumerate}
\end{itemize}
\renewcommand{\labelenumi}{(\roman{enumi})}



\bsp	
\label{lastpage}
\end{document}